\documentclass[11pt,letter]{article} 
\usepackage{jheppub} %

\usepackage{multirow}
\usepackage{hhline}

\usepackage[ps,dvips,matrix,arrow,frame,import,curve,color]{xy}

\usepackage[caption=false]{subfig}

\def\blfootnote{\xdef\@thefnmark{}\@footnotetext}

\long\def\symbolfootnote[#1]#2{\begingroup%
\def\thefootnote{\fnsymbol{footnote}}\footnote[#1]{#2}\endgroup}

\newcommand{\be}{\begin{eqnarray}}
\newcommand{\ee}{\end{eqnarray}}
\newcommand{\ben}{\begin{eqnarray*}}
\newcommand{\een}{\end{eqnarray*}}

\newcommand{\bcent}{\begin{center}}
\newcommand{\ecent}{\end{center}}
\newcommand{\benum}{\begin{enumerate}}
\newcommand{\eenum}{\end{enumerate}}
\newcommand{\bdesc}{\begin{description}}
\newcommand{\edesc}{\end{description}}
\newcommand{\bitem}{\begin{itemize}}
\newcommand{\eitem}{\end{itemize}}
\newcommand{\bquote}{\begin{quote}}
\newcommand{\equote}{\end{quote}}
\newcommand{\bhalfp}{\begin{minipage}{0.45\textwidth}}
\newcommand{\ehalfp}{\end{minipage}}
\newcommand{\bhead}{\begin{center}\bf \Large}
\newcommand{\ehead}{\end{center}\bigskip}

 \newcommand{\del}{{\partial}}
 
\def\be{\begin{equation}}
\def\ee{\end{equation}}
\def\ba{\begin{eqnarray}}
\def\ea{\end{eqnarray}}

\newcommand{\roughly}[1]{\mathrel{\raise.3ex\hbox{$#1$\kern-0.85em
\lower1ex\hbox{$\sim$}}}}

\def\2pi{\left(2\pi\right)}

\def\beq{\begin{equation}}
\def\eeq{\end{equation}}
\def\bg{\begin{eqnarray}}
\def\nd{\end{eqnarray}}
\def\bea{\begin{eqnarray}}
\def\eea{\end{eqnarray}}

\def\D3{\overline{\mbox{D3}}}

\preprint{NSF-KITP-13-038, LTH-973}

\title{Gauge/Gravity Duality in Heterotic String Theory} 

\author{Fang Chen${}^1$,}
\author{Keshav Dasgupta${}^2$,}
\author{Joshua M. Lapan${}^2$,}
\author{Jihye Seo${}^{2, 3}$,}
\author{and Radu Tatar${}^4$}
\affiliation{\vskip 0.15cm ${}^1$KITP, University of California, Santa Barbara,
CA 93106-4030, USA}
\affiliation{${}^2$Physics Department, McGill University,
3600 University St, Montr{\'e}al, QC H3A 2T8, Canada}
\affiliation{${}^3$CRM, Universit\'{e} de Montr\'{e}al, C.P. 6128, succ. centre-ville, Montr\'{e}al, QC H3C 3J7, Canada}
\affiliation{${}^4$Department of Mathematical Sciences, Liverpool University, Liverpool, L69 7ZL, UK \vskip 0.15cm}
\emailAdd{fchen@kitp.ucsb.edu, Radu.Tatar@Liverpool.ac.uk}
\emailAdd{\qquad\quad\, keshav, jlapan, \& jihyeseo @ hep.physics.mcgill.ca}

\abstract{Gravity duals for little string theories --- which give rise to four-dimensional theories that undergo permanent confinement in the infrared --- 
have not been studied in great detail.  We address this question in the framework of 
heterotic $SO(32)$ and $E_8 \times E_8$ string theory, constructing these backgrounds by wrapping heterotic five-branes on calibrated two-cycles of non-K\"ahler 
resolved conifolds.  Related to deformations of the 
underlying little string theories, we find numerous analytic solutions preserving
${\cal N} = 1$ supersymmetry in four-dimensions. 
These theories all have non-abelian
global symmetries that generally arise from both the 
heterotic vector bundle and from certain orbifold states.  In the decoupling limit, we argue that the gravity 
duals are 
given by non-K\"ahler manifolds that have both blown-up two-cycles and three-cycles at the origin. 
We argue this following certain duality sequences that include 
M-theory torsional manifolds at an intermediate step, which help us to construct new type ${\rm I}'$ gauge/gravity duality pairs. 
In the M-theory duality frame, we also elucidate new sequences of flips and flops.}

\begin{document}
\maketitle
\flushbottom
 \toccontinuoustrue

\section{Introduction}

Geometric transitions have long been exploited to explore strong-coupling regimes of 
field theories, beginning with \cite{vafaGT} in the context of type II string theories where powerful topological string methods were available.  The basic setup is to wrap branes on compact two- or three-cycles of Calabi--Yau threefolds; in the limit where the number of branes is very large, we can shrink the wrapped cycle until it vanishes so that the manifold develops a singularity.  That singularity is then resolved (or deformed) via an extremal transition, where a different vanishing cycle grows and where the branes are replaced by fluxes that thread a dual cycle.  In terms of the gauge theory living on the original branes, the size of the new cycle is determined by strong-coupling effects, such as gluino condensation.

A natural question to ask is how this generalizes to heterotic string theory where, in principle, one could have worldsheet descriptions of both sides of the duality.  However, there are a few immediate 
hurdles that must be overcome.
The first is that the
gauge theory side, described by little string theory (LST), is quite mysterious.  This is in contrast to the type II case, where the gauge theories are simply given by SYM theories in various dimensions with well-understood dynamics. 
The second hurdle is on the gravity side, where we generally expect the manifolds to have non-K\"ahler metrics. For these kinds of manifolds,
it is much more difficult to solve the supersymmetry conditions and Bianchi identity, as well as to single out the relevant cycles.  Finally, while in the type II cases we
could gain more information by using topological methods, the heterotic theory generally only admits a ``half-twist'' which is not nearly as powerful.

Nevertheless, some recent progress has been made in a series of papers \cite{chen1,chen2}, as well as \cite{caris}, starting with the program initiated in \cite{anke1}.  In fact, the first hurdle regarding the mysterious gauge theories can actually be turned to our advantage: having a gravity dual to a more mysterious LST could be a fruitful way to extract information about these theories.  In earlier works of some of the present authors \cite{chen1, chen2}, we adopted this viewpoint and argued for the 
gravity duals using various consistency arguments stemming from the duality cycle advertised in \cite{anke1,anke2,anke3}.  These consistency arguments also helped us overcome the second hurdle, finding non-K\"ahler backgrounds, without the need to appeal to half-twisted topological theories.  However, no {\it direct} proof of the gauge/gravity duality was provided in \cite{chen1, chen2} and, furthermore, most of 
 the analysis focussed on the $SO(32)$ heterotic string.

In the current work, we will begin to address these shortcomings in a more unified approach, setting up a geometric transition route in the $E_8 \times E_8$ heterotic theory. Our  construction 
will exploit the F-theory and the $E_8\times E_8$ heterotic duality by using D3-branes to probe a system of
two orthogonal sets of D7/O7 branes/orientifolds that wrap 
a ${\bf P}^1$. This additional ${\bf P}^1$ makes the issue of supersymmetry much more subtle since we need to switch on additional fluxes to preserve supersymmetry. 
In sections \ref{sec:het}, we show
that these fluxes force the heterotic geometries to be non-K\"ahler resolved conifolds, with the 
NS5-branes wrapping calibrated two-cycles.  In section \ref{sec:hetslns}, we determine metric, three-form, and dilaton, presenting three distinct cases that correspond to different configurations of wrapped NS5-branes.  In section \ref{sec:hetbundle}, we present simple non-abelian vector 
bundles that, together with the other fields, preserve supersymmetry and satisfy the Bianchi identity.  We also discuss the
asymptotic behavior and range of applicability of these solutions.
To our knowledge, this is the first time most of these backgrounds have
been determined for the $E_8\times E_8$ theory. 
   
Through the course of section \ref{sec:transition}, we argue for the gravity duals by following a duality chain that incorporates type I, type I$^\prime$, and M-theory.  Compared to some of our earlier works \cite{chen1,chen2}, this is a more direct way to integrate these theories into a single duality chain.  
Using this chain, we obtain two distinct limits: a non-K\"ahler
resolved conifold and a non-K\"ahler deformed conifold.    For the present work,
we focus on the resolved conifold side.  The LSTs have global degrees of freedom, and we argue that they reside {\it both} in the heterotic vector 
bundle as well as in certain orbifold states.  Of course, vector bundle degrees of freedom can also be viewed as small instantons in certain limits, which are intimately connected to singularities in the torsion and, hence, NS5-branes \cite{Witten:1995gx}.  In all three cases that we study in section \ref{sec:hetslns}, the torsion asymptotes to the constant value required to maintain the non-K\"ahlerity, as well as zero potential \eqref{potcar}, 
of the resolved conifold backgrounds --- this is shown in section \ref{sec:torsion}.
In the dual gravity description, we don't expect these singularities to be present, which we know can happen if the small instantons dissociate into smooth gauge flux.
This raises the intriguing possibility that the gauge/gravity duality in the heterotic theory is directly related to a small instanton transition.  We discuss this further in sections \ref{sec:torsion} and \ref{sec:hetgravity}, where we also show how to look for the confining strings and new exotic states in the dual LST.

\section{Heterotic and Little String Theory Background}

As shown in \cite{stromtor}, the supersymmetric variations
in heterotic string theory allow for a complex but non-K\"ahler internal manifold, satisfying:
\begin{eqnarray}\label{BE}
&& d^\dag J ~=~ i(\partial-\bar{\partial})\ln \parallel\Omega\parallel\nonumber\\
&& J^{a\bar{b}}F_{a\bar{b}}~ =~ F_{ab} ~= ~ F_{\bar{a}\bar{b}} = 0\nonumber\\
&& i\partial\bar{\partial}J ~=~ \alpha'\left[\textrm{tr} (R_+\wedge R_+) - \frac{1}{30}\textrm{tr} (F\wedge F)\right]
\end{eqnarray}
where $J$ is the fundamental $(1, 1)$ hermitian form on the internal manifold, $\Omega$
is the holomorphic $(3,0)$-form, $F$ is the $SO(32)$ or $E_8\times E_8$
gauge field strength, and $R_+$ is the Ricci two-form constructed from the $\omega_+$ connection: $\omega_{+\mu}^{ij} \equiv \Gamma_\mu^{ij} + \frac{1}{2}{\cal H}_\mu^{ij}$, where 
${\cal H}_\mu^{ij} \equiv {\cal H}_{\mu\nu\rho}e^{\nu i}e^{\rho j}$ and $e^{\nu i}$ is the vielbein. 
The NS three-form flux ${\cal H}$ and dilaton $\phi$ are related via:
\begin{eqnarray}
{\cal H}~=~ -e^{2\phi}\ast d\left(e^{-2\phi}J\right) ~ = ~ i(\bar{\partial}-\partial)J \ ,
\end{eqnarray}
so the final equation in \eqref{BE} is the Bianchi identity for ${\cal H}$.   Heterotic NS5-branes
appear on the right-hand side of the Bianchi identity in \eqref{BE}, as delta-function contributions to $d{\cal H}$.  

In \cite{Witten:1995gx}, six-dimensional theories with ${\cal{N}}=1$ supersymmetry, arising from the $SO(32)$ heterotic theory compactified on K3, were studied.  There it was argued that when $k$ small gauge theory instantons on K3 shrink to zero size, they become a stack of $k$ NS5-branes, emerging from the vector bundle mist.\footnote{This can also be seen from a sigma model analysis of the background, see \cite{Douglas:1996uz}.}  In the Bianchi identity \eqref{BE}, this corresponds to tuning vector bundle moduli so that ${\rm tr}(F\wedge F)$ develops delta-function contributions, arising from ``small instantons'' that shrink to zero size, that can be reinterpreted as source terms for NS5-branes.
It was also argued there that there is an $Sp(2k)$ gauge group the lives on the $k$ NS5-branes.
This was generalized to the $E_8 \times E_8$ heterotic theory in
\cite{Ganor:1996mu}, where it was argued that similar zero-size instantons give rise to LSTs. This was further analyzed in
\cite{Seiberg:1996vs}, where the author studied a broad class of six-dimensional theories with ${\cal{N}}=1$ superconformal symmetry --- which could arise, for example, from compactification of the $E_8 \times E_8$ heterotic theory on K3 --- and found singularities arising from tensionless strings at boundaries between phases with different instanton numbers.\footnote{For an early history and progress report on LSTs, see the excellent review \cite{Aharony:1999ks}.}

When compactified on tori, LSTs enjoy T-duality symmetry, arising from the fact that they are non-local theories 
\cite{LST}.\footnote{This is different from D-branes, which have local QFTs because the LSTs, being given by the NS5-branes, do not change dimension under 
longitudinal T-dualities. Since the worldvolume theories just shifts from one description to another, they don't have a well-defined 
energy-momentum tensor \cite{Seiberg:1996vs}.}
They also have no gravitational degrees of freedom.
Most familiar examples of LSTs are obtained by studying the dynamics of multiple five-branes in various limits of string theory. A LST is labeled by $N$, the number of five-branes, but it does not become weakly coupled at large $N$, i.e., there is no $1/N$ expansion \cite{Giveon:1999px}.   LSTs can also arise from the dynamics of string theory at a singularity; for example, they can arise in type IIA and IIB theories on an orbifold singularity $\mathbb{C}^2/\Gamma_G$, as reviewed in \cite{Aharony:1999ks,Intriligator:1997dh}.

Holographic duals for LSTs from multiple coincident NS5-branes in type IIA were studied in \cite{Aharony:1998ub}. Along these lines, in
\cite{Gremm:1999hm} the authors found the holographic dual of a single heterotic NS5-brane.  In \cite{Dorey:2004iq}, 
another LST was studied using a technique called deconstruction, obtaining a six-dimensional LST on a torus as a special limit of four-dimensional gauge theory.  There they also briefly discussed confinement, but a more detailed study of confinement is still 
lacking.\footnote{The theories that we are most interested in are the LSTs compactified on a two-sphere, giving rise to four-dimensional theories below a certain 
energy scale.  In this paper, we will be particularly interested in the confining behavior in the deep infrared, so the
UV completion will be beyond the scope of our discussions.}

In the rest of our paper, we consider similar setups, but without studying the LSTs directly.  Instead, we aim to
utilize geometric transitions in heterotic string theory to shed light on LSTs.  The usual problem with LSTs is that they 
cannot be treated on the same footing as many of the systems studied using AdS/CFT techniques.
On the other hand, geometric transitions work predominantly in non-conformal cases, 
so they can be used to study strong-coupling effects in setups that are not covered by AdS/CFT.  Our goal, then, is twofold: one is
to obtain a large set of supergravity solutions corresponding to wrapped NS5-branes in $E_8 \times E_8$ heterotic string theory, and two is to conjecture 
possible gravity duals of these theories, obtained via geometric transitions.

\section{Type II Duality Frame}

\begin{figure}[htb]
\begin{center}
\includegraphics[width=\textwidth]{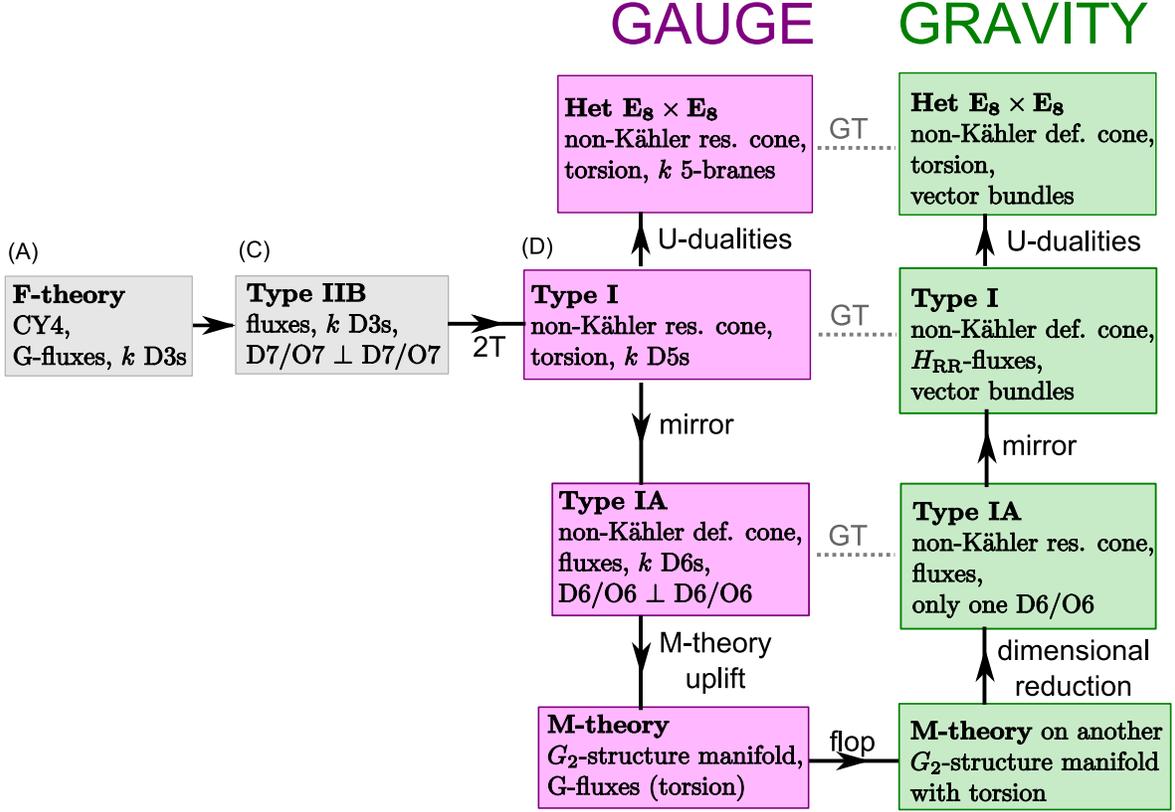}   \end{center}
\caption{The duality sequence that will be used in this paper to argue for the gravity duals in heterotic, type I and M-theories. The boxes in red 
in the left-hand column are related
to the theories on the branes, i.e., the strongly-coupled gauge theories in the IR, except for the M-theory case. 
The boxes in green in the right-hand column are the possible gravity duals.
F-theory origins of these theories are depicted by the boxes in white. The alphabetical ordering of the left-most boxes are 
related to a refinement of the parameter regimes of the theories, described in {\bf figure \ref{varyTau}} below.}
        \label{hetGT}
        \end{figure}

\begin{figure}[htb]
\begin{center}
\includegraphics[width=\textwidth]{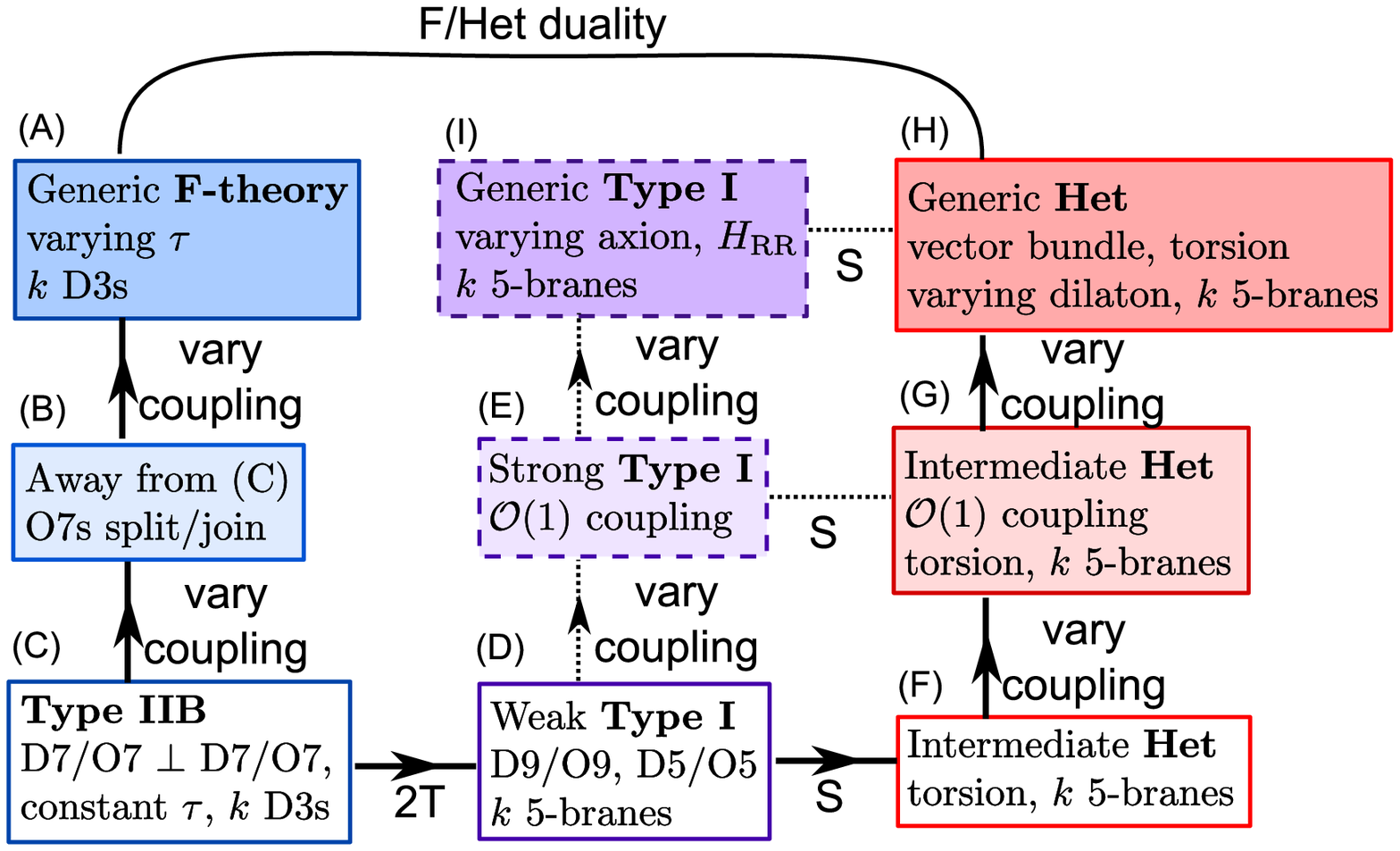}   \end{center}
\caption{This flow diagram depicts the various refinements and regimes of interest of the theories that we study here. The duality that we study in the early
sections of the paper are related to the boxes marked A, C, and D. These are the regions where the orientifold descriptions are most useful. The heterotic analysis 
that we perform later in the text are related to the boxes marked H and G. Other interconnections between the theories are depicted by arrows.}
        \label{varyTau}
        \end{figure}

Our aim  is to extend five-brane gauge/gravity duality to the $E_8\times E_8$ heterotic theory. 
In section \ref{sec:het}, we will return to the heterotic setting, but for now we begin in type II where we have a better handle on gauge/gravity duality.  We begin with a large number of five-branes wrapped on a calibrated two-cycle of a non-K\"ahler resolved conifold.  The duality chain that we will follow is depicted in {\bf figure \ref{hetGT}}.  The solutions that we obtain using the duality chain will have ranges of validity that we need to keep track of, so some of these details are illustrated in {\bf figure \ref{varyTau}}. 

The resolved and the deformed conifold that we will consider support non-K\"ahler metrics that will be relevant to our work. The main feature we want to highlight is the presentation of them as ${\bf P}^1$ fibrations over an ALE space, so we describe the geometry next.

\subsection{The resolved and deformed conifolds, revisited}

The standard deformed conifold, as embedded in $\mathbb{C}^4$, has equation
\bg\label{deco}
z_1^2 + z_2^2 + z_3^2 + z_4^2 = \mu^2
\nd
for any $\mu \in \mathbb{C}^*$.  For fixed $\mu$, this can be viewed as a fibration over the $z_4$ plane, with fiber
\bg\label{fibrose}
z_1^2 + z_2^2 + z_3^2 = \mu^2 - z_4^2 \, .
\nd
If $z_4$ is not $\pm\mu$, this fiber is a noncompact smooth
complex surface, a K3 surface in the sense that its canonical bundle is
trivial.  If $z_4$ equals $\pm\mu$, the fiber is an ALE space, a
singular K3.  In other words, this is essentially the same fibration as for the conifold case, with the parameter values altered.  This realizes the conifold as an ALE fibration over ${\bf P}^1$.\footnote{We thank Sheldon Katz for explaining these details to us.}  
In our case, as we will momentarily see, it will be more useful to instead view the warped deformed and warped resolved conifolds as ${\bf P}^1$ fibrations over a warped ALE space.

First let us show this for the warped deformed conifold.  The metric
of the warped deformed conifold can be written in the following way:
\bg\label{defconw}
ds^2 =  ds^2_{\rm ALE} + d{s}^2_{\rm fiber}  \ ,
\nd
where the metrics of the ALE base and two-sphere fiber are given by
\bg\label{shuor}
ds^2_{\rm ALE} = && f_1 dr^2 + f_2(d\psi + \cos \theta_1 d\phi_1)^2 + \alpha_3 (d\psi + \cos \theta_1 d\phi_1) \nonumber\\
&& + f_3(d\theta_1 + \epsilon_1)^2 + f_3(\sin \theta_1 d\phi_1 + \epsilon_2)^2\nonumber\\
d{s}^2_{\rm fiber}  =  && \left(f_3 - {f_4^2\over 4f_3}\right)\left(d\theta_2^2 + \alpha_4 \sin^2\theta_2 d\phi^2_2\right)
\nd
where we have defined various variables, $f_i$, $\epsilon_i$, and $\alpha_i$, to be:
\bg\label{fieiai}
&&f_1 = {r^2\hat\gamma'-\hat\gamma\over r^2} + {r^2 \, \hat\gamma\over r^4-\mu^4}, ~~~~~~~~ f_2 = \frac{r^4-\mu^4}{4r^2}f_1,   ~~~~~~~~ f_3 = {\hat\gamma\over 4}, ~~~~~~~~ f_4 =  \frac{2\mu^2}{r^2} f_3, \nonumber\\
&&  \epsilon_{1} = {f_4\over 2f_3} ( \cos \psi \, d\theta_2 + \sin\psi \, \sin \theta_2\, d\phi_2 ) ,   ~~~~~~~~~  \epsilon_{2} = {f_4\over 2f_3} ( \sin\psi \, d\theta_2 - \cos\psi\,\sin \theta_2\, d\phi_2 ) ,   \nonumber\\
&& \alpha_3 = 2\, \alpha(\theta_1, \phi_1, \psi) \, f_2 \cos 
\theta_2\,  d\phi_2, ~~~~~~~~~~~~ \alpha_4 = 1 + {4f_2 f_3\over 4f_3^2 - f_4^2}~{\rm cot}^2\theta_2 ,
\nd
and where $\hat\gamma$ is given by:
\begin{equation}
\hat\gamma \equiv r^{\frac{4}{3}} \frac{\left(\sqrt{1-\frac{\mu^4}{r^4}} - \frac{\mu^4}{r^4} \cosh^{-1}\left(\frac{r^2}{\mu^2}\right)\right)^{\frac{1}{3}}}{\sqrt{1-\frac{\mu^4}{r^4}}} \, ,    \qquad \hat\gamma' \equiv \frac{\partial\hat\gamma}{\partial r^2} \ .
\end{equation}
Finally, the function $\alpha(\theta_1, \phi_1, \psi)$ appearing in \eqref{alpdef} is typically an even or odd function of the variables $(\theta_1, \phi_1, \psi)$ under a
$\mathbb{Z}_2$ reflection, i.e., $\alpha(-\theta_1, -\phi_1, -\psi) = \pm \alpha(\theta_1, \phi_1, \psi)$.  
For the standard deformed conifold with K\"ahler metric we have $\alpha = 1$, but for a warped deformed conifold, $\alpha$ will be non-trivial.

To see the fibrational structure, fix a point on the ${\bf P}^1$ with coordinates $(\theta_2, \phi_2)$, then
\bg\label{fixbco}
\epsilon_1 =  \epsilon_2 = \alpha_3 = 0
\nd
and so the metric \eqref{defconw} becomes:
\bg\label{alef}
ds^2_{\rm ALE} = f_1 dr^2 + f_3(d\theta_1^2 + \sin^2\theta_1 d\phi^2_1)  +f_2(d\psi + \cos \theta_1 d\phi_1)^2 \ .
\nd
Defining a radial variable $\rho \equiv \sqrt{\frac{3}{2}}\,  r^{2/3}$, we see that the metric has the asymptotic form
\bg
ds^2_{\rm ALE} \longrightarrow d\rho^2 + \tfrac{1}{6} \rho^2 ( d\theta_1^2 + \sin^2\theta_1\, d\phi_1^2 ) + \tfrac{1}{9} \rho^2 (d\psi + \cos\theta_1\, d\phi_1)^2  \, .
\nd
This is not quite an ALE space because the constant factors $\frac{1}{6}$ and $\frac{1}{9}$ would have to both be $\frac{1}{4}$.  Instead, we call this a \emph{warped} ALE space, though we may drop the ``warped'' adjective at times.
Next, fixing a point on the warped ALE space, we see that the fiber metric is that of a squashed ${\bf P}^1$, with squashing $\alpha_4$ depending on where we are in the warped ALE space.

In a similar vein, the warped resolved conifold can be presented as a ${\bf P}^1$ fibration over a warped ALE.  The metric of the resolved conifold can be written as
\bg\label{resconw}
ds^2 & = & \gamma'dr^2 + {\gamma + a^2\over 4} (d\theta_1^2 + \sin^2\theta_1 d\phi_1^2) + {\gamma' r^2\over 4} (d\psi + \cos \theta_1 d\phi_1)^2 
+ \alpha_1 (d\psi + \cos \theta_1 d\phi_1)\nonumber\\
 && ~~+ ~{\gamma \over 4}\left(d\theta_2^2 + \alpha_2 \sin^2\theta_2 d\phi_2^2\right)\nonumber\\
& \equiv & ds^2_{\rm ALE} + ds^2_{\rm fiber}
\nd
where $a$ is the resolution parameter; $ds^2_{\rm ALE}$ denotes the metric of the warped ALE space plus mixed terms with the fiber; $ds^2_{\rm fiber}$ denotes the metric of a squashed two-sphere with coordinates $(\theta_2,\phi_2)$; $\gamma$ is given by \cite{pandozayas}
\begin{equation}
\gamma = -2a^2 + 4a^4 N(r)^{-\frac{1}{3}} + N(r)^{\frac{1}{3}} \, , \qquad N(r) = \tfrac{1}{2}\big( r^4 - 16a^6 + \sqrt{r^8-32a^6 r^4}\big) \, ;
\end{equation}
and 
$\alpha_1$ and $\alpha_2$ describe the ALE\footnote{The warping, as mentioned above, takes us 
away from the standard ALE metric \cite{kron} (see also the recent paper \cite{witTN} for more details on ALE and ALF spaces).}  
warping and the squashing of the two-sphere, respectively, given by
\bg\label{alpdef}
\alpha_1 \equiv {\gamma' r^2 \alpha(\theta_1, \phi_1, \psi)\over 2}~ \cos \theta_2 d\phi_2\, , \qquad
\alpha_2 \equiv 1 + {r^2\gamma'\over \gamma} {\rm cot}^2\theta_2 \, , \qquad \gamma' \equiv \frac{\partial \gamma}{\partial r^2} \, ,
\nd
away from the origin of $(\theta_2, \phi_2)$.  At any other point on sphere, we have an ALE space, as seen from the first line of \eqref{resconw}.

Just as for the warped deformed conifold, for the warped resolved conifold we can fix a point on the ${\bf P}^1$ fiber by $(\theta_2, \phi_2)$ to constant values.  This implies $\alpha_1 = 0$ so the metric \eqref{resconw} becomes:
\bg\label{resconow}
ds^2_{\rm ALE} =  \gamma'dr^2 + {\gamma + a^2\over 4} (d\theta_1^2 + \sin^2\theta_1 d\phi_1^2) + {\gamma' r^2\over 4} (d\psi + \cos \theta_1 d\phi_1)^2
\nd
which is the metric for a warped resolved ALE space.

\subsection{F-Theory Picture}

In our previous papers \cite{chen1,chen2},  we have argued for the existence of gauge/gravity duality in the $SO(32)$ heterotic theory by duality chasing the original type IIB
geometric transition. In particular, the technique for going from a local heterotic description to a global description with wrapped five-branes gives us a way to
generate the gravity solution {\it before} the geometric transition.

Another way to study the type IIB transition is through compactification of F-theory on fourfolds. 
For example, consider F-theory on an
elliptically fibered Calabi--Yau fourfold ${\bf X} \stackrel{\pi}{\longrightarrow} {\bf B}$, where ${\bf B}$ is the threefold base. We will assume that ${\bf B}$ contains a smooth curve
${\bf E} \simeq {\bf P}^1$ with normal bundle ${\cal O}(-1) \oplus
{\cal O}(-1)$, which implies that locally, near ${\bf E}$, the base looks like a resolved conifold. 
We can then perform an extremal transition
from ${\bf B}$ to ${\bf B}'$, obtained by contracting the ${\bf P}^1$ to a
point and then smoothing.  Under this transition, we obtain another elliptically fibered
Calabi--Yau fourfold
\bg\label{4fol}
{\bf X}'~ \stackrel{\pi}{\longrightarrow} ~ {\bf B}',
\nd
which is the manifold that, in the presence of both RR and NS fluxes
in type IIB, can be deformed to yield a metrically 
non-K\"ahler manifold (which still has the same topology as ${\bf B}'$) \cite{anke3, chen1}.  This
construction yields the same result as the one discussed earlier,  namely that the branes disappear in the process so that the final result just contains the gravitational dual with fluxes and no extra branes.\footnote{Although, in more general cases the gravity duals can be
{\it non-geometric}.  This feature was discussed in detail in \cite{chen1, chen2}.}

The duality map between F-theory and heterotic theory could now, in principle, help us understand the transition on the heterotic side.  
Unfortunately, this is easier said than done since the dualities between the heterotic theories and F-theory on fourfolds are involved.  One has to tread carefully to find the appropriate duality.

As discussed above, our aim is to connect a specfic F-theory compactification  with a compactification of $E_8 \times E_8$ heterotic theory. 
One way to do this would be to exploit the duality between the Gimon--Polchinski model \cite{gimpol} and F-theory on a particular Calabi--Yau threefold with 
Hodge numbers $(3, 243)$, which admits an elliptic fibration over ${\bf P}^1 \times {\bf P}^1$ \cite{MV1, MV2}. This duality is useful because of its connection to the $E_8 \times E_8$ heterotic theory on a K3 manifold, where the 24 instantons are divided equally between the two $E_8$'s. Our next step then would be to express the resolved conifold as a K3 fibration over a 
${\bf P}^1$. We could then extend the duality to heterotic on a resolved conifold by dualizing along the K3 fiber.\footnote{There are other variants of this story. For example, the F-theory fourfold compactification with a  ${\bf P}^1 \times {\bf P}^1$ is also connected to another Calabi--Yau with
Hodge numbers $(51,3)$ (and probably also to $(3, 51)$), as pointed out by
\cite{gopamu, dabholkarP, blumZ}. The difference between the two
compactifications is related to the number of tensor and charged hypermultiplets.
In this paper, we will concentrate only on the $(3, 243)$ case where the heterotic dual has only one
tensor multiplet.  To get more tensors in six dimensions, we have
to redefine the orientifold operation in the dual type I side. 
A way to achieve this would be to define the
orientifold operation in such a way that, in addition to reversing the
world sheet coordinate $\sigma$ to $\pi - \sigma$, it also flips
the sign of the twist fields at all fixed points (see for example \cite{dabholkarP}, \cite{ASYTqa}). This way
the closed string sectors of both the theories match, but in the twisted
sector we get 17 tensor multiplets instead of hypermultiplets.
Therefore even though the orientifolding action looks similar in
both cases, the massless multiplets are quite different.}

Our starting point is then to study the theory
generated by $k$ D3-branes probing type IIB backgrounds with intersecting seven-branes
and orientifold-planes.  We expect the gauge theory on the $k$ D3-branes to be $Sp(2k) \times Sp(2k)$
in the presence of one set of 
intersecting branes and planes.  Local charge
cancellation then imposes a global $U(4) \times U(4)$ symmetry, where a set
of four D7-branes are placed perpendicular to another set of four D7-branes.
We will discuss soon how other global symmetries may appear.

\subsection{Type II Picture}

Our starting point, as we mentioned above,  
would be to take $k$ D3-branes in the type IIB theory to probe the intersecting D7/O7 system. This is almost like the Gimon--Polchinski setup \cite{gimpol}, but
with one crucial difference: the two intersecting D7/O7 systems wrap a ${\bf P}^1$.  In the absence of the ${\bf P}^1$, we would expect the infrared theory to be an
${\cal N} = 2$, $Sp(2k) \times Sp(2k)$ gauge theory that would, under some special conditions, flow to a conformal fixed point.  In the presence of the ${\bf P}^1$, the
infrared theory should instead be an ${\cal N} = 1$ gauge theory.  The moduli space of the gauge theory --- ignoring the global symmetry, for now --- will be $k$ copies of the
following manifold:
\bg\label{manifold}
{\cal M} ~\equiv~  {{\rm ALE}  ~ \otimes ~ {\bf P}^1\over (-1)^{F_L}\cdot \Omega \cdot \sigma} 
\nd
where $\sigma$ is the Nikulin involution \cite{nikulin} if the ALE space is replaced by a K3 manifold, otherwise it is the usual orbifolding for the ALE case,
and $\otimes$ denotes a \emph{local} product, i.e., there is some non-trivial fibrational structure.  Note that the moduli space is a non-compact manifold. The global
symmetry of the system will be typically ${\cal G} \times {\cal G}$, where we will soon discuss form of ${\cal G}$.

The story now is simple. Under two T-dualities along the two-cycle of the ALE space followed by an S-duality will convert this background to the heterotic theory (a table of the brane setup appears in {\bf table \ref{gpbranes}}).

\subsubsection{Gauss law constraints}

\begin{table}[tb]
 \begin{center}
\renewcommand{\arraystretch}{1.5}
\begin{tabular}{|c||c|c|c|c|c|c|c|c|c|c|}\hline Direction & 0 & 1 & 2
& 3 & 4 & 5 & 6 & 7 & 8 & 9 \\ \hline\hline
D7/O7 & $\surd$  & $\surd$   & $\surd$   & $\surd$  & $\cdot$  & $\cdot$ & $\surd$ & $\surd$ & $\surd$
& $\surd$ \\  \hline
D7${}^\prime$/O7${}^\prime$  & $\surd$  & $\surd$   & $\surd$   & $\surd$  & $\surd$  & $\surd$  & $\surd$  & $\surd$ &
$\cdot$  & $\cdot$\\  \hline
$k$ D3s & $\surd$  & $\surd$   & $\surd$   & $\surd$  & $\cdot$  & $\cdot$ & $\cdot$ & $\cdot$ & $\cdot$  & $\cdot$
\\  \hline
ALE & $\cdot$  & $\cdot$  & $\cdot$  & $\cdot$ & $\surd$  &
$\surd$   & $\cdot$  & $\cdot$ & $\surd$  & $\surd$ \\  \hline
${\bf P}^1$ & $\cdot$  & $\cdot$  & $\cdot$  & $\cdot$ & $\cdot$  &
 $\cdot$  & $\surd$  & $\surd$ &  $\cdot$  &  $\cdot$ \\  \hline
  \end{tabular}
\renewcommand{\arraystretch}{1}
\end{center}
  \caption{The orientations of various branes in the type IIB, intersecting seven-brane duality frame. The check-marks denote the
directions along which we have either the branes or the manifold.}
  \label{gpbranes}
\end{table}

Before moving ahead, let's see how Gauss' law is satisfied.  Let the $k$ D3-branes in type IIB be oriented along $x^{0, 1, 2, 3}$, the ALE space along $x^{4, 5, 8, 9}$, and the ${\bf P}^1$ along $x^{6, 7}$ (see {\bf table \ref{gpbranes}}).  
T-dualizing along a two-cycle of the ALE space that we take to have coordinates $x^{4,5}$, we see that the heterotic NS5-branes will be along the $x^{0,1,2,3,4,5}$ directions, so the $x^{6,7,8,9}$ directions should be noncompact.

Back in the type IIB picture, there are two set of $(p, q)$ seven-branes, one set wrapping $x^{0,1,2,3,4,5,6,7}$, and one set wrapping $x^{0,1,2,3,6,7,8,9}$.  If we didn't have the $k$ D3-branes, this could allow for the $x^{4, 5}$ and $x^{8, 9}$ 
directions to be compact, e.g., ${\bf P}^1 \times {\bf P}^1$, over which the F-theory torus would be non-trivially fibered, giving rise to 
F-theory on a Calabi-Yau threefold with ${\bf F}_n$ base \cite{MV1, MV2}.  
In our case, the D3-brane charge is a problem unless we make the $x^{6,7,8,9}$ directions noncompact, while the $x^{4, 5}$ 
directions need to be compact since we will be T-dualizing along them. 
Thus, we see that the ALE space in \eqref{manifold} really has to be a noncompact, as does the ${\bf P}^1$:
\begin{itemize}
\item  The set of $(p,q)$ seven-branes along $x^{0,1,2,3,4,5,6,7}$ has fewer than 24 branes. This can be implemented by taking only one O7 plane and set of charge canceling D7-branes. 

\item The ${\bf P}^1$ oriented along $x^{6, 7}$ is topologically non-compact, with one or two anti-podal points removed. This can be implemented via 
making $\phi_2$ non-periodic or via changing the definition of $\alpha_2$ and $\alpha_4$ in \eqref{resconw} and \eqref{defconw}, respectively. 
\end{itemize}

\subsubsection{Algebro-geometric picture}

To better understand the IIB/F-theory setup, consider the standard Weierstrass
equation governing the F-theory axio-dilaton as it varies over the ALE space (the ${\bf P}^1$ will be suppressed for this subsection, as will the D3-brane probes):
\bg\label{wseop}
y^2 ~ = ~ x^3 + f(u, v)x + g(u, v),
\nd
where the coordinate 
$u = x^4 + i x^5$ corresponds to the ALE two-cycle along which we will perform T-duality (and is the usual $u$-plane of Seiberg--Witten theory \cite{SW1}), and $v \equiv x^8 + i x^9$ corresponds to the other ALE directions.
$f(u, v)$ is a polynomial of bi-degree $(8, 8)$, and $g(u, v)$ is a polynomial of bi-degree $(12, 12)$.
These polynomials give us the physics not only at the orientifold point, but also away from it. In fact, at the
orientifold point the description can be made a little simpler by choosing the functional forms for $f(u, v)$ and $g(u, v)$ to be
\bg\label{fgndef}
f(u, v)  =  \prod_{i=1}^8 {\cal A}_1^i W_i + \prod_{i=1}^4 {\cal A}_2^i Z_i^2, \quad
g(u, v) =  \prod_{i=1}^{8}  \prod_{j=1}^{4} {\cal A}_3^{ij} W_i Z_j + \prod_{i=1}^4 {\cal A}_4^i Z_i^3 +
\prod_{i=1}^{12} {\cal A}_5^i U_i,
\nd
where ${\cal A}^i_k$ are coefficients that are constrained by using consistency conditions for orientifolds. The other variables $Z_i, W_i$, and 
$U_i$, are defined as
\bg\label{gdass}
&&Z_i ~ \equiv ~ (u - \hat{u}_i)(v - \hat{v}_i),\nonumber\\
&&W_i ~ \equiv ~ (u - {u}_i)(v - {v}_i),\nonumber\\
&&U_i ~ \equiv ~ (u - \widetilde{u}_i)(v - \widetilde{v}_i).
\nd
More details about these polynomials
\eqref{fgndef} can be found in \cite{becker4}, where the coefficients ${\cal A}^i_k$'s were derived.\footnote{We have also corrected a typo in \cite{becker4}.}
The discriminant locus is then given by the curve
\bg\label{dloka}
\sum_{n, p} {\cal C}_{np} \prod_{i, j, k}  Z_i^{6-2n-3p} W_j^n U_k^p ~ = ~ 0 ,
\nd
which can be decomposed as
\bg\label{simcur}
\left({\cal C}_{10} {\cal F}_1 ~ + ~ {\cal C}_{11} {\cal F}_2 ~ + ~ {\cal C}_{02} {\cal F}_3 ~ + ~ {\cal C}_{00} {\cal F}_4
~ + ~ {\cal C}_{30} {\cal F}_5\right)
\left({\cal C}_{20} {\cal F} ~ + ~ {\cal C}_{01} {\cal G}\right) ~ = ~ 0.
\nd
where the polynomials ${\cal F}$ and ${\cal G}$ are degree 16 in $(u, v)$, and where
${\cal C}_{mn} \ne {\cal C}_{nm}$.  Thus, we have two curves spanning the discriminant locus given by
\bg\label{lkeyes}
&& {\cal C}_{20} {\cal F} ~ + ~ {\cal C}_{01} {\cal G} ~ = ~ 0 , \nonumber\\
&&{\cal C}_{10} {\cal F}_1 ~ + ~ {\cal C}_{11} {\cal F}_2 ~ + ~ {\cal C}_{02} {\cal F}_3 ~ + ~ {\cal C}_{00} {\cal F}_4
~ + ~ {\cal C}_{30} {\cal F}_5 ~ = ~ 0 .
\nd
The first curve specifies the orientifold condition under which the type IIB theory
has a heterotic dual, given by a heterotic compactification on a K3 manifold.
The second curve should be interpreted as
$(p, q)$ seven-branes that form a non-dynamical orientifold plane.  Since our concern is mostly the orientifold background specified by the 
first curve in \eqref{lkeyes}, we will ignore
the physics behind the second curve in this paper.

A more general elliptic fibration can be specified, for example, by
\bg\label{deffg}
f_k(u) & = & \prod_{i= 1}^8 A_{ik} (u - a_{ik}) ~ + ~ \prod_{i= 1}^4 B_{ik} (u - b_{ik})^2
~ + ~ \prod_{i= 1}^2 C_{ik} (u - c_{ik})^4 ~ + ~
D_k (u - d_k)^8, \nonumber\\
g_k(u) & = & \prod_{i= 1}^{12} M_{ik} (u - m_{ik}) ~ + ~ \prod_{i= 1}^6 N_{ik} (u - n_{ik})^2
~ + ~ \prod_{i= 1}^4 S_{ik} (u - s_{ik})^3
~ + ~
\prod_{i= 1}^3 P_{ik} (u - p_{ik})^4 \nonumber\\
&& ~~~~~~~~ + ~ \prod_{i= 1}^2 Q_{ik} (u - q_{ik})^6 ~ + ~ R_k (u - r_k)^{12},
\nd
where, as before, the coefficients are allowed to take values determined by the underlying dynamics of F-theory. The polynomials
$f_k(u)$ and $g_k(u)$ are now used to determine $f(u, v)$ and $g(u, v)$ by
\bg\label{fvgv}
f(u, v) ~ \equiv ~ f_1(u) f_2(v), \qquad g(u, v) ~ \equiv ~ g_1(u) g_2(v).
\nd
As we mentioned above, these polynomials give us the physics not only at the orientifold point but also away from it, for example as in \eqref{choofofg}. 
We will discuss some of these
curves later when we study vector bundles.

\vskip.1in

\subsubsection{The orientifold limit and duality chain}

After having provided the algebro-geometric details of the type IIB background, we now ask:
what happens when we allow $k$ D3-branes to probe the orientifold point?  The
orientifold action is similar to the one discussed earlier, and the space involutions are done via
the involution $\sigma$.  The type IIB manifold will be
\bg\label{nikuk3}
{\cal M} \equiv {{\cal C}_{\rm res}\over (-1)^{F_L} \cdot \Omega \cdot \sigma}
\nd
where ${\cal C}_{\rm res} \equiv {\rm ALE} \otimes {\bf P}^1$. For the K3 case,
$\sigma$ describes the Nikulin involution of the form ($r, a, \delta$) $=$ ($2, 2, 0$) on the K3 subspace of ${\cal C}_{\rm res}$ (e.g., see \cite{nikulin} for 
more details). 
Making two T-dualities along the
$x^{4,5}$ directions (which is the resolved ${\bf P}^1$ of the ALE fiber in \eqref{resconw})
will take us to type I on the resolved cone ${\cal C}_{\rm res}$, which is then S-dual to 
heterotic on ${\cal C}_{\rm res}$.

The $k$ D3-branes at the intersecting orientifold point dualize to $k$ small instantons wrapping the $x^{4,5}$ two-cycle of the ALE space
on the heterotic side, hence, wrapping the two-cycle of the resolved conifold.  The resolved conifold will support a non-K\"ahler metric with $SU(3)$ structure.  
Interestingly, as we will soon see, our construction
then brings us to the intrinsic torsion and $G$-structures of
\cite{gauntlett}. Recall that for a manifold with $SU(3)$ structure, we have the two conditions
\bg\label{su2str2u}
0 ~ = ~ d(e^{-2\phi}\Omega)~ = ~ d(e^{-2\phi}\ast J),
\nd
where $J$ is the fundamental two-form,
$\Omega$ is the holomorphic $(3, 0)$ form, and $\phi$ is the dilaton.
The torsion ${\cal H}$ will lie in:
\bg\label{tarkadal}
{\cal H} ~ \in ~ W_3 ~ \oplus ~ {W}_4 ~\oplus ~ {W}_5 \nd 
where $W_i$ are torsion classes that describe what type of manifold we have \cite{carluest}.  In the {\it absence} of the probe D3-branes, the heterotic
dual  ${\cal C}_{\rm res}$ contains a topologically noncompact K3 surface.  Of course, this doesn't imply that the metric
 ${\cal C}_{\rm res}$ is conformally K\"ahler, and in the presence of
the D3-brane probes the situation is even more different.  
Before we describe the non-K\"ahler heterotic geometry,
we should ask what happens to the $Sp(2k) \times Sp(2k)$ gauge symmetry on the
heterotic side.

To do this, we will have to study the orientations of various branes on the type IIB side. The two
${\mathbb Z}_2$ orientifold transformations are generated by $\{1, g\}$ and $\{1, h\}$, where \cite{sengimon, sengimon2}:
\bg\label{gimgen}
g ~ = ~ (-1)^{F_L} \cdot \Omega \cdot {\cal I}_{45}, \qquad h ~ = ~ (-1)^{F_L} \cdot \Omega \cdot {\cal I}_{89},
\nd
where ${\cal I}_{ab}$ denotes orbifold action along $x^{a,b}$. The ${\mathbb Z}_2 \times {\mathbb Z}_2$ action can can be suggestively rewritten as
\bg\label{z2xz2}
\{1, g\} \times \{1, h\} &=& \{1, ~g, ~h, ~gh\} \nonumber\\
 & = & \{1, ~(-1)^{F_L} \cdot \Omega \cdot {\cal I}_{45}, ~ (-1)^{F_L} \cdot \Omega \cdot {\cal I}_{89}, ~ I_{4589}\} \nonumber\\
 & = & \{1, ~(-1)^{F_L} \cdot \Omega \cdot {\cal I}_{45}\}~\times ~ \{1, ~I_{4589}\} .
\nd
This implies the following relation: 
\bg\label{oactinn}
{{\mathbb R}^4\over \{1, g\} \times \{1, h\}} ~ = ~ {{\mathbb R}^4 \over \{1, ~I_{4589}\} \times \{1, ~(-1)^{F_L} \cdot \Omega \cdot {\cal I}_{45}\}} ~ \equiv~ 
{{\rm ALE} \over \{1, ~(-1)^{F_L} \cdot \Omega \cdot {\cal I}_{45}\}} .
\nd 
Now recall that the warped ALE space is obtained precisely at a fixed point $(\theta_2, \phi_2)$ of the ${\bf P}^1$, i.e., we are taking $\alpha$ in \eqref{alpdef} to be 
an even function of ($\theta_1, \phi_1, \psi$).  We can now extend this orientifold action over the full six-dimensional internal space. 
Thus, allowing a global structure of the form 
${\rm ALE} \otimes {\bf P}^1$ leads to the space \eqref{nikuk3}, i.e.,
\bg\label{spaceim}
 {{\cal C}_{\rm res}\over \{1, ~(-1)^{F_L} \cdot \Omega \cdot {\cal I}_{45}\} } .
\nd
In a more general setting with a
slightly different ${\mathbb Z}_2$ action \cite{gopamu, dabholkarP, blumZ},
the type IIB manifold is an orientifold of 
a compact K3 manifold, related to \eqref{deffg}.

T-dualizing along $x^{4,5}$ to type IIB, the orientifold actions \eqref{gimgen} transform into
\bg\label{gimgenow}
(-1)^{F_L} \cdot \Omega \cdot {\cal I}_{45} ~\longrightarrow ~ \Omega, \qquad
(-1)^{F_L} \cdot \Omega \cdot {\cal I}_{89} ~ \longrightarrow ~ \Omega \cdot {\cal I}_{4589},
\nd
where the former would lead to O9-planes and the latter would lead to O5-planes.  The two sets of D7-branes become D9- and D5-branes, while the $k$ probe D3-branes become $k$ D5-branes.  Thus, we actually arrive at the type I theory. This
is the well-known Gimon--Polchinski system \cite{gimpol}, except there is an additional
${\bf P}^1$.
The
orientations of various branes
are given in {\bf table \ref{gpbranesT45}}.
\begin{table}[tb]
\begin{center}
\renewcommand{\arraystretch}{1.5}
\begin{tabular}{|c||c|c|c|c|c|c|c|c|c|c|}\hline Direction & 0 & 1 & 2
& 3 & 4 & 5 & 6 & 7 & 8 & 9 \\ \hline\hline
D9/O9 & $\surd$  & $\surd$   & $\surd$   & $\surd$  & $\surd$  & $\surd$ & $\surd$ & $\surd$ & $\surd$
& $\surd$ \\  \hline
D5${}^\prime$/O5${}^\prime$ & $\surd$  & $\surd$   & $\surd$   & $\surd$  & $\cdot$   & $\cdot$   & $\surd$  & $\surd$ &
$\cdot$  & $\cdot$\\  \hline
$k$ D5s & $\surd$  & $\surd$   & $\surd$   & $\surd$  & $\surd$ & $\surd$ & $\cdot$ & $\cdot$ & $\cdot$  & $\cdot$ \\  \hline
${\cal C}_{\rm res}$ & $\cdot$  & $\cdot$   & $\cdot$   & $\cdot$  & $\surd$ & $\surd$ &  $\surd$  & $\surd$ & $\surd$  & $\surd$ \\ \hline
  \end{tabular}
\renewcommand{\arraystretch}{1}
\end{center}
  \caption{Configuration from {\bf table \ref{gpbranes}} after T-dualizing along $x^{4,5}$. ${\cal C}_{\rm res}$ is the resolved conifold.}
    \label{gpbranesT45}
\end{table}

\begin{figure}[tb]
\begin{center}
\includegraphics[height=5cm]{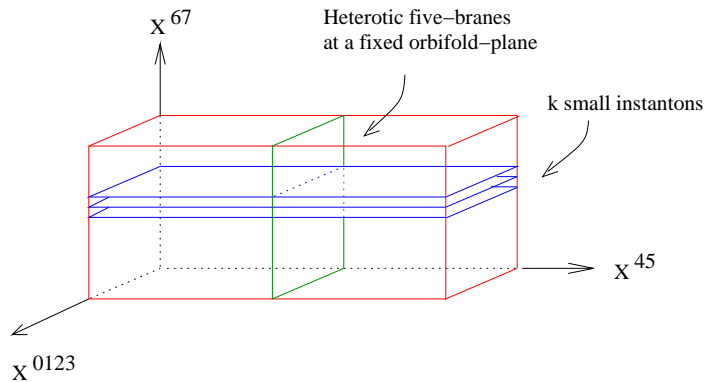}   \end{center}
\caption{The orientations of the NS5-branes on the heterotic side. 
The $k$ D3-branes become heterotic NS5-branes which we refer them as $k$ small instantons here.  One set of type IIB D7/O7 branes/planes becomes the heterotic gauge bundle, while the other set 
generically becomes another set of NS5-branes on an orbifold plane.  In the charge canceled scenario for the second set of D7/O7 branes/planes, which we will concentrate on in this paper, the heterotic dual is just the orbifold five-plane with no NS5-branes on top.}
        \label{sp2kgp}
        \end{figure}

This can now be S-dualized to the
heterotic theory. 
An S-duality
transformation
will convert the D9/O9 system into equivalent vector bundles, and the D5/O9 generically transforms
into NS5-branes stuck to an orbifold plane.  In the case that the D5/O5 system has zero net charge, this will S-dualize to only an orbifold plane \cite{sennbps}.  Geometrically, using the fiber-wise duality, we expect to obtain heterotic on ${\cal C}_{\rm res}$, as discussed in \cite{sensethi}. 
The final heterotic configuration is now expressed in {\bf table \ref{gpbranesnow}} and depicted in {\bf figure \ref{sp2kgp}}. 

\begin{table}[tb]
\begin{center}
\renewcommand{\arraystretch}{1.5}
\begin{tabular}{|c||c|c|c|c|c|c|c|c|c|c|}\hline Direction & 0 & 1 & 2
& 3 & 4 & 5 & 6 & 7 & 8 & 9 \\ & $(x^0)$ & $(x^1)$ & $(x^2)$ & $(x^3)$ & $(\theta_1)$ & $(\phi_1)$ & $(\theta_2)$ & $(\phi_2)$ & $(r)$ & $(\psi)$ \\  \hline\hline
O${}_r5$/NS5 & $\surd$  & $\surd$   & $\surd$   & $\surd$  & $\cdot$   & $\cdot$   & $\surd$  & $\surd$ &
$\cdot$  & $\cdot$\\  \hline
$k$ NS5s & $\surd$  & $\surd$   & $\surd$   & $\surd$  & $\surd$ & $\surd$ & $\cdot$ & $\cdot$ & $\cdot$  & $\cdot$ \\  \hline
${\cal C}_{\rm res}$ & $\cdot$  & $\cdot$   & $\cdot$   & $\cdot$  & $\surd$ & $\surd$ &  $\surd$  & $\surd$ & $\surd$  & $\surd$ \\  \hline
  \end{tabular}
\renewcommand{\arraystretch}{1}
\end{center}
  \caption{Configuration from {\bf table \ref{gpbranesT45}} after a U-duality to heterotic. The orbifold 5-plane is 
denoted by O${}_r5$. The vector bundles are not shown.}
    \label{gpbranesnow}
\end{table}

What about the heterotic gauge symmetry? 
There are two kinds of gauge groups involved here: the type IIB 
$Sp(2k) \times Sp(2k)$ pulled to the heterotic side by  U-dualities (including the transformations in \cite{sensethi}),
and the remnant of the $E_8 \times E_8$ gauge symmetry from the type IIB 
global symmetry.  For example, if we focus on a single probe D3-brane in this background, the generic gauge group is $U(1)$ and {\it not}
$U(1) \times U(1)$.  At the orientifold intersection point, one might expect the gauge symmetry to 
enhance to $Sp(2) \times Sp(2)$, but this is broken to $U(1)$ by quantum corrections. The full theory is then given by an ${\cal N} = 1$ SCFT with a
$U(1)$ vector multiplet and a massless charged hypermultiplet \cite{sengimon, sengimon2}.  If we move a D3-brane along $u=x^4+i x^5$ (the ALE two-cycle), the
massless charged hypermultiplet can be interpreted as the monopole/dyon point for one of the $Sp(2)$ groups.  Similarly, moving the
D3-brane along $v=x^8+i x^9$ (the other ALE directions), the same massless charged hypermultiplet may now be interpreted as the monopole/dyon point of the
other $Sp(2)$ gauge group.  Thus, the non-perturbative effects in this model convert the monopole/dyon point of one
$Sp(2)$ gauge group to the monopole/dyon point of the other $Sp(2)$ gauge group. 

One may expect similar behavior for the heterotic NS5-branes obtained through our U-duality chain.  The discussion of the gauge symmetry in heterotic will appear in section \ref{sec:hetbundle}. 
Note, however, that the physics of type IIB probe D3-branes must be a bit different from that of the heterotic NS5-branes, although they share similar confining properties, 
since we expect LST to play a role in the decoupling limit of the heterotic side \cite{LST}.

\section{Heterotic Duality Frame}
\label{sec:het}

Our next set of questions are related to each other. The first question is how to find the supergravity solution on the heterotic side with a large number $k$ of NS5-branes and 
small string coupling $g_s \to 0$, with $g^2_sk \to \infty$.
The second question is the issue of supersymmetry. Since we have NS5-brane sources on the resolved conifold,
we need to switch on torsion to preserve supersymmetry, which is where the torsion classes $W_i$ from \eqref{tarkadal} enter the picture. 

Let us start with finding the geometric part of the supergravity solution, we will turn to the gauge bundle in section \ref{sec:hetbundle}. 
The setup was described in {\bf table \ref{gpbranesnow}} and depicted in {\bf figure \ref{sp2kgp}}.  We have $k$ small instantons/NS5-branes along
$x^{0, 1, 2, 3, 4, 5}$ --- wrapping the two-cycle of the ALE given by $x^{4,5}$ --- and another (small) set of NS5-branes oriented along $x^{0, 1, 2, 3, 6, 7}$ --- i.e., wrapping the ${\bf P}^1$ given by $x^{6,7}$ --- on top an orbifold five-plane.  See {\bf table \ref{gpbranesnow}} for the mapping of these coordinates to those in \eqref{resconw}.

This is somewhat similar to the scenario studied in \cite{becker4},
where the metric ansatz for this case was called the ``conformal K3 ansatz''. For the present case, we want to use the resolved conifold metric \eqref{resconw}, so the simplified ansatz of \cite{becker4} becomes
\bg\label{nikara}
ds^2 ~ = ~ ds^2_{0123} ~ + ~ \Delta^m ds^2_{\rm ALE} ~ + ~ ds^2_{\rm fiber}, \qquad e^\phi ~ = ~ \Delta ,
\nd
where $\Delta$ is the warp factor and $m$ is an integer. In our case $m = 1$, which appears from the consistency
conditions in \cite{gauntlett}.  Clearly, because of the warp factor $\Delta$, the resolved conifold must have a non-K\"ahler metric on it in this case.

Let us now see if we can derive the heterotic metric directly from type IIB using duality chasing. In type IIB, the moduli space is given by \eqref{manifold}.
Under two T-dualities followed by an S-duality, the orbifolded ALE space is replaced by just an
ALE space. The metric ansatz of the orbifolded space ALE/$\langle (-1)^{F_L} \cdot \Omega \cdot \sigma \rangle$ can be written as
\bg\label{orbale}
ds^2 = F_0 dr^2 + F_1 d\psi^2 + F_2 d\theta_1^2 + F_3 d\phi_1^2 .
\nd
This will solve the equations of motion if we also switch on $B_{\psi\phi_1} \equiv b$, along with a dilaton. Note that this $B$-field is not projected out by the orientifold action. The
heterotic dual is then given by:
\bg\label{hetu}
ds^2 = dr^2 + {F_1\over F_0}\left(d\psi + {b\over F_1} d\phi_1\right)^2 + {1\over F_0}\left({1\over F_3} - {b^2\over F_1}\right)d\phi_1^2 + {1 \over F_2 F_0}d\theta_1^2
\nd
with a dilaton that will be assumed to take the form: $\phi \equiv -{\rm log}~F_0$. 
Note that the metric \eqref{hetu} has a fibration structure somewhat similar to the fibration structure in \eqref{resconw}.
The fibration can be made exactly as in \eqref{resconw} if we make the warp factors functions of $(\theta_1, r)$. The Busher rules don't actually allow this, but there are more refined T-duality rules given in \cite{harvmoo}. 

The ALE space that we have here appears as cross-sections at a fixed point of the ${\bf P}^1$ that is parameterized by $x^{6,7}=(\theta_2,\phi_2)$,  just as in the metric \eqref{resconow}. The ${\bf P}^1$ fibration 
should then convert \eqref{resconow} to \eqref{resconw}. In the same vein, we will then take the following as building blocks for the background metric:
\bg\label{4589}
&&ds^2_{45} \equiv {1\over F_0}\left({1\over F_3} - {b^2\over F_1}\right)d\phi_1^2 + {1\over F_2 F_0} d\theta_1^2 ,  \qquad\quad ds^2_{67} \equiv ds^2_{\rm fiber}, \nonumber\\
&& ds^2_{89} \equiv dr^2 + {F_1\over F_0}\left(d\psi + {b\over F_1} d\phi_1\right)^2 + F_5\left(d\psi + {b\over F_1} d\phi_1\right)d\phi_2 .
\nd
where $F_i$ are functions of ($\theta_1, r$).
The $k$ small instantons along $x^{0,1,2,3,4,5}$ and the other NS5/O${}_r$5 branes/planes make the background more complicated.
Let us assume that the two sets of five-branes come with warp
factors $h_i = h_i(x^8, x^9)$, with $i = 1, 2$, such that $h_1$ denotes the warp factor for $k$ small instantons
and $h_2$ denotes the warp factor for the NS5/O${}_r$5 branes/planes.\footnote{This can be seen from the type I picture more clearly.  If we have $N_2$ D5-branes and $N_1$ O5-planes in type I, then $h_2$ will be proportional to the difference, i.e.,  $h_2 \propto 1 + {\cal O}(N_2 - N_1)$, since the net D5 charge $(N_2-N_1)$ gives the number of NS5-branes in the heterotic picture.\label{n2n1}} 
This gives the geometric background
\bg\label{bgfromm}
ds^2  =  ds^2_{0123} + h_2 (ds^2_{45} + h_1 ds^2_{89}) + h_1 ds^2_{67}, \qquad
{\cal T}_{abi}^{(l)}  = c_l \epsilon_{abik} \partial_k h_{l},
\nd
where the ${\cal T}_{abi}^{(l)}$ are torsion polynomials, $c_l$ are constants, and
$(a, b)$ runs over $(4, 5)$ and $(6, 7)$ with $i =$ ($8, 9$). The $\epsilon$ tensor is related to the volume form on ALE.
Note that if we make $h_1 = 1$ (i.e., remove all the small instantons) and set $\Delta={h_2}$,
then we reproduce \eqref{nikara}, so our starting ansatz \eqref{orbale} is consistent with this limit.  For generic choices of $h_1$, the
metric in \eqref{bgfromm} is non-K\"ahler, as we will see below.

We find it useful to parameterize the background as:
\bg\label{manipul}
ds^2 & = & ~ ds^2_{0123} + h_2(ds^2_{45} + h_1ds^2_{89}) + h_1ds^2_{67} \nonumber\\
& = &~ ds^2_{0123} + h_2\left[a_1 d\phi_1^2 + a_2 d\theta_1^2 + h_1 dr^2 + h_1a_3(d\psi + a_4 d\phi_1)^2  \right] \nonumber\\
&&~~~~~~ + h_1(a_5d\theta_2^2 + a_6 d\phi_2^2) + h_1h_2 a_7(d\psi + a_4 d\phi_1)d\phi_2 \nonumber\\
& = &~ ds^2_{0123} + h_1 h_2 dr^2 + h_1h_2a_3 \left(d\psi + a_4 d\phi_1 + {a_7\over 2a_3}d\phi_2 \right)^2 \nonumber\\
&&~~~ + h_2a_2\left(d\theta_1^2 + {a_1\over a_2}d\phi_1^2\right) + h_1a_5\left[d\theta_2^2 + {1\over a_5}\left(a_6- {a_7^2h_2\over 4a_3}\right)d\phi_2^2\right]
\nd
where $a_1$ through $a_4$ depend on $(\theta_1, r)$, $a_5$ through $a_7$ depend on ($\theta_i, r$) and $a_1$ through $a_4$ are given by:
\bg\label{a12a4}
a_1 = {1\over F_0}\left({1\over F_3} - {b^2\over F_1}\right), ~~~ a_2 = {1\over F_2F_0}, ~~~ a_3 = {F_1\over F_0}, ~~~ a_4 = {b\over F_1} .
\nd
$a_5$ through $a_7$ will be related to the warp factors $h_1$ and $h_2$ via the
torsional equations below, where we will use ${\cal N} = 1$ supersymmetry to impose additional relations between the metric factors and the torsion.

As one would expect, this metric is somewhat similar to
the ones that we considered in
\cite{chen1} and in \cite{chen2}. The difference here is that we also have a set of NS5/O${}_r$5 branes/planes.  The ultimate question would be what happens if one performs a geometric transition, but first we need to fully
understand the solution before we can perform a geometric transition.

\subsection{Supersymmetry and torsion classes}
\label{susy}

We will start by rewriting \eqref{manipul} as:
\begin{eqnarray}\label{mopla}
ds^2=ds_{0123}^2+H_2dr^2+H_1(d\psi+c_4d\phi_1+c_7d\phi_2)^2+H_3(d\theta_1^2+c_1^2d\phi_1^2)+H_4(d\theta_2^2+c_5^2d\phi_2^2)\nonumber\\
\end{eqnarray}
but now with one crucial difference from \eqref{manipul}: we will be away from the orientifold point. This aspect has already been addressed in {\bf figure \ref{varyTau}}
where the orientifold regimes were depicted by the boxes marked $A, C$ and $D$. The metric \eqref{manipul} is derived in this regime. However we can go to a 
more generic set-up, marked by boxes $G$ and $H$ in {\bf figure \ref{varyTau}}, where we are no longer restricted by the type I orientifold constraints.   

In this regime, we can allow $H_i$ to depend only $r$ while the $c_i$ depend on $\theta_1$ and $\theta_2$. Note that 
none of the parameters depend on ($\phi_1, \phi_2, \psi$), which is important since we will perform three T-dualities along these directions to study the mirror later in this paper. 
Therefore to proceed, we choose the following vielbein \cite{gwynknauf}:
\begin{eqnarray}\label{vielgood}
&&e_1=\sqrt{H_3}(\cos \psi_1d\theta_1+\sin\psi_1c_1d\phi_1),\quad e_2=\sqrt{H_3}(-\sin \psi_1d\theta_1+\cos\psi_1c_1d\phi_1),\nonumber\\
&&e_3=\sqrt{H_4}(\cos \psi_2d\theta_2+\sin\psi_2c_5d\phi_2),\quad e_4=\sqrt{H_4}(-\sin \psi_2d\theta_2+\cos\psi_2c_5d\phi_2),\nonumber\\
&&e_5=\sqrt{H_1}(d\psi+c_4d\phi_1+c_7d\phi_2),\quad\quad~~ e_6=\sqrt{H_2}dr
\end{eqnarray}
This choice suggests a fundamental two form:
\begin{eqnarray}
J&=&e_1\wedge e_2+e_3\wedge e_4+e_5\wedge e_6\nonumber\\
&=&H_3c_1d\theta_1\wedge d\phi_1+H_4c_5d\theta_2\wedge d\phi_2 +\sqrt{H_1H_2}(d\psi+c_4d\phi_1+c_7d\phi_2)\wedge dr .
\end{eqnarray}
The torsion ${\cal H}$ then follows from supersymmetry:\footnote{Note that the sign for 
${\cal H}$ follows the convention of \cite{carluest2} which, in turn, differs from the sign choice of \cite{stromtor}. Additionally, our choice of the dilaton 
$\phi$ is $-4$ times the choice of the dilaton in \cite{carluest2}. \label{dilchoice}}
\begin{eqnarray}\label{toren}
{\cal H}= - e^{2\phi}\ast d(e^{-2\phi}J) \equiv 2 \ast d\phi \wedge J - \ast dJ
\end{eqnarray}
along with a dilaton $\phi$ that generically may be a function of the internal coordinates $(r,\theta_1,\theta_2)$.

The torsion is {\it not} closed and must additionally satisfy the Bianchi identity:
\bg\label{torclosed}
d{\cal H} ~ = ~ {\rm sources} + \alpha' \left[{\rm tr} (R_+ \wedge R_+) - {1\over 30}{\rm tr} (F \wedge F)\right] \equiv
\alpha' \left[{\rm tr} (R_+ \wedge R_+) - {1\over 30}{\rm tr} ({\widetilde F} \wedge {\widetilde F})\right] \qquad
\nd 
where by {\it sources} we mean contributions from all the NS5-branes.
The second equality in \eqref{torclosed} is an alternative way to view the background where
the {\it sources} are directly absorbed into the 
definition of the vector bundle (i.e., viewing the NS5-branes as small instantons of the heterotic gauge theory). 
We will use the latter interpretation for the background throughout the paper. 

Interestingly, both from our choice of absorbing the sources into the definition of the vector 
bundle \eqref{torclosed} and from the torsion \eqref{toren}, it is easy to argue that the total scalar potential of the effective four-dimensional theory, obtained by compactification, is given by the following expression \cite{carluest2}:
\bg\label{potcar}
V = -{1\over 2} \int e^{-2\phi} \left({\cal H} + {1\over 2} e^{2\phi}\ast d(e^{-2\phi}J)\right) \wedge \ast \left({\cal H} + {1\over 2} e^{2\phi}\ast d(e^{-2\phi}J)\right)\nd
plus a D-term.
The fact that extra five-brane sources {\it do not} break any supersymmetry is related to the
concept of {\it generalized calibration} from \cite{gauntlett}: unbroken supersymmetry is restored by five-branes wrapping the two-cycle that is calibrated by the 
same invariant form as 
the one calibrating the solution involving the full backreactions. This way, the potential $V$ would be exactly zero for the background satisfying 
\eqref{toren}.
Note that although the vanishing of the potential \eqref{potcar} is necessary to have supersymmetry preserved, it is not sufficient, so we will show the vanishing of the $F$- and the $D$-terms directly.\footnote{Indeed, a contradiction was shown in \cite{sully} where  
ISD fluxes were switched on to break supersymmetry without generating a potential.}

Now, using the torsional equation \eqref{toren}, we find:
\begin{eqnarray}\label{torsa}
{\cal H} &=&(2B_6-A_1)H_4\sqrt{H_1}c_5d\theta_2\wedge d\phi_2\wedge d\psi\nonumber\\ &&-\Big((A_1-2B_6)H_4\sqrt{H_1}c_5c_4\nonumber\\ &&-2(B_1\cos\psi_1+B_2\sin\psi_1)H_4\sqrt{H_3}c_1c_5\Big)d\theta_2\wedge d\phi_2\wedge d\phi_1\nonumber\\ &&-(A_2-2B_6)H_3\sqrt{H_1}c_1d\theta_1\wedge d\phi_1\wedge d\psi\nonumber\\ &&-\Big((A_2-2B_6)H_3\sqrt{H_1}c_1c_7-(A_4 +2B_3\cos\psi_2\nonumber\\ &&-2B_4\sin\psi_2)H_3\sqrt{H_4}c_1c_5\Big)d\theta_1\wedge d\phi_1\wedge d\phi_2\nonumber\\ &&-2\sqrt{H_1H_2H_4}(B_4\cos\psi_2+B_3\sin\psi_2)d\theta_2\wedge (d\psi+c_4d\phi_1+c_7d\phi_2)\wedge dr\nonumber\\ &&-2\sqrt{H_1H_2H_4}(B_4\sin\psi_2+B_3\cos\psi_2)c_5d\phi_2\wedge (d\psi+c_4d\phi_1)\wedge dr\nonumber\\ &&-2\sqrt{H_1H_2H_3}(B_1\sin\psi_1-B_2\cos\psi_1)d\theta_1\wedge (d\psi+c_4d\phi_1+c_7d\phi_2)\wedge dr\nonumber\\ &&+2\sqrt{H_1H_2H_3}(B_1\cos\psi_1-B_2\sin\psi_1)c_1d\phi_1\wedge (d\psi+c_7d\phi_2)\wedge dr\nonumber\\ &&-2H_3\sqrt{H_4}(B_3\cos\psi_2-B_3\sin\psi_2)c_1d\theta_1\wedge d\theta_2\wedge d\phi_1\nonumber\\ &&-2H_4\sqrt{H_3}(B_1\sin\psi_1-B_2\cos\psi_1)c_1d\theta_1\wedge d\theta_2\wedge d\phi_2\nonumber\\ &&+B_5\sqrt{H_2}dr\wedge (c_1 d\theta_1\wedge d\phi_1-c_5 d\theta_2\wedge d\phi_2)\nonumber\\&&-\frac{H_1c_5c_{4\theta_2}}{c_1}d\theta_1\wedge d\phi_2\wedge (d\psi+c_4d\phi_1+c_7d\phi_2)\end{eqnarray}
where to preserve the isometry along $\psi$ direction we will take $\psi_1 = \psi_2 = {\langle\psi\rangle\over 2}$ with $\langle\psi\rangle$ being a constant. 
Furthermore we will demand all the coefficients above to be completely independent of ($\phi_1, \phi_2, \psi$) although they could be functions of ($\theta_1, \theta_2$) 
in addition to being functions of $r$. 
The coefficients $A_i$ are defined as:  
\begin{eqnarray}\label{torsacoef}
A_1&=&\frac{(H_{3}c_1)_r+\sqrt{H_1H_2}c_{4\theta_1}}{H_3\sqrt{H_2}c_1},~~~
A_2=\frac{(H_{4}c_5)_r+\sqrt{H_1H_2}c_{7\theta_2}}{H_4\sqrt{H_2}c_5}\nonumber\\
A_3&=&\frac{(H_{3}c_1)_{\theta_2}}{\sqrt{H_4}H_3c_1},~~~
A_4=\frac{(H_1H_2)_{\theta_2}}{2H_1H_2\sqrt{H_4}}
\end{eqnarray}
and the $B_i$ coefficients are defined as:
\begin{eqnarray}\label{torsacoef2}
B_1&=&\Big(\frac{\phi_{\phi_1}\sin\psi_1}{\sqrt{H_3}c_1}+\frac{\phi_{\theta_1}\cos\psi_1}{\sqrt{H_3}}-\frac{\phi_{\psi}\sin\psi_1c_4}{\sqrt{H_1H_3}c_1}\Big)\nonumber\\
B_2&=&\Big(\frac{\phi_{\phi_1}\cos\psi_1}{\sqrt{H_3}c_1}-\frac{\phi_{\theta_1}\sin\psi_1}{\sqrt{H_3}}-\frac{\phi_{\psi}\cos\psi_1c_4}{\sqrt{H_1H_3}c_1}\Big)\nonumber\\
B_3&=&\Big(\frac{\phi_{\phi_2}\sin\psi_2}{\sqrt{H_4}c_5}+\frac{\phi_{\theta_2}\cos\psi_2}{\sqrt{H_4}}-\frac{\phi_{\psi}\sin\psi_2c_7}{\sqrt{H_1H_4}c_5}\Big)\nonumber\\
B_4&=&\Big(\frac{\phi_{\phi_2}\cos\psi_2}{\sqrt{H_4}c_5}-\frac{\phi_{\theta_2}\sin\psi_2}{\sqrt{H_4}}-\frac{\phi_{\psi}\cos\psi_2c_7}{\sqrt{H_1H_4}c_5}\Big)\nonumber\\
B_5&=&\frac{\phi_{\psi}}{\sqrt{H_1}},\,
B_6=\frac{\phi_r}{\sqrt{H_2}}
\end{eqnarray} 
The subscript on $f_\alpha$, where $\alpha$ is a coordinate, means derivative $\del_\alpha f$. The torsion contains all the information of the heterotic five-branes, 
as well as information about the vector bundle 
via the relation \eqref{torclosed}. From above, we also see that the 
torsion has the following nonzero components:\footnote{Note that the heterotic torsion comes from two different sources in type IIB. The first one is from the 
$B_{\rm RR}$ fields that have one leg along the orientifolding direction (so that they are not projected out by the orientifold operation). The second one is from the 
axion and four-form fields that survive the orientifold operation. Some part of the torsion, along with the size of the two-sphere on which we
have the wrapped heterotic five-branes, will eventually be responsible for generating the RG flow in the theory. This will be clearer from the gravity dual. Furthermore,
the powerful machinery of the torsion classes that we are going to use to justify many of the subsequent results can be compared with the interesting work of 
\cite{butti}. It would certainly be interesting to make a precise comparison with the
results of \cite{butti}, but we think that this comparison would be more
appropriately addressed by a separate work.}
\begin{eqnarray} 
&&{\cal H}_{\theta_1\theta_2\phi_1},~~{\cal H}_{\theta_1\theta_2\phi_2},~~{\cal H}_{\theta_1\phi_1\phi_2},~~{\cal H}_{\theta_1\phi_1 r},~~{\cal H}_{\theta_1\phi_2 r},
~~{\cal H}_{\theta_1\phi_2 \psi},~~{\cal H}_{\theta_1\psi r}, ~~{\cal H}_{\theta_2\phi_1\phi_2},\nonumber\\ 
&& {\cal H}_{\theta_2\phi_1r},~~{\cal H}_{\theta_2\phi_2\psi},~~{\cal H}_{\theta_2\phi_2r},
~~{\cal H}_{\theta_2\psi r},~~{\cal H}_{\phi_1\phi_2r},~~{\cal H}_{\phi_1\psi r},~~{\cal H}_{\phi_2\psi r}
\end{eqnarray}
All the five-brane components of the torsion will receive ${\cal O}(\alpha'^{0})$ contributions from the sources in addition to the ${\cal O}(\alpha')$ anomaly term. 
All other components 
will only have the ${\cal O}(\alpha')$ anomaly term. These can be worked out for the generic case. In {\bf section \ref{sec:hetbundle}} we will study a
scenario --- with the torsional components as functions of the radial coordinate $r$ only --- where the Bianchi identity to ${\cal O}(\alpha')$ will be satisfied 
by switching on vector bundles on the internal manifold.


Next, we note the relation to the torsion classes in \eqref{tarkadal}.  They are given by:
\begin{eqnarray}\label{torkada}
W_1&=& \frac{c_{4\phi_2}-c_{7\phi_1}-(c_{4\psi} c_7-c_{7\psi}c_4)+ic_{4{\theta_2}}}{\sqrt{H_2H_3H_4}}\nonumber\\
W_4&=&\frac{1}{4}\Big[\frac{1}{c_1H_3}\Big((c_1H_3)_r dr+(c_1H_3)_{\psi}d\psi+(c_1H_3)_{\phi_2}d\phi_2+(c_1H_3)_{\theta_2}d\theta_2\Big)\nonumber\\
&&+\frac{1}{c_5H_4}\Big((c_5H_4)_rdr+(c_5H_4)_{\psi}d\psi\Big)+\frac{1}{\sqrt{H_1H_2}}\Big((\sqrt{H_1H_2})_{\psi}(c_4d\phi_1+c_7d\phi_2+2d\psi)\nonumber\\
&&\textstyle+(\sqrt{H_1H_2})_{\phi_2}\left({d\psi+c_4d\phi_1\over c_7}+2d\phi_2\right)+ (\sqrt{H_1H_2})_{\theta_2}d\theta_2\Big) + \sqrt{H_1H_2}dr 
\left({c_{7\theta_2} \over H_4c_5} + {c_{4\theta_1} \over H_3c_1}\right)\Big]\nonumber\\
{\rm Re}~{W_5}&=&\textstyle\frac{1}{12}\Big\{\frac{4}{\sqrt{H_3H_4}}\Big[{1\over 2}(\sqrt{H_3H_4})_rdr+(\sqrt{H_3H_4})_{\psi}d\psi+(\sqrt{H_3H_4})_{\phi_2}d\phi_2
+ (\sqrt{H_3H_4})_{\theta_2}d\theta_2\Big]\nonumber\\
&&+\frac{2(\sqrt{H_1}_rdr+\sqrt{H_1}_{\phi_2}d\phi_2+\sqrt{H_1}_{\theta_2}d\theta_2+\sqrt{H_1}_{\psi}d\psi)}{\sqrt{H_1}}\nonumber\\
&&+\frac{2(\sqrt{H_2}_{\phi_2}d\phi_2+\sqrt{H_2}_{\theta_2}d\theta_2+\sqrt{H_2}_{\psi}d\psi)}{\sqrt{H_2}} -2\sqrt{H_2\over H_1} dr\Big\}
\end{eqnarray}
and we can similarly determine the $W_2$ torsion class from above.\footnote{The expression for $W_2$ is very long, 
so we will not write it explicitly.}
In the language of torsion classes, the supersymmetry conditions can be written
as:\footnote{In our conventions, the torsion classes ${\cal W}_4$ and ${\cal W}_5$ of \cite{carluest2} are related to $W_4$ and $W_5$ of \eqref{torkada} 
as ${\cal W}_4 = -2 W_4$ and ${\cal W}_5 = 2 W_5$.  This means that the supersymmetry condition $2{\cal W}_4 + {\cal W}_5= 0$ will become $2W_4 - W_5 = 0$, which is 
\eqref{suco} above. 
Furthermore, ${\cal W}_4 \equiv 4 d\Phi$, where $\Phi$ is the dilaton. In our conventions, the
dilaton $\phi$ is $\phi = - 4\Phi$ (see {\bf footnote \ref{dilchoice}}), this implies that
$d\phi = 2W_4$. For more details see {\bf Appendix \ref{conventioncheck}}. \label{convention}}
\bg\label{suco} W_1 = W_2 = 0 , \qquad 2W_4~ =~ {\rm Re}~W_5. \nd  
Next, we will find solutions to these conditions.

\subsection{An infinite class of solutions}
\label{sec:hetslns}

Some recent studies (for example, \cite{caris}) have found similar types of backgrounds of heterotic five-branes 
wrapped on a resolved conifold. 
What we will find here is that there is a huge class of solutions related to various possible LSTs \cite{LST} on the heterotic five-branes. The story then 
is similar to what we encountered in the $SO(32)$ case \cite{chen2}: there is an infinite class of LSTs. A small subset of these theories are dual to 
geometric backgrounds of the type studied in \cite{anke2, chen1, chen2}. Most of these theories will be dual to non-geometric backgrounds \cite{chen1, chen2}.   

Keeping this in mind, let us now fix the starting coefficients $F_0,\ldots, F_3$ in \eqref{orbale} assuming we are away from the orientifold point as discussed 
earlier and independent of ($\phi_1, \phi_2, \psi$) coordinates. 
A hint may come from the heterotic metric \eqref{hetu} because
we expect this to be a warped ALE space of the form:
\bg\label{alu}
ds^2 = dr^2 + a_3(d\psi + \cos \theta_1 d\phi_1)^2 + {r^2f^2\over 6} (d\theta_1^2 + \sin^2\theta_1 d\phi_1^2)
\nd
where $a_3$ and $f$ are generic functions whose values will be determined later.\footnote{The radial coordinate chosen here is not quite the same as in \eqref{resconow}. Abusing notation, if we call the radial coordinate in \eqref{resconow} $\widetilde{r}$, then $r = \int \sqrt{\gamma'} d\widetilde{r}$. Similarly, $f^2$ in 
\eqref{alu} can be related to $\gamma$ in \eqref{resconow}.}
This suggests that we choose:
\bg
\label{Fivalues}
F_0 = {b\over a_3 \cos \theta_1}, ~~~~ F_1 = {b\over \cos \theta_1}, ~~~~ F_2 =  {6 a_3\cos \theta_1 \over b r^2 f^2}, ~~~~
F_3 =  {6 a_3\cos \theta_1 \over 6 b a_3\cos^2\theta_1 + br^2 f^2 \sin^2\theta_1},~~~~
\nd  
which implies
\bg\label{aivalues}
a_1 = {r^2 f^2 \sin^2\theta_1\over 6}, ~~~~ a_2 = {r^2 f^2\over 6}, ~~~~ a_4 = \cos \theta_1
\nd
with $a_3$ thus far unfixed.  Note that the values for $F_i$ (or, equivalently, for $a_i$) cannot be determined from T-duality 
since that would require they be independent of $(\theta_1, \phi_1)$. 

Now looking at the resolved conifold metric of \eqref{manipul}, we can argue from the fibrational structure for $d\psi$ that:
\bg\label{a4etc}
a_4 = \cos \theta_1, \qquad a_7 = 2 a_3 \cos \theta_2,
\nd
where the $a_4$ is consistent with the value quoted in \eqref{aivalues}. 
Similarly, the dependence of the background \eqref{manipul} on the resolution parameter $a^2$  implies we should set:
\bg\label{a1a5}
a_2 h_2 - a_5 h_1 = a^2 .
\nd
Since $a$ is a constant, it is easy to see that:
\bg\label{akhta}
a_5 = {h_2a_2 - a^2\over h_1} \equiv {\Delta \over h_1} ,~~~~~~ a_6 - {a_7^2h_2 \over 4a_3} = {\Delta ~\sin^2\theta_2\over h_1}, 
~~~~~~ a_1 = {(\Delta + a^2)~\sin^2\theta_1\over h_2} .
\nd
Comparing $a_1$ from above with $a_1$ from \eqref{aivalues} and using $h_2=1$ when there are no NS5-branes on top of the orbifold five-plane (see {\bf footnote \ref{n2n1}}), we see that:
\bg\label{gol}
h_2 = {6(\Delta + a^2) \over r^2 f^2} = 1
\nd
which implies that
\bg
f = {\sqrt{6(\Delta + a^2)}\over r}, \nd
where $\Delta$ and $a_3$ are still undetermined functions of $r$.
Defining
\bg\label{f1234} 
{\cal G}_1 \equiv e^{-\phi} H_2, ~~~~~~~~ {\cal G}_2 \equiv e^{-\phi} H_1, ~~~~~~~~ {\cal G}_3 \equiv e^{-\phi} H_3, ~~~~~~~~ {\cal G}_4 \equiv e^{-\phi} H_4,\nd
the supersymmetry condition \eqref{suco} becomes:
\bg\label{sucobe}
{5\over 6} \left({{\cal G}_{3r}\over {\cal G}_3} + 
{{\cal G}_{4r}\over {\cal G}_4}\right) -{1\over 6} {{\cal G}_{2r}\over {\cal G}_2} + {3\over 2} \phi_r 
= \left({1\over {{\cal G}_3}} + {1\over {\cal G}_4}\right) \sqrt{{\cal G}_1{\cal G}_2} - {1\over 3} \sqrt{{\cal G}_1\over {\cal G}_2}. \nd
We already know that $H_3 = \Delta + a^2$ and $H_4 = \Delta$, and now we find that
\bg\label{dilh1}
 e^\phi = {4\Delta\over r^2}, ~~~~~~~~~~ h_1 = {2 \Delta \over r\sqrt{a_3}}. \nd
We will also find it useful to define
\bg\label{otcoef} f_1(r) ~ \equiv  e^{-\phi}\frac{h_1 a_3}{r^2} , ~~~~~~~~~~~  f_2(r) ~\equiv~ e^{-\phi} h_1 \nd
which, using \eqref{dilh1}, we see satisfy\footnote{See {\bf Appendix \ref{proof}} for a proof for \eqref{dilh1} and \eqref{a3ans}.}
\bg\label{a3ans}
a_3  = {4r^2f_1^2} , \qquad f_1 = {1\over 4 f_2} . \nd
Then the supersymmetry equation \eqref{sucobe} becomes:
\bg\label{bekibeta}
{\del\Delta\over \del r} - \left[{{1\over 2}{\del\over \del r}({\rm log}~a_3) -{2\over \sqrt{a_3}} + {1\over r}\left(11 + {12\Delta\over \Delta + a^2}\right)
\over 4 + {5\Delta \over \Delta + a^2}}\right]\Delta = 0\nd
Note that in the limit that $a$ is much smaller than any 
other scale in the theory, the differential equation \eqref{bekibeta} simplifies to: 
\bg\label{beki}
r {\del\Delta\over \del r} - \left[{r\over 18} {\del ({\rm log}~a_3)\over \del r} - {2r\over 9\sqrt{a_3}} + {23\over 9}\right]\Delta + {\cal O}(a^2) = 0 ,
\nd
where the ${\cal O}(a^2)$ terms involve powers of $\Delta$ and its first derivative.\footnote{In fact, the ${\cal O}(a^2)$ term is given by 
$-{a^2\over 9}\left[{a_{3r}\over 2a_3}-{2\over\sqrt{a_3}} +{11\over r} - 4{\del \over \del r}({\rm log}~\Delta)\right]$.
This is followed by ${\cal O}(a^4)$ terms as can be derived from \eqref{bekibeta}. It is now easy to see that 
the $n$-th term can be derived from $\left(1+{a^2\over \Delta}\right)^{-1} \left[\left(1+{4a^2\over 9\Delta}\right){\del \over \del r}({\rm log}~\Delta) 
- {12\Delta\over r}\right]$, 
therefore, no higher powers of $\Delta_r$ appear in the 
series.}
The solution for $\Delta$ from \eqref{beki} then is:
\bg\label{deltav}
\Delta(r) = \Delta_0~ {\rm exp}\int dr \left[{1\over 18} {\del ({\rm log}~a_3)\over \del r} - {2\over 9\sqrt{a_3}} + {23\over 9r}\right], ~~~~~~~ r > 0
\nd   
where $\Delta_0$ is a constant. Then the background metric is:
\bg\label{bgmetDf}
ds^2 = && ds^2_{0123} + {2\Delta \over r \sqrt{a_3}} dr^2 + {2 \Delta \sqrt{a_3}\over r}(d\psi + \cos \theta_1 d\phi_1 + \cos \theta_2 d\phi_2)^2 \nonumber\\
&& + (\Delta + a^2)\left(d\theta_1^2 + \sin^2\theta_1 d\phi_1^2\right) + \Delta \left(d\theta_2^2 + \sin^2\theta_2 d\phi_2^2\right).\nd
Calculating the torsion classes $W_1$ and $W_2$ from \eqref{torkada}, we
see that they vanish and, therefore, that the manifold admits a complex structure.

The heterotic torsion can now be read off from \eqref{torsa}. This simplifies 
quite a bit in the limit where all the $H_i$ in \eqref{mopla} are just functions of the radial coordinate $r$ because all $B_i = 0$ except $B_6$ 
in \eqref{torsacoef}. The result is:
\bg\label{hettros}
{\cal H}  &=& -\sqrt{H_1\over H_2}\left(G_1 ~\sin \theta_2 d\theta_2 \wedge d\phi_2 
+ G_2 ~ \sin \theta_1 d\theta_1 \wedge d\phi_1\right)\wedge e_\psi\\
 & = -& \sqrt{a_3} \left({\del \Delta \over \del r}-\frac{2\Delta}{r}\right)\left(\sin \theta_2 d\theta_2 \wedge d\phi_2 
~ + ~ \sin \theta_1 d\theta_1 \wedge d\phi_1\right)\wedge e_\psi\nonumber \nd
where $e_\psi \equiv \left(d\psi + \cos \theta_1 d\phi_1 + \cos \theta_2 d\phi_2\right)$ and 
we see that the torsion is asymmetric over the two-spheres because of the $G_1$ and $G_2$ factors.
The precise form shows an amazing simplification\footnote{Consistent with the fact that for both the conifold as well as the resolved conifold, where 
$\Delta = {r^2\over 6}$ and $a_3 = {r^2\over 9}$, the torsion \eqref{hettros} vanishes, as expected for a Calabi--Yau geometry.}, and to ${\cal O}(a^2)$ 
$G_1 = G_2$ as:
\bg\label{epsidef}
G_1 & = & {H_4\over H_3}\left(\sqrt{H_1 H_2} + 2 \phi_r H_3 - H_{3r}\right) ~ = ~ {\del \Delta \over \del r}- {2\Delta\over r} \nonumber\\
G_2 & = & {H_3\over H_4}\left(\sqrt{H_1 H_2} + 2 \phi_r H_4 - H_{4r}\right)~ = ~ {\del \Delta \over \del r}- {2\Delta\over r} \nd
A simple way to relate $\Delta$ and $a_3$ is to relate $h_1$ to the five-brane
harmonic function --- though when $r$ and $a$ are small, we will instead use the three-brane harmonic function since the $k$ NS5-branes will wrap a collapsed cycle, appearing as $(3+1)$-dimensional sources --- i.e.:
\bg\label{demand} h_1 = {2\Delta \over r\sqrt{a_3}} \equiv f_3(r)\left[1 + {\alpha' k\over r^2} \right]\nd
with $f_3$ being a dimensionless function. 
Plugging this into the supersymmetry condition \eqref{sucobe} leads to the following differential equation for $a_3$:
\bg\label{difeku}
&& {da_3\over dr}\left[1 + {3 a^2 r \over 4 f_3 \sqrt{a_3}(e_0 + r^2)}\right] 
+ {\sqrt{a_3}\over 2} - {a_3\over 2r} 
\left({7r^2 + 16 e_0\over r^2 + e_0}\right)  + {9\over 4} {f_{3r} a_3 \over f_3}\nonumber\\
&&~~~~~~~~ + a^2\left[1 + {2f_{3r}\sqrt{a_3}\over f_3} - {\sqrt{a_3}\over 2r}\left({7r^2 + 15 e_0\over r^2 + e_0}\right)\right]{r\over f_3(r^2 + e_0)} = 0 
\nd
where we have defined
\bg e_0 \equiv \alpha' k. \nd
In the limit that $a^2 \ll \alpha'$, then \eqref{difeku} simplifies to:
\bg\label{difsim}
{da_3\over dr} + {\sqrt{a_3}\over 2} - {a_3 \over 2r}\left({7r^2 + 16e_0\over r^2 + e_0}\right) + {9\over 4} {f_{3r} a_3 \over f_3} + {\cal O}(a^2) = 0 \nd 
whose solution can be easily determined if the functional form for $f_3$ is known. In general, however, to solve \eqref{difeku} we will analyze different choices for $f_3$. 

\vskip.1in

\noindent ${\bullet}$ {\bf Case I}: $f_3 = 1$

\vskip.1in

\noindent This is the simplest case where $h_1$ in \eqref{demand} is exactly the five-brane harmonic function for $k$ coincident five-branes. In this case, $a_3$ will 
satisfy:
\bg
\label{case1}
{da_3\over dr}\left[1 + {3a^2 r\over 4 \sqrt{a_3}(r^2 + e_0)}\right] 
+ {\sqrt{a_3}\over 2} - {a_3\over 2r} 
\left({7+{16e_0\over r^2}\over 1 + {e_0\over r^2}}\right)
+ {a^2 r\over r^2 + e_0}\left[1 - {\sqrt{a_3}\over 2r}\left({7r^2 + 15 e_0\over r^2 + e_0}\right)\right] = 0  \!\!\! \nonumber \\
\nd
whose solution will determine the full metric of the system. In the limit $a^2 \ll \alpha'$, this reduces to \eqref{difsim}, with $f_{3r}\equiv \partial_r f_3 =0$, of course. In fact, 
$a_3(r)$ can be solved for exactly and, in the limit of small $a^2$, the result is:
\bg\label{a3r1}
a_3(r) = {r^2\over (192)^2}\left[{9r^4{\cal F}_{1, 2}\over e_0^{7/8}(r^2+e_0)^{9/8}}-\left({9\sqrt{\pi}\Gamma(3/8)r^3\over e_0^{3/8}(r^2+e_0)^{9/8}} + 
{8\Gamma(7/8)(4e_0 + 13r^2)\over e_0+r^2}\right)\right]^2 + {\cal O}(a^2) , \nonumber \\ \nd
where
\bg\label{hyperg}
{\cal F}_{1, 2} \equiv {}_2F_1\left({1\over 2}, {7\over 8}, {3\over 2}, -{r^2\over e_0}\right) . \nd
We can also now solve for $\Delta$, which gives
\bg\label{delu}
\Delta(r) &=& {r^2\over 384}\left(1+{e_0\over r^2}\right)\left[{9r^4{\cal F}_{1, 2}\over e_0^{7/8}(r^2+e_0)^{9/8}}-\left({9\sqrt{\pi}\Gamma(3/8)r^3\over e_0^{3/8}(r^2+e_0)^{9/8}} + 
{8\Gamma(7/8)(4e_0 + 13r^2)\over e_0+r^2}\right)\right]. \nonumber\\ \nd
The small $r$ and large $r$ behaviors of $\Delta$ and $a_3$ are given by
\bg\label{bval} {e_0\over 24} ~ ~\stackrel{0 \leftarrow r}{\longleftarrow}~\Delta ~ \stackrel{r\rightarrow\infty}{\longrightarrow}~~ {r^2\over 6}\nd 
\bg\label{asibe} \frac{r^2}{144} ~ ~\stackrel{0 \leftarrow r}{\longleftarrow}~ a_3 ~ \stackrel{r\rightarrow\infty}{\longrightarrow}~~  \frac{r^2}{9} \nd 
which would imply that at $r = 0$, the two two-spheres have radii ${e_0\over 24}$ and ${e_0\over 24} + a^2$.\footnote{In fact, to state the latter radius, we really need to know $\Delta$ at order $a^2$, 
not just $a^0$. This can easily be done by solving \eqref{case1} to the next order.}
  So, the gravity dual should be given by 
a warped resolved deformed conifold with torsion.\footnote{Interestingly, this case somewhat resembles the type IIB case studied in \cite{klebmuru}, where the authors studied the wrapped five-brane scenario with closed three-form fluxes. Additionally one may refer to \cite{murthy1, murthy2} where the $d = 4$ example studied therein fits into
one of our large class of models.}

However, as mentioned earlier, very close to the origin $r\to 0$ and in the limit that $a^2$ is small, the $k$ five-branes are wrapped on an almost vanishing cycle and therefore appear as three-brane sources.  While
the ansatz \eqref{demand} suffices in a delocalized limit, in the localized limit a 
better ansatz would be to use a three-brane harmonic function rather than five-brane harmonic function --- i.e., to replace the $\frac{e_0}{r^2}$ term in \eqref{demand} with $\alpha'\frac{e_0}{r^4}$. Indeed we can also replace our above ansatze \eqref{demand} with a more generic one of the form:
\bg\label{demandnow}
h_1 \equiv 1 + {e_0\over r^2} + {\alpha' e_0\over r^4}\nd
where we see that for $r \gg \sqrt{\alpha'}$ we recover the warp factor \eqref{demand} while for small $r$, $r \ll \sqrt{\alpha'}$, this will convert to the localized three-brane
ansatze. Taking this into account  
converts \eqref{case1}
to the following differential equation for $a_3$ (with $\alpha' = 1$ for convenience) near $r \to 0$:
\bg\label{case1change}
{da_3\over dr} + {\sqrt{a_3}\over 2} - {a_3\over 2r}\left({7r^4 + 25 e_0\over r^4 + e_0}\right) + {\cal O}(a^2) = 0\nd
As expected, the large $r$ behavior is unaffected, but the small $r$ behavior does change. Now,
\bg\label{a3vicall}
a_3 ~ \stackrel{r\to0}{\longrightarrow} ~~ {r^2\over 441} , \qquad
\Delta ~ \stackrel{r\to0}{\longrightarrow} ~~ {e_0\over 42 r^2} ,
\nd
which means that the dilaton diverges at the origin and that the metric is affected by the branes near the origin. Naturally, this change also affects the torsion and the 
vector bundle, as we will see later. 

To complete the story, of course, we still must find a vector bundle that satisfies the Donaldson--Uhlenbeck--Yau equations and the Bianchi identity, which we postpone until section \ref{sec:hetbundle}.

\vskip.1in

\noindent ${\bullet}$ {\bf Case II}: $f_3 = f_2 \equiv {r\over 2\sqrt{a_3}}$

\vskip.1in

\noindent In this case, the dilaton and $\Delta$ are simply
\bg\label{delno}
e^\phi = 1 + {\alpha' k\over r^2} + {\cal O}\left(\frac{1}{r^3}\right),   \qquad\quad  \Delta(r) = {r^2 + e_0 \over 4}. \nd
As in case I, the two two-cycles at $r = 0$ will have sizes 
${e_0\over 4}$ and ${e_0\over 4} + a^2$, until we replace the five-brane harmonic function with the three-brane harmonic function.
Thus, the gravity dual should also be a resolved warped deformed conifold with torsion. 

The differential equation for $a_3$, \eqref{difeku}, becomes:
\bg\label{case2}
{da_3\over dr}\left(1 + {a^2\over \Delta}\right) - 4\sqrt{a_3} + {2a_3\over r}\left({5r^2 + 23e_0\over r^2+e_0}\right) - {16a^2\over r^2 + e_0}\left[\sqrt{a_3} 
-{a_3\over 2r}\left({3r^2 + 11 e_0 \over r^2 + e_0}\right)\right] = 0 . \nonumber\\ \nd
Again, this equation can be solved exactly in terms of Appell hypergeometric functions, but for simplicity we will focus on the small $a^2$ limit.  In this case, the $a_3$ equation \eqref{case2} is
\bg\label{case2sim} 
{da_3\over dr} - 4\sqrt{a_3} + {2 a_3\over r}\left({5r^2 + 23e_0\over r^2 + e_0}\right) + {\cal O}(a^2) = 0 \nd
which appears quite different from the $a^2 \ll \alpha'$ case for the first scenario.  The value for $a_3(r)$ at zeroth order in $a^2$ can now be written as:
\bg\label{a3now}
a_3(r) & = & {r^2\over (168)^2}\left(1+{e_0\over r^2}\right)^2\Bigg[1 + {495}\left({e_0\over r^2}\right)^{10} + {7425\over 2}\left({e_0\over r^2}\right)^9 
+{12045}\left({e_0\over r^2}\right)^8 + {87945\over 4}\left({e_0\over r^2}\right)^7\nonumber\\
&& + 24519 \left({e_0\over r^2}\right)^6 + {33789\over 2}\left({e_0\over r^2}\right)^5 + {47619\over 7}\left({e_0\over r^2}\right)^4 
+ {75339\over 56}\left({e_0\over r^2}\right)^3 
+ {55}\left({e_0\over r^2}\right)^2\nonumber\\ 
&& ~~~~~~~~~~~~~ - {11\over 2}\left({e_0\over r^2}\right) 
- {495}~\left({e_0\over r^2}\right)^3\left(1+{e_0\over r^2}\right)^8 {\rm log}\left(1 + {r^2\over e_0}\right)\Bigg]^2 + {\cal O}(a^2) \nonumber \nd
Surprisingly, the leading behavior at both large and small
$r$ is precisely the same as in case I!

\begin{figure}[tb]
\begin{center}
\includegraphics[
width=0.5\textwidth
]{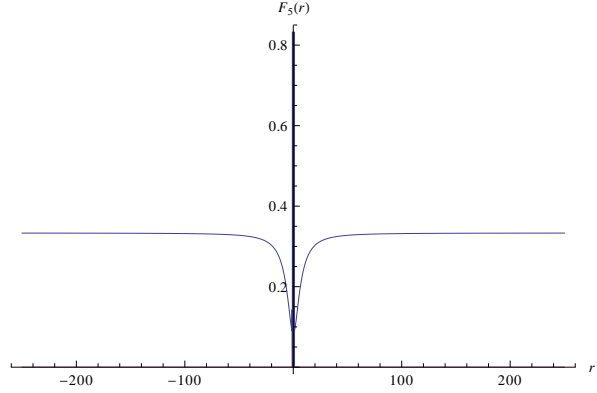}
\end{center}
\caption{{The behavior of the function $F_5(r)$ plotted with respect to the radial coordinate.
Note that the $F_5$ function asymptotes to ${1\over 3}$ at large $r$, while there is an expected singularity at the origin $r = 0$.  For the plot, we have chosen $e_0 = 10$ in units of $\alpha'$.}}
\label{f5plot}
\end{figure}

To get a better feel for how the metric behaves, let us define a function $X(r)$ in the following way:
\bg\label{xrdefina}
X(r) =&&  {11\over 2}\left({e_0\over r^2}\right)- {495}\left({e_0\over r^2}\right)^{10} - {7425\over 2}\left({e_0\over r^2}\right)^9 
-{12045}\left({e_0\over r^2}\right)^8 - {87945\over 4}\left({e_0\over r^2}\right)^7\nonumber\\
&& - 24519 \left({e_0\over r^2}\right)^6 - {33789\over 2}\left({e_0\over r^2}\right)^5 - {47619\over 7}\left({e_0\over r^2}\right)^4 
- {75339\over 56}\left({e_0\over r^2}\right)^3 
- {55}\left({e_0\over r^2}\right)^2\nonumber\\ 
&& ~~~~~~~~~~~~~~~~~~ + {495}~\left({e_0\over r^2}\right)^3\left(1+{e_0\over r^2}\right)^8 {\rm log}~\left(1 + {r^2\over e_0}\right)\nd
using which we can define another function $F_5(r)$ as:
\bg\label{F5f5}
F_5(r) \equiv {1\over 168}-{1\over 168}\left[X(r) + {e_0X(r)\over r^2} - {e_0\over r^2}\right].\nd
With these definitions, we can write the metric as:
\bg\label{metnowmet}
ds^2 = && ds^2_{0123} + \left(1+ {e_0\over r^2}\right)\Bigg[{dr^2\over 2 F_5} + {r^2\over 2}F_5 \left(d\psi 
 + \cos \theta_1 d\phi_1 + \cos \theta_2 d\phi_2\right)^2 \nonumber\\
&& ~~~~~~~~~~~~ + \left({r^2\over 4}+ {\widetilde a}^2\right)\left(d\theta_1^2 + \sin^2\theta_1 d\phi_1^2\right) 
+ {r^2\over 4}\left(d\theta_2^2 + \sin^2\theta_2 d\phi_2^2\right)\Bigg]\nd
where ${\widetilde a} \equiv {2a\over \sqrt{1+{e_0\over r^2}}}$ and the behavior of $F_5(r)$ is plotted in {\bf figure \ref{f5plot}}. It's interesting that even for 
very large $r$, the manifold doesn't quite
become a Calabi--Yau resolved conifold as the coefficients differ (see \cite{candelas} for details on the Calabi--Yau resolved conifold).

As in case I, to be careful about the limit that $r \to 0$, we should employ the three-brane harmonic function instead of the five-brane harmonic function. behavior using similar point of view as employed for case I.  This also applies to the 
string coupling and to $\Delta$, which will now be (with $\alpha'=1$)
\bg
e^\phi = 1+\frac{e_0}{r^4} , \qquad\quad  \Delta = \frac{r^2}{4} + \frac{e_0}{4r^2}. \nd
Then the equation for $a_3$ \eqref{case2sim} becomes:
\bg\label{case2change}
{da_3\over dr} - 4\sqrt{a_3} + {2a_3\over r}\left({5r^4+41e_0\over r^4 + e_0}\right) + {\cal O}(a^2) = 0 . \nd 
The components of the torsion will blow up near the origin because of the sources, but 
will asymptote to a constant value for large $r$.

We will solve for the vector bundle in section \ref{sec:hetbundle}.

\vskip.1in

\noindent ${\bullet}$ {\bf Case III}: $f_3 \ne f_2$

\vskip.1in

\noindent Now the equation for $a_3$ is simply \eqref{difeku}. In the limit with small resolution parameter, \eqref{difsim} tells us that
\bg\label{f3f3} \vert f_{3r} \vert  < |f_3| , \nd
where the inequality should be viewed order by order in a ${1\over r}$ expansion, 
the result will be similar to case I studied above.

\begin{figure}[tb]
\begin{center}
\includegraphics[
width=0.5\textwidth
]{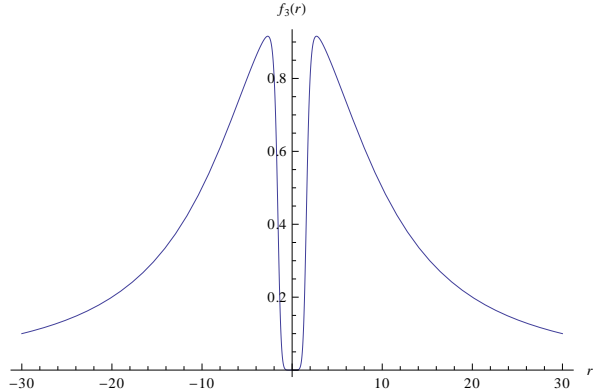}
\end{center}
\caption{{The behavior of the function $f_3(r)$ for case III plotted along the direction orthogonal to the wrapped five-branes without taking into account the 
source contributions and asymptotic modification as in \eqref{wfa0}.  
Note that the $f_3$ function asymptotes to $0$, whereas at 
the origin $r = 0$ it vanishes. For simplicity we have chosen $e_0 = 100$ and $b = 50$ in units of $\alpha'$. If we include \eqref{wfa0}, the behavior at $r \to 0$ and 
at $r \to \infty$ will change.}}
\label{f3behavior}
\end{figure}

For a class of examples in this case, we consider a form for $h_1$ that has the following piece-wise behavior:
\bg\label{wfa0}\label{h1CI}
h_1 = \left\{ \begin{array}{ccc}   1+ \frac{e_0}{r^2}\left(1+\frac{b}{r^8}\right)^{-1} , & \qquad &  r^8 \gg \widetilde{b}   \\   \frac{e_0}{r^2}\left(1+\frac{\widetilde{b}}{r^8}\right), & \qquad &    r^8 \ll \widetilde{b}   \end{array} \right.
\nd
where $\widetilde{b} \ll b$ are two parameters defining the class.\footnote{Alternatively, 
we could have followed a consistent set of conventions for $h_1$ across the cases and instead defined $f_3$ in this of piece-wise fashion. See {\bf figure \ref{f3behavior}}
for details.}

The ansatz \eqref{h1CI} is similar to the one considered in \cite{caris}, where the $r\to \infty$ behavior was like case I studied above. 
The difference is in the intermediate $r$ behavior.
The equation for $a_3$ in the intermediate region ${\widetilde{b}}^{1\over 8} \ll r \ll b^{\frac{1}{8}}$ becomes:
\bg\label{tdina}
{da_3\over dr} + {\sqrt{a_3}\over 2} - {2 a_3\over r}\left({4r^8 - 5b\over r^8 + b}\right) + {\cal O}(a^2) = 0 . \nd
Note that $e_0$ doesn't appear in the $a_3$ equation but does appear in the definition of $\Delta$ because of the relation \eqref{h1CI}.  In the limit that $a^2$ much 
smaller than any other scale in the theory, the solution for $a_3$ becomes:
\bg\label{a3kutta}
a_3(r) = \frac{\left(b+r^8\right)^2 \left[24 c\sqrt[8]{b+r^8} - {6 r^6\over b}\left(1 -{5\over 6}  \sqrt[8]{1+{r^8\over b}}
   ~{\cal D}_{1,2}\right)\right]^2}{576~r^{10}}\nd
where we have defined another hypergeometric function ${{\cal D}_{1,2}}$ and a constant $c$ in the following way:
\bg\label{anhype}
{{\cal D}_{1,2}} \equiv {}_2F_1\left({1\over 8}, {3\over 4}, {7\over 4}, - {r^8\over b}\right), ~~~~~~~~~
c = {5\Gamma(5/8)\cdot\Gamma(7/4)\over 24 ~b^{3/8}~\Gamma(5/8)} . \nd
The large $r$ behavior of $a_3$ is consistent with our earlier ansatz in \eqref{a3ans}, with leading behavior proportional to $r^2$.   On the other hand, for $\widetilde{b}^{1\over 8} \ll r \ll b^{1\over 8}$, the leading behavior of $a_3(r)$ is proportional to $\frac{b^{3/2}}{r^{10}}$, as 
one might have expected from our ansatz \eqref{h1CI}. 

Once we know $a_3$, we can readily get $\Delta$ from the ansatz \eqref{h1CI} and find:
\bg\label{deltush}
\Delta(r) ~ = ~ {e_0r^2\over 24} \left[24 c\sqrt[8]{b+r^8} - {6 r^6\over b}\left(1 -{5\over 6}  \sqrt[8]{1+{r^8\over b}}
   ~{\cal D}_{1,2}\right)\right] . \nd
$\Delta$ has the following asymptotics:
\bg\label{ranofdel}
&&\Delta(r)\big\vert_{{\widetilde{b}}^{1/8} \ll r \ll b^{1/8}} =  
{5e_0\over 48}\Bigg\{{\Gamma(5/8)\Gamma(7/4)\over \Gamma(1/4)}\left({r^8\over b}\right)^{1/4}\left[1 +
{1\over 8}\left({r^8\over b}\right)\right] -{1\over 5}\left({r^8\over b}\right)\Bigg\} + {\cal O}(r^{11})\nonumber\\ 
&& \Delta(r)\big\vert_{r \gg b^{1/8}} = {\Gamma(7/4)\over \Gamma(3/4)}\left[
{e_0\over 18} -  {e_0\over 99} \left({b\over r^8}\right) + {10 e_0\over 1881} \left({b\over r^8}\right)^2 - {20 e_0\over 5643} \left({b\over r^8}\right)^3\right]
+ {\cal O}\left({1\over r^{32}}\right) \qquad\quad  \nd
Since $\Delta$ is vanishing at small $r$, behaves like a resolved conifold in this regime, but at large $r$, $r\gg b^{\frac{1}{8}}$, both cycles attain finite sizes. 

\begin{figure}[tb]
\begin{center}
\includegraphics[
width=0.5\textwidth
]{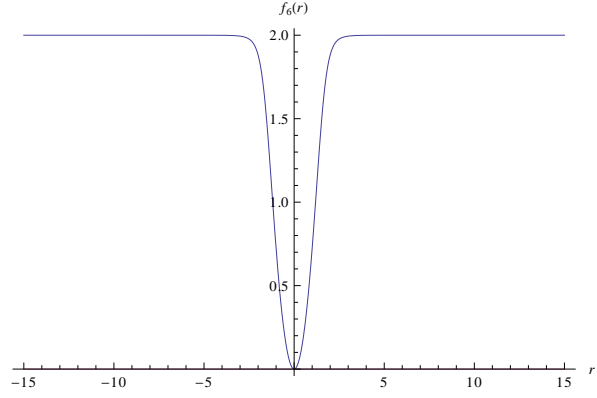}
\end{center}
\caption{{The behavior of the function $f_6(r)$ for case III that determines the metric \eqref{backmetd}.  
Note that the $f_6$ function asymptotes to ${8\Gamma(7/4)\over 3\Gamma(3/4)}$, while at 
the origin $r = 0$ it vanishes. Of course the behavior at the origin, as well as at large $r$, will be different if we include \eqref{wfa0}. 
For simplicity we have chosen $b = 100$ in units of $\alpha'$.}}
\label{f6plot}
\end{figure}

To simply encapsulate the metric behavior, 
we define another dimensionless function $f_6(r)$ as (see {\bf figure \ref{f6plot}}):
\bg\label{dimfuc}
f_6(r) \equiv  r^2\left[24 c\sqrt[8]{b+r^8} - {6 r^6\over b}\left(1 -{5\over 6}  \sqrt[8]{1+{r^8\over b}}
   ~{\cal D}_{1,2}\right)\right] ,
\nd
\noindent which asymptotes to ${8\Gamma(7/4)\over 3\Gamma(3/4)} + {\cal O}\left({1\over r^{8}}\right)$ at large $r$.
Then the background metric for this case can be expressed as:
\bg\label{backmetd}
ds^2 & = & ds^2_{0123} + {e_0\over r^2}\Bigg[{dr^2\over 1+{b\over r^8}} + {r^2\over 576}\left(1+{b\over r^8}\right)f_6^2 
(d\psi + \cos \theta_1 d\phi_1 + \cos \theta_2 d\phi_2)^2 \nonumber\\
&& ~~~~~~~~ + r^2\left({f_6 \over 48}+ a^2\right) \left(d\theta_1^2 + \sin^2\theta_1 d\phi_1^2\right) + 
 r^2\left({f_6 \over 48}\right)\left(d\theta_2^2 + \sin^2\theta_2 d\phi_2^2\right)\Bigg]\nd
where for $r \gg b^{1\over 8}$ the metric resembles the one of \cite{caris} at the leading order, albeit with a positive value for $b$. 
The metric \eqref{backmetd} does not cover the patches $r \to 0$ and $r \to \infty$, as the two asymptotes are given by the warp factor \eqref{wfa0} and the one 
studied for case I, respectively. 

If we now consider the small $r$ behavior, $r\ll \widetilde{b}^{\frac{1}{8}}$, the equation for $a_3$ becomes:
\bg\label{tdiyona}
{da_3\over dr} + {\sqrt{a_3}\over 2} - {2 a_3\over r}\left({4r^8 + 13\widetilde{b}\over r^8 + \widetilde{b}}\right) + {\cal O}(a^2) = 0 \nd
The behavior of ($a_3, \Delta, e^\phi$) near the origin $r\to 0$ is then given by:
\bg\label{a3deep}
&&a_3 ~ \to ~ {r^2\over 2304} + {r^2\over 512} \left({r^8\over \widetilde{b}}\right) + {c_1r^2\over 4096} \left({r^8\over \widetilde{b}}\right)^{3/2} 
+ {\cal O}(r^{16})\nonumber\\
&& e^\phi ~ \to ~ {13e_0\over 96 r^2} + {e_0\over 24 r^2}\left({\widetilde{b}\over r^8}\right) + {3c_1e_0\over 256 r^2}\left({r^8\over \widetilde{b}}\right)^{1/2} 
+ {\cal O}(r^6)\nonumber\\  
&& \Delta ~ \to ~ {13e_0\over 384} + {e_0\over 96}\left({\widetilde{b}\over r^8}\right) + {3 c_1 e_0\over 1024} \left({r^8\over \widetilde{b}}\right) ^{1/2} 
+ {\cal O}(r^6)\nd
where $c_1 = {\sqrt{\pi}\Gamma(3/8)\over \Gamma(7/8)}$. We see that 
the dilaton diverges in an expected fashion.

\subsection{Analysis of the vector bundles: global and local symmetries}
\label{sec:hetbundle}

Now that we have determined the background geometry, we must find a suitable vector bundle. In heterotic theories there are two sources of vector bundles related to the global and the local symmetries. From the original F-theory
perspective, the two symmetries are easy to see: the global symmetries come from the intersecting system of D7/O7s and the local symmetries come from the probe D3-branes.
In the following we will determine these symmetries explicitly.

\subsubsection{Global symmetries and the torsional backgrounds}

F-theory models have many enhanced global symmetry points.
In the absence of the second set of D7/O7s in the type IIB set-up, there would be multiple points with constant couplings \cite{senF, DM1}. One of the simplest ones is 
the $D_4$ point, leading to a $SO(8)$ global symmetry \cite{senF}. One would then think that, in the presence of the second set of D7/O7s, the enhanced symmetry group would be 
given by the $D_4 \times D_4$ point from Tate's algorithm \cite{tate}.  One 
might worry that there would be tensionless strings at such a colliding $D_4$ singularity \cite{bikmsv}, but this doesn't happen in our case because the orbifold singularity
associated with the generator $gh$ in \eqref{gimgen} hides half a unit a
$B_{\rm NS}$ flux wrapping the collapsed two-cycle (for example, see \cite{aspinwall}).

Beyond this, the colliding $D_4$ singularities actually do not even survive the orientifolding operation that we 
performed in earlier sections.  What we get instead of colliding $D_4$ singularities is colliding $A_3$ singularities, leading to a  
$U(4) \times U(4)$ global symmetry. 
This is because the orientifold projection is coupled with a gauge transformation, so the surviving symmetry group is the subgroup of $SO(8)$ that commutes with this gauge transformation:
\begin{equation}
\xymatrix@1@C=25mm{
D_4~~\ar[r]^{\left(\begin{smallmatrix}~{\cal M}&0\\~0 & -{\cal M}\end{smallmatrix}\right)}
\ar[r]&~~ A_3 ,
}\label{eq:freeX}
\end{equation}
where the $4\times 4$ matrix ${\cal M}$ is given by:
\bg\label{matmat}
{\cal M} ~ \equiv ~ \left(\begin{matrix}~0&~0&~1&~0\\~0&~0&~0&~1\\-1&~0&~0&~0\\~0&-1&~0&~0\end{matrix}\right).
\nd
The adjoint hypermultiplet of $SO(8)$ picks up a minus sign when conjugated with the matrix ${\cal M}$, leading to two hypermultiplets in the 
${\bf 6}$ of $SU(4)$. Thus, putting four D7 branes on top of the O7 plane, and including the necessary $U(1)$ factors, we get a
$U(4)$ global symmetry.  The full global symmetry group at the constant coupling point is then given by $U(4)_u^4 \times U(4)_v^4$. 

Of course, if we arrange the branes slightly differently, we can have other global symmetry groups.  For example, we could go to break to $SU(2)_u^8 \times SU(2)_v^8$ by allowing the seven-branes to move in pairs \cite{sengimon, sengimon2}, though this will no longer correspond to constant coupling. This is exemplified by the curve \eqref{deffg} with the choice 
given by \eqref{fvgv}. 
An example of this would be the following choices:
\bg\label{choices}
&& f(u,v) = (u-u_1) F(u, v)\nonumber\\
&&g(u,v) = (u-u_1) G(u, v)\nd
where $F$ and $G$ are chosen not to have additional zeros at $u = u_1$.  In terms of the coefficients in \eqref{deffg}, this is equivalent to the following choices:
\bg\label{choofofg}
a_{11}= b_{11}= c_{11}= d_{1}= m_{11}= n_{11}= s_{11}= p_{11}= q_{11}= r_1 \equiv u_1, ~~~~~ (A_{11}, M_{11}) \ne 0 \nd 
with all other coefficients not equal to $u_1$. 
The discriminant locus will then be given by (the discriminant $\Delta$ should not be confused with the parameter in the metric of the same name)
\bg\label{dislok}
\Delta ~ = ~ (u-u_1)^2 \left[4(u-u_1)F^3 + 27 G^2\right] ~ \equiv ~ (u-u_1)^2 \widetilde{\Delta}(u, v)\nd
so that we have only a pair of seven-branes together, resulting in a classical $U(2)$ global symmetry \cite{gimpol}.   This is the type of singularity we will study in this paper.\footnote{Of course, we could have moved in the opposite direction, enhancing the global symmetry instead. For example, 
if we modify \eqref{dislok} to:
\begin{equation*}
\Delta ~ = ~  (u-u_1)^{10} (v-v_1)^{10} \widetilde{\Delta}(u, v) , \end{equation*}  
such that the vanishing of $\widetilde{\Delta}(u, v)$ leads to no new enhanced symmetry points (other than the $U(1)$'s), 
then we could achieve the full $E_8 \times E_8$ global symmetry in our setup, though it 
is believed that such a point does not occur along the constant coupling branches of F-theory (see for example \cite{jatkar}).}

Of course, we expect that the variety of global symmetries that arise in the F-theory construction can be reproduced by varying moduli on the heterotic side.  Actually verifying this, of course, is quite involved since we have to satisfy both the Donaldson--Uhlenbeck--Yau equations as well as the Bianchi identity:
\bg\label{vecbund}F_{ab} = F_{\bar a \bar b} = g^{a \bar
b}F_{a \bar b} = 0, \qquad {1\over 30}{\rm tr}~F \wedge F = {\rm tr}~R_+ \wedge
R_+ - {1\over \alpha'} d{\cal H}
\nd
where, again, $R_+$ is the Ricci two-form constructed from a metric-compatible connection with torsion --- the ``plus'' connection.  When we have torsion, the dilaton is not constant and so on the type II side we will not 
be at the constant coupling point of the Gimon--Polchinski
model.  The functional form for the axio-dilaton would determine the resulting positions of the
branes and planes in this scenario, and therefore the gauge
bundle. 

Another reason this is challenging on the heterotic side is because the global symmetry generically will also arise from the orbifold five-planes.  This is different from the $SO(32)$ case studied earlier \cite{chen1}, where the global symmetries
appeared only from the usual heterotic vector bundles with appropriate Wilson lines.  In our case, we expect part of the non-abelian global symmetry to appear from the 
twisted sectors states of the orbifold.\footnote{Unfortunately, this non-abelian enhancement is not visible from string perturbation theory.  An alternative way 
to see this would be to dualize to a singular type IIA geometry, where the symmetry enhancement could be 
computed using the techniques of \cite{senE1, senE2}. We thank Ashoke Sen for clarifying this point.}

In this paper, we will study only $U(2)$ global symmetries arising from a nontrivial vector bundle, and none from the orbifold.  This corresponds to an F-theory curve of the form \eqref{choices} with discriminant locus given by \eqref{dislok}, 
leading to a $U(2) \times U(1)^n$ symmetry.  We then decouple the $U(1)^n$ so that the heterotic global symmetry is just 
$U(2)$, all coming from the vector bundle and none from the orbifold planes.\footnote{Actually, there could still be a single localized $U(1)$ 
that would be difficult to decouple.  We will ignore this subtlety and only consider a $U(2)$ bundle for simplicity.} 

We now work out an explicit example.  We take the metric ansatz \eqref{mopla} 
with $c_4=\cos\theta_1$, $c_7=\cos\theta_2$, $c_1=\sin\theta_1$, and $c_5=\sin\theta_2$, and as before $H_i$ only depends on the radial coordinate $r$ (we could just as well have chosen the metric ansatz \eqref{bgmetDf}). The torsion polynomial \eqref{hettros} then gives us:
\begin{eqnarray}\label{binki}
d{\cal H} = \frac{1}{\sqrt{H_2}}\Big(A_de_3\wedge e_4+B_de_1\wedge e_2\Big)\wedge e_5\wedge e_6
\end{eqnarray}
where $e_i$ are the vielbeins \eqref{vielgood}, and $A_d$ and $B_d$ are defined as:
\bg\label{adbddef}
A_d~\equiv ~  A {\del\over \del r} {\rm log}(H_4\sqrt{H_1}) - A_r, ~~~~~~~ B_d~\equiv ~ B {\del\over \del r} {\rm log}(H_3\sqrt{H_1}) - B_r\nd
with the subscript $r$ being the derivative with respect to $r$, as before.
The two functions $A$ are $B$ are now defined from the torsion polynomial as:
\begin{eqnarray}\label{abra}
A&=& -\frac{2\phi_r}{\sqrt{H_2}} + \frac{H_{3r}}{H_3\sqrt{H_2}} - \frac{\sqrt{H_1}}{H_3}\nonumber\\
B&=& - \frac{2\phi_r}{\sqrt{H_2}} + \frac{H_{4r}}{H_4\sqrt{H_2}} - \frac{\sqrt{H_1}}{H_4} .
\end{eqnarray}
The Bianchi identity in \eqref{torclosed} then implies that the RHS of \eqref{binki} should have three contributions: one from the $k$ five-brane sources, one from 
${\rm tr}~R_+ \wedge R_+$, and one from ${\rm tr}~F \wedge F$. The torsional connections have been worked out 
in {\bf appendix \ref{app3}}. Using these, we obtain:
\begin{eqnarray}\label{rashi}
\textrm{Tr}~R_+\wedge R_+ ~= ~ (R_1e_1\wedge e_2+R_2e_3\wedge e_4)\wedge e_5 \wedge e_6
\end{eqnarray}
where $R_1$ and $R_2$ are given by:
\bg\label{r1}
R_1&=&\frac{3}{2}\Bigg[\left(8+6B\sqrt{H_2}-\frac{3H_2}{H_3}+B^2H_3^2+\frac{H_3'}{H_1H_3}\right)
\left(\frac{2H_2'}{H_3}-\frac{BH_2'}{\sqrt{H_2}}-\frac{2H_2H_3'}{H_3^2}-2\sqrt{H_2}B'\right)\nonumber\\
&&+\frac{4(\sqrt{H_2}-AH_4)}{H_4^3}\left(2H_2H_4^2A'-2\sqrt{H_2}H_2'H_4+AH_4^2H_2'+2H_2^{3/2}H_4'\right)\nonumber\\
&&+\frac{2}{H_1^2H_2^2H_3}\Big(H_2H_3'+H_3(2H_2'+B\sqrt{H_2}H_3')\Big)\Big(H_1H_2^{'2}+H_2(H_1'H_2'-2H_1H_2'')\Big)\Bigg]\nd
\bg\label{r2}
R_2&=&\frac{3}{2}\Bigg[\left(8+6A\sqrt{H_2}-\frac{3H_2}{H_4}+A^2H_4^2+\frac{H_4'}{H_1H_4}\right)
\left(\frac{2H_2'}{H_4}-\frac{AH_2'}{\sqrt{H_2}}-\frac{2H_2H_4'}{H_4^2}-2\sqrt{H_2}A'\right)\nonumber\\
&&+\frac{4\left(\sqrt{H_2}-BH_3\right)}{H_3^3}\left(2H_2H_3^2B'-2\sqrt{H_2}H_2'H_3+BH_3^2H_2'+2H_2^{3/2}H_3'\right)\nonumber\\
&&+\frac{2}{H_1^2H_2^2H_4}\Big(H_2H_4'+H_4(2H_2'+A\sqrt{H_2}H_4')\Big)\Big(H_1H_2^{'2}+H_2(H_1'H_2'-2H_1H_2'')\Big)\Bigg]
\nd
Away from the origin of $r$, $d{\cal H}$ receives no contributions from delta-function sources in \eqref{torclosed}, so we next work on the vector bundle, which must satisfy:
\begin{eqnarray}\label{mukhtar}
\frac{1}{30}\textrm{tr}~F\wedge F = \Bigg[\left(R_1-\frac{A_d}{\alpha'\sqrt{H_2}}\right)e_1\wedge e_2
+\left(R_2-\frac{B_d}{\alpha'\sqrt{H_2}}\right)e_3\wedge e_4\Bigg]\wedge e_5 \wedge e_6
\end{eqnarray}
Our aim now is to determine the $U(2)$ bundle, which we will assume comes from the remnants of the seven branes and planes in the $u \equiv x^4 + ix^5$ plane, 
ignoring $U(1)$s from the orbifold states.  The $U(2)$ bundle can then be expressed in terms of the Pauli matrices, and one simple ansatz will be:
\begin{eqnarray}\label{ftko}
F=(\mathfrak{f}_1e_1\wedge e_2+\mathfrak{f}_2e_3\wedge e_4+\mathfrak{f}_3e_5\wedge e_6)I+(\mathfrak{f}_4e_1\wedge e_2+\mathfrak{f}_5e_3\wedge e_4)\sigma^1
\end{eqnarray}
where $\sigma^1$ is the first Pauli matrices and $\mathfrak{f}_i$ are real functions of $r$ that
will be determined shortly. The choice \eqref{ftko} then
immediately implies the following value for ${\rm tr}~F \wedge F$:
\begin{eqnarray}\label{muksin}
\frac{1}{30}\textrm{tr}~F\wedge F&=&\frac{2}{15}\Big[(\mathfrak{f}_1\mathfrak{f}_2-\mathfrak{f}_4\mathfrak{f}_5)e_1\wedge e_2\wedge e_3\wedge e_4+\mathfrak{f}_2\mathfrak{f}_3~e_3\wedge e_4\wedge e_5 \wedge e_6\nonumber\\
&&+\mathfrak{f}_1\mathfrak{f}_3~ e_1\wedge e_2\wedge e_5\wedge e_6\Big]
\end{eqnarray}
Comparing \eqref{mukhtar} with \eqref{muksin}, we see the following conditions on the $\mathfrak{f}_i$:
\begin{eqnarray}\label{jomader}
\mathfrak{f}_1\mathfrak{f}_2=\mathfrak{f}_4\mathfrak{f}_5, \qquad \mathfrak{f}_1\mathfrak{f}_3=R_1-\frac{A_d}{\alpha'\sqrt{H_2}}, \qquad \mathfrak{f}_2\mathfrak{f}_3=R_2-\frac{B_d}{\alpha'\sqrt{H_2}} \ .
\end{eqnarray}
We still have to satisfy the Donaldson--Uhlenbeck--Yau conditions, which are equivalent to
$F_{mn}J^{mn}=0$.  Imposing this on our ansatz immediately implies the following two additional constraints on $\mathfrak{f}_i$:
\begin{eqnarray}\label{cond2}
\mathfrak{f}_1+\mathfrak{f}_2+\mathfrak{f}_3=0, \qquad \mathfrak{f}_4+\mathfrak{f}_5=0 .
\end{eqnarray}

These turn out to be enough to determine the functional forms for $\mathfrak{f}_i$ uniquely. Combining \eqref{cond2} and \eqref{jomader} gives us the following:
\begin{eqnarray}\label{joishila}
&&\mathfrak{f}_1=\pm\frac{R_1-A_d/\alpha'\sqrt{H_2}}{\sqrt{-R_1-R_2+(A_d+B_d)/\alpha'\sqrt{H_2}}},\nonumber\\
&&\mathfrak{f}_2=\pm\frac{R_2-B_d/\alpha'\sqrt{H_2}}{\sqrt{-R_1-R_2+(A_d+B_d)/\alpha'\sqrt{H_2}}},\nonumber\\
&&\mathfrak{f}_3=\pm\sqrt{-R_1-R_2+(A_d+B_d)/\alpha'\sqrt{H_2}},\nonumber\\
&&\mathfrak{f}_4=-\mathfrak{f}_5=\pm\sqrt{\frac{(R_1-A_d/\alpha'\sqrt{H_2})(R_2-B_d/\alpha'\sqrt{H_2})}{R_1+R_2-(A_d+B_d)/\alpha'\sqrt{H_2}}}
\end{eqnarray}
with $R_1$ \& $R_2$, $A$ \& $B$, and the warp factors $H_i$, given by \eqref{rashi}, \eqref{abra}, and \eqref{mopla}, respectively. 
Note that $\mathfrak{f}_i$ are defined for $r > 0$.  At the origin, we have to account for the sources in \eqref{torclosed} to determine corrections to $\mathfrak{f}_i$. Therefore with \eqref{ftko} and \eqref{joishila}, we have the $U(2)$  
global symmetries for all the three cases discussed earlier. We now go to the issue of local symmetries.

\subsubsection{Local symmetries and gauge groups in the strong coupling limit}

To study the local symmetries, or the gauge groups on the wrapped heterotic five-branes, it will be easier to study them without going to the decoupling limit. 
As we  discussed briefly in the introduction, once we are away from the resolved conifold point, there are two possible theories in six-dimensions:
theories with six-dimensional vector multiplets and theories with six-dimensional tensor multiplets. These two theories are related to $SO(32)$ and $E_8 \times E_8$ 
heterotic theories, respectively. However our heterotic theories appeared in conjunction with other string theories that were related by a series of dualities. In fact, 
we considered three different theories which are related by T- or S-dualities:
\begin{itemize}
\item Type IIB theory with $k$ D3-branes probing two sets of D7/O7 branes/orientifolds, as in {\bf table \ref{gpbranes}}.
\item Type I theory with $k$ D5 branes and D5${}^\prime$/O5${}^\prime$ branes/orientifolds, as in {\bf table \ref{gpbranesT45}}.
\item Heterotic $E_8 \times E_8$ with a set of $k$ NS5-branes and a set of orbifold five-planes O${}_r$5 (on top of which we could also layer NS5-branes), as in {\bf table \ref{gpbranesnow}}.
\end{itemize}
In the type IIB model, we can displace the D3-branes along the $u=x^4+ix^5$ direction and along the $v=x^8+ix^9$ direction.
The gauge group on the $k$ D3-branes is $Sp(2k)\times Sp(2k)$. What happens in the strong coupling regime? We know that when the type IIB model is lifted to an elliptically fibered F-theory model, the
orientifold planes split into two sets of
non-local seven-branes and so we would expect to have four different local 7-branes.

Nevertheless, the story is different as we know from \cite{sengimon}. There it was shown that the location of a set of seven-branes for large 
$v$ is identical to the the location of
the other set of seven-branes for large $u$ which implies that the two seven-branes in which the $u$ plane splits join the two seven-branes in which 
the $v$ plane splits. In the
non-perturbative regime, the breaking of the $Sp(2k) \times Sp(2k)$
group is related to the existence of massless hypermultiples which are identified with the monopoles/dyons for either one of the two
 $Sp(2k)$ groups. The interpretation of the results
in \cite{sengimon} is that these two monopoles can be deformed into each other.

Our question is how we translate the results of \cite{sengimon} into type I and then to the heterotic picture.  To do this, we are going to invoke the discussion of
\cite{bds}. Since $u$ and $\bar{u}$ are the directions that must be T-dualized to relate the type IIB frame to the type I frame, the IIB position of the D3-branes in the $u$-plane maps into Wilson
lines that are switched
 on on both the type I D5-branes as well as the background D9/O9 branes/planes. 
For generic Wilson lines, one of the $Sp(2k) \subset Sp(2k)\times Sp(2k)$ groups is completely broken while the other $Sp(2k)$ is kept intact; the
gauge group on the D9/O9 branes/planes is broken to
$U(1)^{16}$.

After an S-duality, one obtains a heterotic string a generic $U(1)^{16}$ vector bundle  and with NS5-branes at the O${}_r$5-plane,
implying an $Sp(k)$ local gauge group.  
In the previous section, we saw that we then preserved only a $U(2)$ subgroup (ignoring other $U(1)$ factors) of the full global group.

What happens now if one displace the D3-branes in the $v$ and $\bar{v}$ directions? The D3-branes would then be moved away from the D7${}^\prime$ branes and it would then be expected that
the only orientifold projection would be due to the other set of D7/O7 branes/planes. This then implies that the gauge group is just $Sp(2k)$. 
After T-dualizing in the $x^{4, 5}$ directions, the
$k$ type I D5-branes would again be separated from the D5${}^\prime$/O5${}^\prime$ brane/plane system, 
and the only orientifold projection would now be due to the D9/O9 branes/planes. In this case, we again have
a gauge group $Sp(2k)$ on the D5-branes with the gauge group on the D9/O9 system appearing as a global symmetry on the D5-branes (of which again we can only keep the 
$U(2)$ subgroup as an illustrative example).

The continuous transformation between the monopoles of the two $Sp(2k)$ groups in the  $Sp(2k) \times Sp(2k)$ theory on the type IIB D3-branes, is mapped in the type I language into 
a deformation of the group on the D9/O9 branes/planes from the unbroken subgroups that are invariant under orientifold operations to the completely broken $U(1)^{16}$. In
this scenario the $U(2)$ global symmetry may be assumed to come from the twisted sector states alone.

\section{Geometric Transitions and Heterotic Gauge/Gravity Duality}
\label{sec:transition}

Our analysis in the previous sections yield a heterotic background with wrapped NS5-branes on a non-K\"ahler warped resolved 
conifold. The full background is\footnote{A summary of the various backgrounds 
before and after geometric transitions is given in {\bf Appendix \ref{bocksol}}.}:
\bg\label{fulbag}
&& ds^2 = ds^2_{0123} + {2\Delta \over r \sqrt{a_3}} dr^2 + {2 \Delta \sqrt{a_3}\over r}(d\psi + \cos \theta_1 d\phi_1 + \cos \theta_2 d\phi_2)^2 \nonumber\\
&& ~~~~~~~~~~~~ + (\Delta + a^2)\left(d\theta_1^2 + \sin^2\theta_1 d\phi_1^2\right) + \Delta \left(d\theta_2^2 + \sin^2\theta_2 d\phi_2^2\right)\nonumber\\
&& {\cal H} = \sqrt{H_1\over H_2}\left(G_1 ~\sin \theta_2 d\theta_2 \wedge d\phi_2 
+ G_2 ~ \sin \theta_1 d\theta_1 \wedge d\phi_1\right)\wedge e_\psi\nonumber\\
&& e^\phi = {4\Delta\over r^2}\nd  
where $G_1$ and $G_2$ are given in \eqref{epsidef} in terms of $\Delta$ and the resolution parameter $a^2$.  All we need to do is substitute the values for 
$a_3$ and $\Delta$ for the three cases that we studied in section \ref{sec:hetslns}.  For case I, $a_3$ and $\Delta$ are given respectively in \eqref{a3r1} and \eqref{delu}; whereas for 
case II, $a_3$ and $\Delta$ are given respectively in \eqref{a3now} and \eqref{delno}. For both cases, we see that the string coupling $e^\phi$ has standard
behavior near the NS5-branes, namely it blows up as $\frac{1}{r^4}$.  Thus, the core of the 5-brane is described by a strongly coupled theory.

\begin{figure}[htb]
\begin{center}
\includegraphics[width=0.5\textwidth]{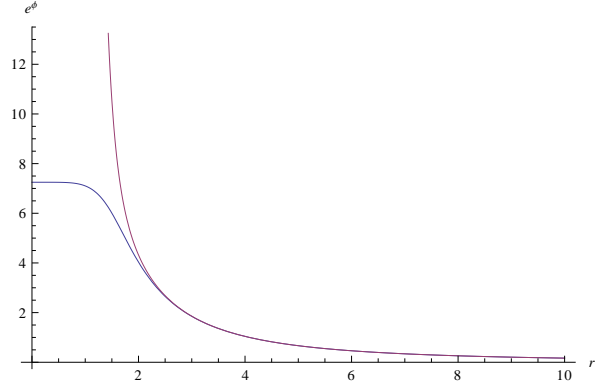}
\end{center}
\caption{{The behavior of the string coupling $e^\phi$ along the direction orthogonal to the wrapped five-branes for case III. If we only use
the warp factor \eqref{h1CI} then near the origin, with $e_0 = 100$ and $b = 50$ in units of $\alpha'$, the string coupling becomes 
$e^\phi = {5e_0\over 12 b^{1/4}} {\Gamma(5/8)\Gamma(7/4)\over \Gamma(1/8)}$ which is large but finite. This is where the regime of validity of the curve terminates.
The modification near the origin is shown by the red curve where we take the warp factor \eqref{wfa0}. We see that the dilaton blows up near the core of the 
five-branes in an expected fashion.}}
\label{dildil}
\end{figure}

For case III, the situation is slightly different because there are three regimes of interest. For region I, which is close to the origin, the values for  
$a_3$ and $\Delta$ can be read off from \eqref{a3deep}. For the intermediate region, $a_3$ and $\Delta$ can be read off from
\eqref{a3kutta} and \eqref{deltush} respectively. Finally for the asymptotic region $r \to \infty$, $a_3$ and $\Delta$ can be read from the asymptotic region for
case I studied earlier. If we extrapolate the value of the string coupling in the intermediate region to the origin, then it will appear as though 
the string coupling $e^\phi$ doesn't blow up at the origin, but attains the following finite value:
\bg\label{dilapna}
e^\phi = {5e_0\over 12 b^{1/4}} {\Gamma(5/8)\Gamma(7/4)\over \Gamma(1/8)}\nd
whereas it vanishes at infinity. However from \eqref{a3deep} we know that $e^\phi$ blows up when $r\to 0$, so the two curves for the two regions have to be attached 
at the value where the string coupling is \eqref{dilapna}. 
The plot for the dilaton is depicted in {\bf figure \ref{dildil}}. 

In the {\it absence} of the five-brane sources, the behavior of the dilaton is completely governed by the warp factor \eqref{h1CI}. 
One of the main reason for such a behavior of dilaton may stem from the fact that the $f_3(r)$ function for case III vanishes at the origin (as can be seen from 
{\bf figure \ref{f3behavior}}). This is reminiscent of the configuration studied in \cite{kiritsis} where the string coupling has somewhat similar behavior. The 
difference therein is that the five-branes are distributed over some orthogonal $S^3$ in the case of \cite{kiritsis}, whereas in our case this effect is captured by the $f_3(r)$ function that 
is {\it distributed} over the radial direction. This however doesn't mean that the five-branes in our case are distributed along $r$, but 
it implies only that the effective 
warp factor has the distribution given by \eqref{h1CI}.\footnote{For example $f_3 = 1$ for case I doesn't imply that the five-branes are distributed equally along the radial 
direction.}

\subsection{The torsion in the heterotic theory}
\label{sec:torsion}

\begin{figure}[htb]
                \begin{center}
		\vspace{-1 cm}
		\centering
		\subfloat[Behavior near the origin]{\includegraphics[width=0.5\textwidth]{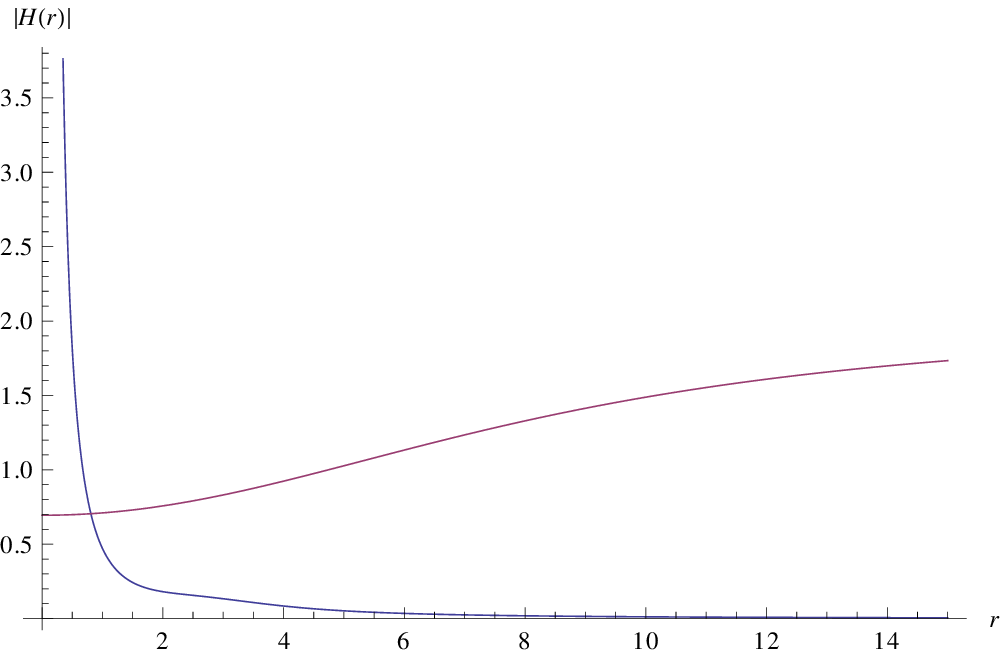}}
		\subfloat[Behavior far from the origin]{\includegraphics[width=0.5\textwidth]{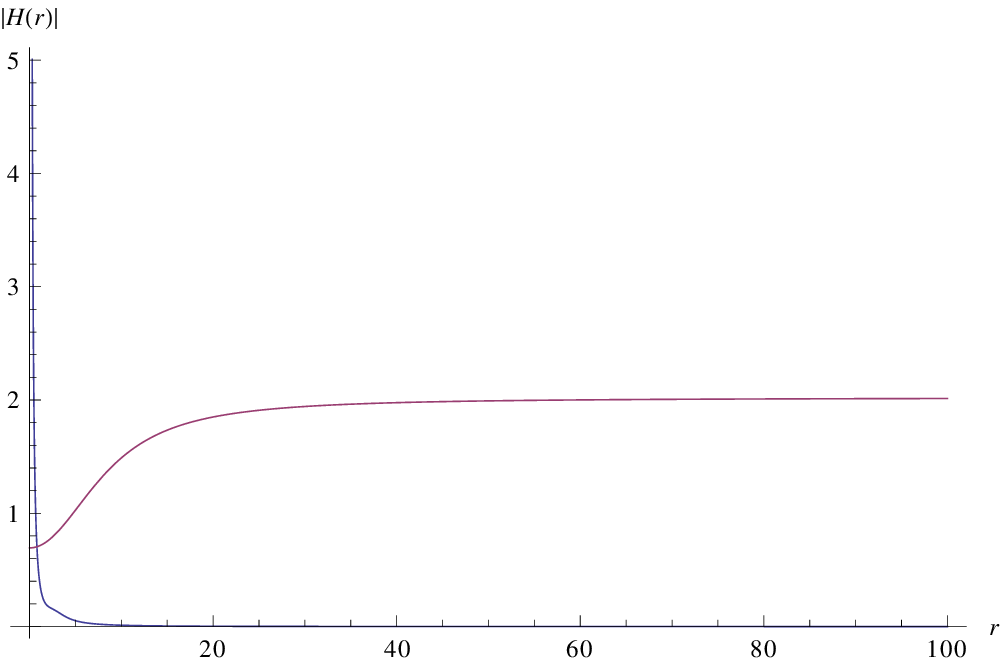}}
			\end{center}
		\caption{{The plot of the torsion coefficient $\vert H(r)\vert$ for case I, i.e., $H(r) = -\sqrt{a_3}\left(\partial_r\Delta - 2\Delta/r\right)$.  The curve 
in blue represents the localized three-brane ansatz 
\eqref{case1change} while the curve in red is the standard ansatz without the modification at small $r$. At large $r$, the effect of the blue curve is negligible
and the red curve dominates, asymptoting to a constant value,
while near the origin the blue curve captures the source contribution.  
For simplicity, we choose $\alpha'=1$ and $e_0 = 100$.}}
\label{Hcase1tot}
\end{figure}

The torsion for all the three cases can be computed from \eqref{hettros} by including the asymmetry factor \eqref{epsidef}. However it will be instructive to analyze 
the functional form of the torsion to gain more information of the background. If we ignore the ${\cal O}(a^2)$ correction in the expression for the dilaton, $\phi$, then the torsion can be written in the following form:
\bg\label{turna}
{\cal H} &=& -\sqrt{a_3}\left(\Delta_r - {2\Delta\over r}\right)\left(\Omega_1+\Omega_2\right)\wedge e_\psi\nonumber\\
&=& -\sqrt{a_3}\Delta \phi_r\left(\Omega_1+\Omega_2\right)\wedge e_\psi + {\cal O}(a^2) \nd 
where $\Omega_i = \sin \theta_i d\theta_i \wedge d\phi_i$ and $\phi_r$ is the derivative of the dilaton without the $a^2$ contribution, i.e., 
$\phi_r = {\Delta_r\over \Delta} - {2\over r} + {\cal O}(a^2)$.  This form of the torsion, with a derivative of the dilaton, is reminiscent of the standard five-brane background.
 
\begin{figure}[tb]
\begin{center}
\includegraphics[width=0.5\textwidth]{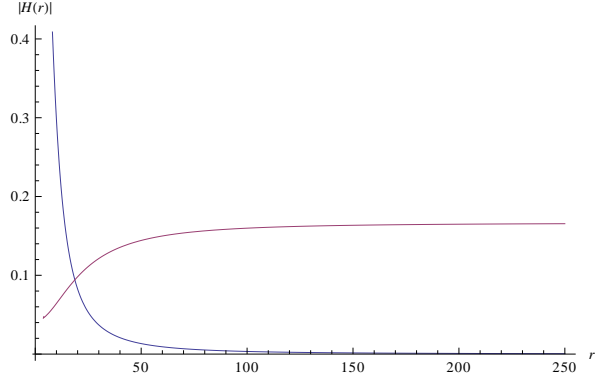}
\end{center}
\caption{{The plot of the torsion coefficient $\vert H(r)\vert$ for case II. As before, note that the torsion becomes finite at infinity. 
We have again chosen $e_0 = 100$ in units of $\alpha'$. 
The 
blue curve is plotted using the $r \to 0$ modification as \eqref{case2change}, while the red curve is for \eqref{case2sim}.}}
\label{Hcase2}
\end{figure}

\begin{figure}[tb]
\begin{center}
\includegraphics[width=0.5\textwidth]{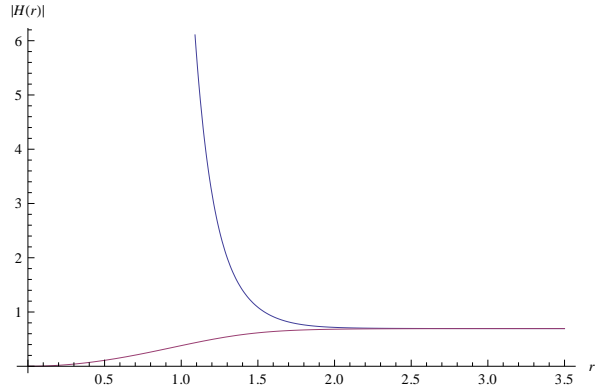}
\end{center}
\caption{{The plot of the torsion coefficient $\vert H(r)\vert$ for case III plotted for the $r\to 0$ and the
intermediate region. Near the origin the torsion blows up, signifying the presence of the 
sources, as depicted by the blue curve using \eqref{wfa0}. For large $r$ the blue curve coincides with the red curve drawn using \eqref{h1CI}, confirming 
our ansat\"ze that the warp factor for small $r$, which is \eqref{wfa0}, should smoothly transform to the warp factor for $r > \vert{\widetilde b}\vert$, which is 
\eqref{h1CI}. Notice that for $r \to \infty$ the torsion becomes finite as in cases I and II. Again, this behavior is necessary for maintaining 
zero energy in the system \eqref{potcar}. We have chosen $e_0 = 100,~ \widetilde{b} = 50$ and $b = -50$ in units of $\alpha'$.}}
\label{Hcase3}
\end{figure}

\begin{figure}[tb]
\begin{center}
\includegraphics[width=0.5\textwidth]{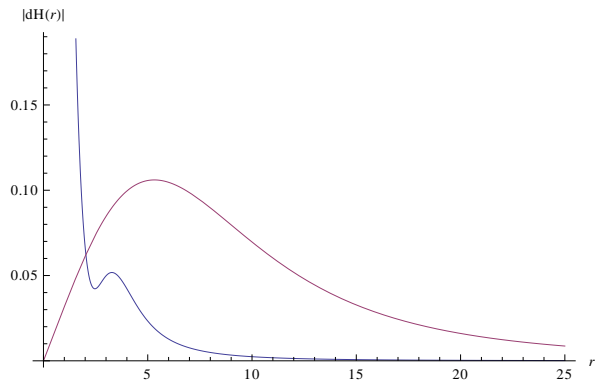}
\end{center}
\caption{{A plot for the coefficient of $d{\cal H}$ for case I, with $e_0 = 100$ as before. The blue curve clearly depicts the presence of the sources, while the 
red curve vanishes near the origin. The color codings are same as in the earlier plots.}}  
\label{dHcase1}
\end{figure}

The plots of the three cases are given in {\bf figure \ref{Hcase1tot}}, {\bf figure \ref{Hcase2}}, and 
{\bf figure \ref{Hcase3}}.  Note that in all three cases, the torsion blows up near the origin, signaling the existence of the wrapped five-branes sources, while
they become constants at large $r$. The reason for this asymptotic behavior is because of the potential \eqref{potcar}, which needs to vanish for all three cases. 
We can also plot the coefficient of $d{\cal H}$ and can see the presence of delocalized sources in {\bf figure \ref{dHcase1}}. 

Note that while we performed this analysis at leading order in small $a^2$, it can be performed to arbitrary orders in the 
resolution parameter.

\subsection{Type I duality frame}
\label{sec:typeI}

Under an S-duality, we go back to the type I background that we studied earlier. The metric in type I is the usual S-dual transform of the metric \eqref{manipul}, or 
of the metric \eqref{mopla}:
\begin{eqnarray}
\textstyle ds^2=e^{-\phi}[ds_{0123}^2+H_2dr^2+H_1(d\psi+c_4d\phi_1+c_7d\phi_2)^2+H_3(d\theta_1^2+c_1^2d\phi_1^2)+H_4(d\theta_2^2+c_5^2d\phi_2^2)] \!\!\!\!\!\!\!\! \nonumber \\
\end{eqnarray}
where the torsion ${\cal H}$ becomes the RR three-form $F_3$. The type I background is useful because it is the closest to the well known type IIB background with wrapped 
D5-branes on the two-cycle of a resolved conifold, namely the one studied in \cite{Vafa:2000wi, anke1, chen1}. The difference now is that we have an additional set of 
D5-branes, which we can trace through a geometric transition 
and obtain the S-dual of the gravity dual of the heterotic side configuration.  Finding the heterotic gravity dual this way also involves understanding the following:
\begin{itemize}
\item The type I D9/O9 system will undergo some changes on its world-volume after the geometric transition, which corresponds to changing the vector bundle.
\item The set of type I D5-branes wrapped on the two-cycle ${\bf P}^1$, parameterized by $x^{6, 7}$, will also dissolve into geometry and flux. This 
means that after the geometric transition, we will only see torsion and all the heterotic five-branes will have dissolved into geometry and flux.
\end{itemize}

We can make the duality more precise as follows: the type I background that we are interested in, under 
a mirror transformation, leads to a type I${}^\prime$ background given by:
\bg
{\rm Type ~IIB~~on}~~~~{{\cal C}_{\rm res}\over \{1, \Omega\}}~~~{}^{\rm mirror}_{{~~\longrightarrow}}~~~
{\rm Type ~IIA~~on}~~~~
{{\cal C}_{\rm def}\over\{1, \Omega\cdot {\cal I}_{\phi_1 \phi_2 \psi}\cdot (-1)^{F_L}\}}\nd
where ${\cal C}_{\rm def}$ is the deformed conifold and the mirror symmetry is defined in the usual way by three T-dualities along the isometry directions 
$\phi_1$, $\phi_2$, and $\psi$ \cite{SYZ}.  The fixed points of the orientifold action in the mirror deformed conifold are two set of O6 planes with bound
D6-branes. This D6/O6 system intersect the other
set of $k$ wrapped D6-branes to form an intersecting brane/plane system. As we saw in the heterotic case, the system is supersymmetric in the absence of the 
deformed conifold background. In the presence of the deformed conifold, supersymmetry is achieved by turning on fluxes (which are mirror to the torsion in the heterotic setup).
The brane configuration is shown in {\bf table \ref{gpbranesmirr}}.

\subsection{Type I${}^\prime$ and type IIA duality frames}
\label{sec:typeIIA}

Our next step would be 
to find the mirror type IIA configuration. 
Naively, this can be obtained by performing three T-dualities along $\psi$, $\phi_1$, and $\phi_2$, but this would lead to an erroneous 
result \cite{anke1, anke2, anke3, chen1, chen2}. The subtlety lies in 
making the base of the manifold, parameterized by $\theta_1$, $\theta_2$, and $r$, very large. 
The simplest way to do this would be to make the following replacements in the
background:
\begin{eqnarray}\label{tufee}
&& d\psi\mapsto d\psi+f_1c_4d\theta_1+f_2c_7d\theta_2\nonumber\\
&& d\phi_1\mapsto d\phi_1-f_1d\theta_1,\quad d\phi_2\mapsto d\phi_2 - f_2d\theta_2
\end{eqnarray}
assuming that $f_1, c_4$ are functions of $\theta_1$ and $f_2, c_7$ are functions of $\theta_2$. 
 Furthermore, as shown in \cite{anke1, chen1}, we need to change the $d\psi$ fibration structure in \eqref{mopla} to:
\bg\label{fibchan}
H_1\left[(1-\sqrt{\epsilon})d\psi + c_4 d\phi_1 + c_7 d\phi_2\right] \left[(1+\sqrt{\epsilon})d\psi + c_4 d\phi_1 + c_7 d\phi_2\right]
\nd
where we have to take the limit where ($f_1, f_2$) is very large and $\epsilon$ is very small, which we can do 
on the mirror metric.  Performing the SYZ mirror transformation, we find the type IIA mirror metric:
\begin{eqnarray}\label{2ametu}
ds^2&=&e^{-\phi}\bigg[ds_{0123}^2+H_2dr^2+\frac{e^{2\phi}A}{H_1H_3H_4c_1^2c_5^2}\left(d\phi-\frac{H_1H_4c_4c_5^2}{A}d\phi_1
-\frac{H_1H_3c_1^2c_7}{A}d\phi_2\right)^2\nonumber\\
&&+\frac{e^{2\phi}(H_1c_7^2+H_4c_5^2)}{A}d\phi_1^2
+\frac{e^{2\phi}(H_1c_4^2+H_3c_1^2)}{A}d\phi_2^2-\frac{2H_1c_4c_7}{A}d\phi_1d\phi_2\nonumber\\
&&+(H_3 - \epsilon H_1f_1^2c_4^2)d\theta_1^2+ (H_4 - \epsilon H_1 f_2^2 c_7^2)d\theta_2^2 - {2H_1 c_4 c_7}\left(\epsilon f_1 f_2 \right)d\theta_1 d\theta_2\bigg]
\end{eqnarray}
with a dilaton $\phi_{(b)}$, and with $A$, $f_1$, and $f_2$ defined by:
\bg\label{Af1f2}
A = H_3 H_4 c_1^2 c_5^2 + H_1 H_4 c_4^2 c_5^2 + H_1 H_3 c_1^2 c_7^2,~~~~~~f_i = {\beta_i\over \sqrt{\epsilon}} .
\nd
Similarly, the $B_{\rm NS}$ field is given by:
\bg\label{bns}
B_{\rm NS} = && \alpha f_1 H_3 c_1^2\left(H_1 c_7^2 + H_4 c_5^2\right) d\theta_1 \wedge d\phi_1 + \alpha f_2 H_4 c_5^2\left(H_1 c_4^2 + H_3 c_1^2\right) d\theta_2 
\wedge d\phi_2\nonumber\\
&& + \left(1- {\epsilon\over \alpha H_1 H_4 c_1^2 c_5^2}\right)\left(f_1c_4 d\theta_1 + f_2 c_7 d\theta_2\right)\wedge d\psi
\nd
where $\alpha \equiv A^{-1}$. In the limit $\epsilon \to 0$ the second line is a pure gauge, but the other two components are 
large. This is expected as the $B_{\rm NS}$ field appears because we made the base of the SYZ $T^3$ fibration large.\footnote{We will discuss another case later 
where we can gauge away such a $B_{\rm NS}$ field.} 

In addition to metric and $B$-field,
we also have gauge flux $F_2$ as well as four-form flux $F_4$.   
The nonzero components of $F_2$ are given by:
\begin{eqnarray}\label{twoform}
&&{F}_{\psi\theta_1}={\cal H}_{\phi_1\phi_2\theta_1},\quad
{F}_{\psi\theta_2}={\cal H}_{\phi_1\phi_2\theta_2},\quad
{F}_{\psi r}={\cal H}_{\phi_1\phi_2 r},\nonumber\\
&&{F}_{\phi_1r}={\cal H}_{r\phi_2\psi}+\frac{2H_1 c_4 c_7}{H_1 c_4^2 + H_3 c_1^2}{\cal H}_{\phi_1
r \psi}+\frac{2H_1 c_4}{H_1 c_4^2 + H_3 c_1^2}{\cal H}_{r \phi_1\phi_2},\nonumber\\
&&{F}_{\phi_2 r}={\cal H}_{\phi_1 r\psi}+2\alpha
\left[H_1 c_7 (H_1 c_4^2 + H_3 c_1^2) - H^2_1 c^2_4 c_7 \right]{\cal H}_{r\phi_1\phi_2},\nonumber\\
&&{F}_{\phi_1\theta_2}= {\cal H}_{\psi\theta_2\phi_2}+2
\frac{H_1 c_4}{H_1 c_4^2 + H_3 c_1^2}H_{\phi_1\phi_2\theta_2},\nonumber\\
&&{F}_{\phi_1\theta_1}=2\frac{H_1 c_4 c_7}{H_1 c_4^2 + H_3 c_1^2}{\cal H}_{\phi_1\theta_1\psi}
+2\frac{H_1 c_4}{H_1 c_4^2 + H_3 c_1^2}{\cal H}_{\phi_1\phi_2\theta_1},\nonumber\\
&&{F}_{\phi_2\theta_1}={\cal H}_{\psi\phi_1\theta_1}+2\alpha
\left[H_1 c_7(H_1 c_4^2 + H_3 c_1^2)-H^2_1 c^2_4 c_7\right]{\cal H}_{\phi_1\phi_2\theta_1},\nonumber\\
&&{F}_{\phi_2\theta_2}=2\alpha
\left[H_1 c_7 (H_1 c_4^2 + H_3 c_1^2)-H^2_1 c^2_4 c_7 \right]{\cal H}_{\phi_1\phi_2\theta_2}
\end{eqnarray}
where one may read off the components ${\cal H}_{mnp}$ from \eqref{torsa}.  Similarly, the nonzero components of $F_4$ are:
\begin{eqnarray}\label{4form}
&&F_{\theta_1\theta_2\phi_1\psi},~~F_{\theta_1\phi_1\phi_2r},~~F_{\theta_1\phi_1\psi r},~~F_{\theta_1\phi_2\psi r},~~F_{\theta_1\theta_2\phi_2\psi},~~F_{\theta_2\phi_1\phi_2r},\nonumber\\
&&F_{\theta_2\phi_2\psi r},~~F_{\theta_2\phi_1\psi r},~~F_{r\psi\theta_1\theta_2},~~F_{\theta_1\theta_2\phi_1\phi_2}, ~~ F_{r\phi_1\theta_1\theta_2},~~F_{r\phi_2\theta_1\theta_2}.
\end{eqnarray}
The whole configurations preserves ${\cal N}=1$ supersymmetry in four-dimensions.

The metric \eqref{2ametu}
can also be written in a more suggestive way:
\begin{eqnarray}\label{2ametuRW}
ds^2&=&e^{-\phi}\bigg\{ds_{0123}^2+H_2dr^2+\frac{e^{2\phi}A}{H_1H_3H_4c_1^2c_5^2}\left(d\psi-\frac{H_1H_4c_4c_5^2}{A}d\phi_1
-\frac{H_1H_3c_1^2c_7}{A}d\phi_2\right)^2\nonumber\\
&+& \textstyle  \left[(H_4 - H_1 \beta_2^2 c_7^2)d\theta_2^2 +\frac{e^{2\phi}(H_1c_7^2+H_4c_5^2)}{A}d\phi_1^2\right]
+ \left[(H_3 - H_1\beta_1^2c_4^2)d\theta_1^2 +\frac{e^{2\phi}(H_1c_4^2+H_3c_1^2)}{A}d\phi_2^2\right] \nonumber\\
&&  - {2H_1 c_4 c_7} \left(\beta_1 \beta_2 d\theta_1 d\theta_2 + {1\over A} d\phi_1d\phi_2\right) \bigg\}
\end{eqnarray}
Now we identify the two two-spheres with the sets of coordinates
($\theta_1, \phi_2$) and ($\theta_2, \phi_1$). 
The supersymmetry variations again lead to
\bg\label{btfut}
c_4 = \cos \theta_1, ~~~~~~c_7 = \cos \theta_2, ~~~~~~c_5 = \sin \theta_2, ~~~~~~c_1 = \sin \theta_1 \nd
with ($H_1, H_2, H_3, H_4$) all being functions of $r$, the radial coordinate. This means the metric along the ($\theta_1, \phi_2$) direction will 
become:\footnote{Note that we will always stay away from the points ($\theta_1, \theta_2$) = ($0, 0$), $(\pi,\pi)$ since the metric is singular at those points and the $T^3$ fibration degenerates.}
\bg\label{metat1p2}
d\theta_1^2 + {e^{2\phi}\over A}\left(\frac{H_1c_4^2+H_3c_1^2}{H_3 - H_1 \beta_1^2 c_4^2}\right)d\phi_2^2 &=& d\theta_1^2 
+ {e^{2\phi}\over A}\left(\frac{H_1 \cos^2\theta_1 + H_3\sin^2\theta_1}
{H_3 -H_1 \beta_1^2 \cos^2\theta_1}\right)d\phi_2^2\\
&=& d\theta_1^2  + {e^{2\phi}\over A}\left[\frac{H_1 + (H_3 - H_1)\sin^2\theta_1}
{H_3 -H_1 \beta_1(\theta_1)^2 + H_1\beta_1(\theta_1)^2 \sin^2\theta_1}\right]d\phi_2^2\nonumber\nd
meaning that the length of the $\phi_2$ cycle will vary between $e^\phi \sqrt{H_1\over H_3-H_1\beta(0)^2}$  and  $e^\phi$ as $\theta_1$ varies as 
$0 < \theta_1 \le {\pi\over 2}$. Similarly, the metric along ($\theta_2, \phi_1$) directions will become:
\bg\label{metat2p1}
d\theta_2^2  + {e^{2\phi}\over A}\left[\frac{H_1 + (H_4 - H_1)\sin^2\theta_2}
{H_4 -H_1 \beta_2(\theta_2)^2 + H_1\beta_2(\theta_2)^2 \sin^2\theta_2}\right]d\phi_1^2 . \nd
Note that we can now absorb $A$ into a redefinition of the dilaton\footnote{Not to be confused with the dilaton $\phi_{(b)}$ in that frame.} 
as $e^{\phi} = e^{\hat{\phi}}\sqrt{A}$.

The brane setup is composed of $k$ D6 branes wrapping the 3-cycle parametrized by ($\theta_1, \phi_2, \psi$), one set of D6/O6 oriented along ($\phi_1, \theta_2, \psi$), and the other set of D6/O6 (coming from type I D9/O9) oriented along ($\theta_1, \theta_2, r$).  This is summarized in {\bf table \ref{gpbranesmirr}}.

\begin{table}[tb]
\begin{center}
\renewcommand{\arraystretch}{1.5}
\begin{tabular}{|c||c|c|c|c|c|c|c|c|c|c|}\hline Direction & 0 & 1 & 2
& 3 & 4 & 5 & 6 & 7 & 8 & 9 \\ \hline\hline
D6/O6 & $\surd$  & $\surd$   & $\surd$   & $\surd$  & $\surd$  & $\cdot$ & $\surd$ & $\cdot$ & $\surd$
& $\cdot$ \\  \hline
${\rm D6}^\prime/{\rm O6}^\prime$ & $\surd$  & $\surd$   & $\surd$   & $\surd$  & $\cdot$   & $\surd$  & $\surd$  & $\cdot$ &
$\cdot$  & $\surd$ \\  \hline
$k$ D6 & $\surd$  & $\surd$   & $\surd$   & $\surd$  & $\surd$ & $\cdot$ & $\cdot$ & $\surd$ & $\cdot$  & $\surd$ \\  \hline
${\cal C}_{\rm def}$ & $\cdot$  & $\cdot$   & $\cdot$   & $\cdot$  & $\surd$ & $\surd$ &  $\surd$  & $\surd$ & $\surd$  & $\surd$ \\  \hline
  \end{tabular}
\renewcommand{\arraystretch}{1}
\end{center}
  \caption{Configuration after taking a mirror transformation of {\bf table \ref{gpbranesT45}}. The deformed conifold is 
denoted as ${\cal C}_{\rm def}$.}
    \label{gpbranesmirr}
\end{table}

\subsection{A type IIA detour to brane constructions}
 \label{sec:IIA2}

 At this point, let us take a short detour 
 to discuss geometrical interpretations of the cycles where we wrap our three types of D6-branes.  
 In the language of \cite{givchur}, there are several Lagrangian submanifolds that we can wrap our D6-branes on.  
Specifying the deformed conifold, as before, by\footnote{Since, for the specific purpose of this section we don't need the added complication of non-K\"ahlerity, we will 
analyze the branes wrapped on cycles using K\"ahler deformed conifold. The analysis can be easily extended to include non-K\"ahlerity.}  
 \bg
 z_1^2 + z_2^2 + z_3^2 + z_4^2 = \mu^2,
 \nd
 then the base $S^3$ can be identified with the fixed point set of the antiholomorphic involution 
 $z_i \rightarrow \bar{z}_i$ , which for $z_i = x_i + i y_i$ and real $\mu$ is given by,
 \bg
 y_i = 0,\qquad x_1^2 + x_2^2 + x_3^2 + x_4^2 = \mu^2 .
 \nd
 The $k$ D6-branes are wrapped on this $S^3$ with coordinates $(\theta_1, \phi_2, \psi)$, i.e., this is $S^3_{(1)}$ above.

 What about the two pairs of D6/O6? One of the D6/O6 systems is along $(\phi_1, \theta_2, \psi)$ and the other is along 
 $(\theta_1, \theta_2, r)$. In the language of \cite{givchur}, there are other Lagrangian submanifolds identified as the fixed point sets of involutions like 
$(z_1, z_2) \rightarrow (\bar{z}_1, \bar{z}_2)$ and $(z_3, z_4) \rightarrow (- \bar{z}_3, - \bar{z}_4)$. This is given by 
 \bg
 y_1 = y_2 = x_3 = x_4 = 0, \qquad  x_1^2 + x_2^2 = \mu^2 + y_3^2 + y_4^2 .
 \nd
 The unconstrained values for $y_3, y_4$ and the phase of $x_1 + i x_2$ implies that we have a 3-cycle of topology $\mathbb{C} \times S^1$.   Note that the $S^3$ and the  $\mathbb{C} \times S^1$ intersect along
 \bg
 x_3 = x_4 = y_i = 0, \qquad x_1^2 + x_2^2 = \mu^2,
 \nd
 which represents a cycle $S^1$. 

 The next question then is: out of the two distinct D6/O6 systems, which one is wrapped on the geometric cycle? From the discussion of \cite{givchur}, 
 the branes wrapped on $\mathbb{C} \times S^1$ survive the geometric transition, so they must be the ones wrapped on the $(\theta_1, \theta_2, r)$ three-cycle. 
 On the other hand, the D6${}^\prime$/O6${}^\prime$ system wrapped along $(\phi_1, \theta_2, \psi)$ become geometry and flux after the geometric transition, so 
 they cannot be wrapped on $\mathbb{C} \times S^1$. From equations \eqref{metat1p2} and \eqref{metat2p1} above, we see that $\psi$ combines with $\phi_1$ and 
 $\phi_2$ so that the $k$ D6-branes and the D6${}^\prime$/O6${}^\prime$ system are each wrapped on a Hopf fibration of $\psi$ over a two-cycle given by 
 $(\phi_1, \theta_2)$ and $(\theta_1, \phi_2)$, respectively.  To 
 complete the story, two additional ingredients are required:
\begin{itemize}
\item Two-form fluxes through the bases of the Hopf fibrations: 
 these are indeed present, as we see from the nonzero $(\phi_1, \theta_2)$ and   $(\theta_1, \phi_2)$ components of $F_2$ in \eqref{twoform}.
\item Supersymmetry: of course, this is true since we determined the background by demanding it preserve supersymmetry \eqref{suco}, in addition to the Bianchi identity.
\end{itemize}

\subsection{M-theory duality frame and new flips and flops}
\label{sec:Mthy}

Our next step is to lift this configuration to M-theory. As we know, in M-theory, the $k$ D6-branes become a $k$-centered Taub--NUT space while the two sets of 
D6/O6 branes/planes become two sets of Atiyah--Hitchin spaces, as shown in {\bf table \ref{upliftbranes}}.  

\begin{table}[tb]
\begin{center}
\renewcommand{\arraystretch}{1.5}
\begin{tabular}{|c||c|c|c|c|c|c|c|}\hline Direction & $\theta_1$ & $\phi_2$ & $\theta_2$
& $\phi_1$ & $\psi$ & $r$ & $x^{11}$  \\ \hline\hline
Taub-NUT (TN) & $\cdot$  & $\cdot$   & $\surd$   & $\surd$  & $\cdot$  & $\surd$ & $\surd$\\  \hline
Atiyah-Hitchin (AH${}_1$) & $\surd$  & $\surd$   & $\cdot$   & $\cdot$ & $\cdot$   & $\surd$  & $\surd$ \\  \hline
Atiyah-Hitchin (AH${}_2$) & $\cdot$  & $\surd$   & $\cdot$   & $\surd$  & $\surd$ & $\cdot$ & $\surd$ \\  \hline
  \end{tabular}
\renewcommand{\arraystretch}{1}
\end{center}
  \caption{The uplift to M-theory of the type IIA configuration {\bf table \ref{gpbranesmirr}}. 
These configurations of Taub--NUT and Atiyah--Hitchin spaces
give rise to a supersymmetric $G_2$-structure manifold.} 
    \label{upliftbranes}
\end{table}

The uplifted geometry of $k$ D6-branes looks like a Taub--NUT space along the $S_{(2)}^3$ given by ($\theta_2, \phi_1, x^{11}$) and stretched along the radial $r$ direction $\mathbb{R}^+$.  Locally, the geometry would then look like ${\mathbb R}^+ \times S_{(1)}^3 \times  S_{(2)}^3$, where $S_{(1)}^3$ is along ($\theta_1, \phi_2, \psi$). Similarly, one of the D6/O6 systems becomes
an Atiyah--Hitchin space with the local geometry ${\mathbb R}^+ \times S_{(3)}^3 \times S_{(4)}^3$, where $S_{(3)}^3$ is along ($\theta_2, \phi_1, \psi$), and 
$S_{(4)}^3$ is along ($\theta_1, \phi_2, x^{11}$). The M-theory metric then takes the following form:
\begin{eqnarray}\label{methmet}
ds^2&=&e^{-\phi -{2\phi_{(b)}\over 3}}\bigg\{ds_{0123}^2+H_2dr^2+\frac{e^{2\phi}A}{H_1H_3H_4c_1^2c_5^2}\left(d\psi-\frac{H_1H_4c_4c_5^2}{A}d\phi_1
-\frac{H_1H_3c_1^2c_7}{A}d\phi_2\right)^2\nonumber\\
&+& \textstyle  \left[(H_4 - H_1 \beta_2^2 c_7^2)d\theta_2^2 +\frac{e^{2\phi}(H_1c_7^2+H_4c_5^2)}{A}d\phi_1^2\right]
+ \left[(H_3 - H_1\beta_1^2c_4^2)d\theta_1^2 +\frac{e^{2\phi}(H_1c_4^2+H_3c_1^2)}{A}d\phi_2^2\right] \nonumber\\
&& ~~~~~~~~~~~ ~~~~~~~~~~~~~ - {2H_1 c_4 c_7} \left(\beta_1 \beta_2 d\theta_1 d\theta_2 + {1\over A} d\phi_1d\phi_2\right) \bigg\}\\
&+& e^{4\phi_{(b)}\over 3}(dx_{11}+A_rdr+A_{\theta_1}d\theta_1+A_{\theta_2}d\theta_2+A_{\phi_1}d\phi_1+A_{\phi_2}d\phi_2+A_{\psi}d\psi)^2\nonumber
\end{eqnarray}
which is a noncompact $G_2$-structure manifold with $G$-fluxes.\footnote{One could make further local rotations to the M-theory metric to bring it
into a more standard form (see \cite{chen1}), but we will not do so here.}

Next we will perform a flop.  This
will be similar to the one in \cite{AMV}, in the sense that we will have to exchange three-cycles, but it will be slightly different. We impose the following flop operation:
\bg\label{flop1}
S_{(1)}^3 ~~ \longleftrightarrow~~ S^3_{(2)}  
\nd
and simultaneously
\bg\label{flop2} S^3_{(3)} ~~ \longleftrightarrow~~ S^3_{(4)} \nd
where the $S_{(3),(4)}^3$ are the same as $S_{(1),(2)}^3$, but with Hopf fibers exchanged.  In coordinates,
\bg\label{floppy}
\theta_1 ~ \leftrightarrow ~ \theta_2 , \qquad  \phi_1 ~ \leftrightarrow ~ \phi_2 , \qquad  \psi ~ \leftrightarrow ~ x^{11} .
\nd
Under this flip and flop\footnote{By {\it flip} we mean the exchange $\psi ~ \leftrightarrow ~ x^{11}$, whereas {\it flop} is the standard 
flop operation of a three-sphere.}, the Taub--NUT space 
will be along $S^3_{(1)}$ and the first Atiyah--Hitchin space will be along $S^3_{(3)}$. The second Atiyah--Hitchin space is actually unchanged under the flop, 
which
means that when we dimensionally reduce on $x^{11}$, it will convert back to the same 
D6/O6 system it came from.  This is depicted in {\bf table \ref{afterflop}}. The dilaton in this frame is $\phi_{(c)}$ which is different from $\phi_{(b)}$ because 
of \eqref{floppy}. 
  
\begin{table}[tb]
\begin{center}
\renewcommand{\arraystretch}{1.5}
\begin{tabular}{|c||c||c||c|}\hline Type IIA & M-theory before flop & M-theory after flop & Type IIA reduction \\ \hline\hline
$k$ D6 & $k$ centered TN along $S^3_{(2)}$ & $k$ centered TN along $S^3_{(1)}$ & geometry $+$ fluxes \\  \hline
D6${}^\prime$/O6${}^\prime$ & AH${}_1$ along $S^3_{(4)}$ & AH${}_1$ along $S^3_{(3)}$ & geometry $+$ fluxes  \\  \hline
D6/O6 &  AH${}_2$ along ($\phi_1, \phi_2, \psi, x^{11}$) & AH${}_2$ along ($\phi_1, \phi_2, \psi, x^{11}$) & D6/O6\\  \hline
  \end{tabular}
\renewcommand{\arraystretch}{1}
\end{center}
  \caption{Type IIA and M branes/planes/geometry before and after the flop \eqref{flop1}--\eqref{flop2}.}
    \label{afterflop}
\end{table}

In the flopped setup, the D6/O6 system arising from the second Atiyah--Hitchin space converts to a D9/O9 system after mirror symmetry, returning to a type I model. 
All other branes/planes in the original type I configuration have dissolved into geometry to become fluxes, consistent with the 
predictions in \cite{anke2}, \cite{anke3}, and \cite{chen2}.  S-dualizing to heterotic then yields the gravity dual that is composed only of geometry and fluxes, with no localized sources.  Following the procedure in \cite{chen1}, we deduce that the gravity dual is also a non-K\"ahler warped resolved conifold.

\subsection{Gravity duals in the heterotic theories}
\label{sec:hetgravity}

So, after dimensional reduction back to type IIA, we will have a manifold that is topologically a resolved conifold with two key differences from the usual 
geometric transition of
\cite{Vafa:2000wi} and \cite{Cachazo:2001jy}: the first is already known from \cite{anke1}, namely that the metric should be non-K\"ahler, and the second is the 
appearance of a D6/O6 brane/plane system (see {\bf table \ref{afterflop}}).  Following the steps and notation in \cite{chen1}, we find the metric:
\bg\label{resanswer} 
ds^2_{\rm IIA} &= &e^{-{\widetilde\phi}}\left[ds^2_{0123} + {H}_2 dr^2\right] + \widetilde{H}_1(d\psi + \Delta_1 \cos \theta_1 d\phi_1 
+ \Delta_2 \cos \theta_2 d\phi_2)^2 \nonumber\\
&& ~~~~~~~~~~ + \left(\widetilde{H}_{3a} d\theta_1^2 + \widetilde{H}_{3b}d\phi_2^2\right) + 
\left(\widetilde{H}_{4a} d\theta_2^2 + \widetilde{H}_{4b}d\phi_1^2\right)\nd
where ${\widetilde\phi}$ is the remnant of the dilaton factor from M-theory after flop and $H_2$ is the same as in \eqref{mopla}. For the other coefficients,
one may look up their values in section 4.4 of \cite{chen1}.\footnote{Note that one has to set to zero all of the $b_{ij}$ 
fields that appear in the 
fibrational structure in \cite{chen1}.} The various components of the fluxes could be traced from the original type I side or from \eqref{twoform} and \eqref{4form}. 
If we start-off with only two components of torsion in the heterotic side as in \eqref{turna} then the field contents in the subsequent theories will be simple. The list 
of all the field contents are depicted in {\bf table \ref{fieldc1}} where we see that in the final type I$'$ theory the field contents are the two-form and the 
NS three-form fields alongwith the dilaton $\phi_{(c)}$ and the metric $\widetilde{G}_{\mu\nu}$ given by \eqref{resanswer}. 

\begin{table}[htb]
  \centering
\renewcommand{\arraystretch}{1.5}
  \begin{tabular}{|c||c|c|c|c|c|c|c|c|}
    \hline
    {Original (I)} & \multicolumn{1}{c||}{$g_{\mu\nu}^{(a)}$} & \multicolumn{1}{c||}{$\phi$}  & ${\cal H}_{\theta_1\phi_1\psi}^{\rm RR}$ 
& ${\cal H}_{\theta_2\phi_2\psi}^{\rm RR}$  & ${\cal H}_{\theta_1\theta_2\phi_1}^{\rm RR}$ & 
\multicolumn{1}{c||}{${\cal H}_{\theta_1\theta_2\phi_2}^{\rm RR}$}& \multicolumn{2}{c|}{$g_{\mu\nu}^{(b)}$} \\
    \hline 
{Mirror (I$'$)} & \multicolumn{1}{c||}{$g_{\mu\nu}$} & \multicolumn{1}{c||}{$\phi_{(b)}$}  & ${F}_{\theta_1\phi_2}^{(2)}$ 
& \multicolumn{1}{c||}{${F}_{\theta_2\phi_1}^{(2)}$}  & ${F}_{\theta_1\theta_2\phi_2\psi}^{(4)}$ & 
\multicolumn{1}{c||}{${F}_{\theta_1\theta_2\phi_1\psi}^{(4)}$}& ${\cal H}_{\theta_i\phi_i r}^{\rm NS}$ & ${\cal H}^{\rm NS}_{\theta_i\phi_i\theta_k}$ \\
    \hline 
{11D lift (M)} & \multicolumn{4}{c||}{$G_{\mu\nu}$} & $G^{(4)}_{\theta_1\theta_2\phi_2\psi}$ & $G^{(4)}_{\theta_1\theta_2\phi_1\psi}$ 
& $G^{(4)}_{\theta_i\phi_i r, 11}$ & \multicolumn{1}{c|}{$G^{(4)}_{\theta_i\phi_i\theta_k, 11}$}\\
    \hline
{Flop (M)} & \multicolumn{4}{c||}{$G'_{\mu\nu}$} & $G^{'(4)}_{\theta_1\theta_2\phi_1,11}$ & $G^{'(4)}_{\theta_1\theta_2\phi_2,11}$ 
& $G^{'(4)}_{\theta_k\phi_k r \psi}$ & \multicolumn{1}{c|}{$G^{'(4)}_{\theta_k\phi_k\theta_i\psi}$}\\
    \hline
{Reduce (I$'$)} & \multicolumn{2}{c||}{${\widetilde G}_{\mu\nu}, \phi_{(c)}$} &${\widetilde F}^{(2)}_{\theta_1\phi_1}$ 
&\multicolumn{1}{c||}{${\widetilde F}^{(2)}_{\theta_2\phi_2}$}
& ${\cal H}^{\rm NS}_{\theta_1\theta_2\phi_1}$ & \multicolumn{1}{c||}{${\cal H}^{\rm NS}_{\theta_1\theta_2\phi_2}$} 
& $G^{'(4)}_{\theta_k\phi_k r \psi}$ & \multicolumn{1}{c|}{$G^{'(4)}_{\theta_k\phi_k\theta_i\psi}$}\\
    \hline
  \end{tabular}
\renewcommand{\arraystretch}{1}
  \caption{Field content traced through much of the duality chain depicted in {\bf figure \ref{hetGT}}, starting from the type I theory and ending with the type I$'$ theory.}
  \label{fieldc1}
\end{table}

The type I$'$ configuration that we get is on a non-K\"ahler resolved conifold, and therefore we need to perform coordinate transformations of the 
form \eqref{tufee} before performing SYZ mirror transformation. There are a few subtleties now compared to \eqref{tufee} that we performed earlier. First, 
the coefficients $\Delta_i$ appearing in the metric \eqref{resanswer} may not necessarily be a function of $\theta_i$ only. Secondly, even if $\Delta_i$ are 
functions of $\theta_i$, the coordinate transformation \eqref{tufee} is not feasible because this will generate a $B_{\rm NS}$ field like \eqref{bns} after a mirror
transformation to type I, which should have been projected out by the orientifold action.\footnote{Note that this subtlety was 
absent in the case studied in \cite{chen1} because the mirror 
was type IIB theory where such a $B_{\rm NS}$ field can exist.} The solution to this puzzle is rather simple and illuminating. There are two ways to make the base of the 
$T^3$ fibration bigger in \eqref{resanswer}: one, by making a coordinate transformation of the form \eqref{tufee}, and two, by changing the complex structures of the 
two base spheres. Performing the second operation implies that the metric \eqref{resanswer} picks up additional terms of the form:
\bg\label{addterm} 
{\widetilde f}_1~ d\theta_1 d\phi_1~ +~ {\widetilde f}_2 ~ d\theta_2 d\phi_2\nd
where ${\widetilde f}_i$, unlike the $f_i$ before in \eqref{tufee}, do not have to be functions of $\theta_i$ only. This means that we can 
define the functional forms for the ${\widetilde f}_i$ in such a way so as to make the $B_{\rm NS}$ field in the mirror type I theory to be a pure gauge. 
The additional warpings of the 
coefficients $d\theta_i$ and $d\phi_i$ of the base spheres in the type I$'$ frame, 
can be absorbed in the definition of ${\widetilde H}_{3a, 3b,4a, 4b}$. 
Of course such a change will also affect the background fields, which in turn are shown in the 
list of fields depicted in {\bf tables \ref{fieldc1}} and {\bf \ref{fieldc2}}. 

In {\bf table \ref{fieldc1}} 
we represented the NS three-form fields in the mirror type I$'$ theory,
coming from the cross-term $g_{\mu\nu}^{(b)}$ of the original type I metric, by ${\cal H}^{\rm NS}_{\theta_i\phi_i r}$ and 
${\cal H}^{\rm NS}_{\theta_i\phi_i\theta_k}$. This is the simplest scenario. In general we could have additional components like
\bg\label{newcomer}
{\cal H}^{\rm NS}_{\theta_1\phi_1 \psi}, ~~~~ {\cal H}^{\rm NS}_{\theta_2\phi_2 \psi}, ~~~~ {\cal H}^{\rm NS}_{\theta_1\phi_1 \phi_2}, ~~~~ {\cal H}^{\rm NS}_{\theta_2\phi_2 \phi_1}.\nd
The latter two are projected out by orientifolding operation. For the first two fields\footnote{These components 
do not appear from \eqref{bns} that we got using T-dualities. We however expect that a more generic choice of the warp factors for the 
original type I background may lead to 
these components.}  
---
under M-theory lift, flop, and 
dimensional reduction ---  we will get back three-form fields with the same components but with different magnitudes, in addition to the 
ones mentioned in {\bf table \ref{fieldc1}}. Changing the complex structure of the base spheres will further change the values of the fields in type I$'$ theory 
similar to the ones
depicted in {\bf table \ref{fieldc2}}. Finally, after the mirror and U-duality we get our required heterotic background with a dilaton and torsion 
as shown in {\bf table \ref{fieldc2}}. 
From this table and \eqref{newcomer}, we can easily see that there are four possible components of the $B_{\rm NS}$ field in the type I$'$ reduction
\bg\label{biness}
B_{\theta_1\phi_1} \equiv b_1, ~~~~~~~ B_{\theta_2\phi_2} \equiv b_2, ~~~~~~~~ B_{\psi\phi_1} \equiv b_3, ~~~~~~~~ B_{\psi\phi_2} \equiv b_4,
\nd
where all other components can be gauged away. Note that 
we have components like ($b_3, b_4$) which have both legs along the T-duality directions. This 
would imply the possibility of non-geometric background after SYZ mirror transformation. We will however only concentrate on the geometrical aspect of the mirror
and leave the intriguing possibility of non-geometric backgrounds for future works.   

After a SYZ mirror transformation (converting D6/O6 to D9/O9) and a U-duality,  we wind up in heterotic with the metric:
\bg\label{defkoan}
ds^2 & = & \textstyle ds^2_{0123} + {2\Delta\over r\sqrt{a_3}}dr^2 + {1\over {\widetilde H}_1 + {\cal A}}\left[d\psi 
+ {\widetilde \Delta}_1 {\rm cos}~\theta_1 \left(d\phi_1 + \alpha_1 d\theta_1\right) 
+ {\widetilde \Delta}_2 {\rm cos}~\theta_2\left(d\phi_2 + \alpha_2 d\theta_2\right)\right]^2 \nonumber\\
&& + \left[g_{\theta_1\theta_1} d\theta_1^2 + g_{\phi_1\phi_1}\left(d\phi_1 + \alpha_1 d\theta_1\right)^2\right]  
+ \left[g_{\theta_2\theta_2} d\theta_2^2 + g_{\phi_2\phi_2}\left(d\phi_2 + \alpha_2 d\theta_2\right)^2\right]\nonumber\\   
&& ~~~~~~~~ + g_{\theta_1\theta_2} d\theta_1 d\theta_2 + g_{\phi_1\phi_2} \left(d\phi_1 + \alpha_1 d\theta_1\right)\left(d\phi_2 + \alpha_2 d\theta_2\right)\nd
Note that there is no coefficient in front of the space-time part. The coefficient, which appears in \eqref{resanswer}, is infact the dilaton 
in the type I/heterotic frame and is related to the dilaton $\phi_{(c)}$ in type I$'$ frame by mirror/U-duality transformation. Therefore in the heterotic frame 
we see no factor of dilaton in front of the space-time part. Additionally, preserving $G_2$ structure in M-theory and $SU(3)$ structure in 
heterotic theory gives the following relation between the dilaton values in various frames \cite{chen1}:
\bg\label{dilframe}
{\widetilde\phi} = \phi + {2\phi_{(b)}\over 3} - {2\phi_{(c)}\over 3}\nd
which is perfectly consistent with our duality chain, EOMs and the metrics in various frames. 
    
The final metric \eqref{defkoan}
resembles somewhat the metric in \eqref{2ametuRW} but the $\phi_i$'s are non-trivially fibered. The non-trivial fibrations are given in terms 
of $\alpha_i$ which we define as:
\bg\label{dhelku}
&& \alpha_1 = {\alpha\over 2}\left[b_1 b_4^2 + {\widetilde H}_1 b_1 ({\widetilde H}_{3b} + {\widetilde H}_1\Delta_2^2 {\rm cos}^2\theta_2 - 
b_1 {\widetilde H}_1\Delta_2^2 {\rm cos}^2\theta_2)\right] \nonumber\\
&&  \alpha_2 = {\alpha\over 2}\left[b_2 b_3^2 + {\widetilde H}_1 b_2({\widetilde H}_{4b} + {\widetilde H}_1\Delta_4^2 {\rm cos}^2\theta_1 - 
b_2 {\widetilde H}_1\Delta_1^2 {\rm cos}^2\theta_1)\right]\nd
where $b_i$ are the components of the $B$-field given in \eqref{biness} and ${\widetilde H}_k$ are the components of the metric \eqref{resanswer}. The
coefficient $\alpha$ in \eqref{dhelku} is given by:
\bg\label{alahala}
\alpha \equiv {1\over {\widetilde H}_{3b}{\widetilde H}_{4b} + {\widetilde H}_{1}\left({\widetilde H}_{3b}\Delta_1^2 {\rm cos}^2\theta_1 + 
{\widetilde H}_{4b}\Delta_2^2 {\rm cos}^2\theta_2\right)}\nd
with ${\widetilde H}_{3b}$ etc are to be viewed as the modified warp factors. 
Note also that the coefficient of the $d\psi$ term is not just ${\widetilde H}_1^{-1}$, but has a 
correction term given by ${\cal A}$ in the following way:
\bg\label{corrA}
{\cal A} = && \alpha\Big[({\widetilde H}_{4b} + {\widetilde H}_1\Delta_1^2 {\rm cos}^2\theta_1)(b_4^2 - {\widetilde H}_1\Delta_2 {\rm cos}~\theta_2) 
+ ({\widetilde H}_{3b} + {\widetilde H}_1\Delta_2^2 {\rm cos}^2\theta_2)(b_3^2 - {\widetilde H}_1\Delta_1 {\rm cos}~\theta_1) \nonumber\\
&& ~~~~~~~~ 2 \Delta_1 \Delta_2 {\rm cos}~\theta_1 {\rm cos}~\theta_2\left({\widetilde H}_1^2 \Delta_1 \Delta_2 {\rm cos}~\theta_1 {\rm cos}~\theta_2 - b_3b_4\right)\Big]\nd
which is non-zero even in the absence of $b_3$ and $b_4$ components of the $B_{\rm NS}$ field.
The other variables in the metric \eqref{defkoan} are defined in the following way:
\bg\label{othervar}
{\widetilde \Delta}_1 \equiv -\alpha \Delta_1 {\widetilde H}_{3b}, ~~~~~~~~~ {\widetilde \Delta}_2 \equiv -\alpha \Delta_2 {\widetilde H}_{4b}\nd
The rest of the 
$g_{ij}$ defined in \eqref{defkoan} are related to the corresponding type IIA warped resolved conifold metric \eqref{resanswer}
by the relations given in {\bf appendix C} of 
\cite{chen1}. Additionally, we also have nonzero ${\cal H}$-flux, dilaton, and vector bundle as depicted in {\bf table \ref{fieldc2}}.
Our claim then is that the gravity dual of the wrapped heterotic five-branes in the heterotic $E_8 \times E_8$ theory can be extracted from 
\eqref{defkoan} by performing a geometric transition 
on the metric \eqref{mopla}. In the decoupling limit of \eqref{mopla}, \eqref{defkoan} will provide the precise gravity dual. Defining $u = {\alpha'\over r}$, 
we see that the five-dimensional part of \eqref{defkoan} takes the following form:
\bg\label{5ddual}
ds^2 ~ = ~ ds^2_{0123} + {du^2\over G(u)}\nd
where $G(u)$ can be extracted from ${2\Delta\over r\sqrt{a_3}}$ in \eqref{defkoan} in the decoupling limit. The spacetime part is flat, 
so we conclude that, in the string frame, a flat 
Minkowski space is dual to LST compactified to 4 dimensions on a non-K\"ahler resolved conifold. Additionally, irrespective of the details of the internal space, 
the five-dimensional dual metric will always be of the form \eqref{5ddual}. 
 
Note that the metric that we get on 
the heterotic side {\it cannot} be derived using any techniques other than the duality chains given in {\bf figures \ref{hetGT}} and {\bf \ref{varyTau}}, as 
our knowledge of the dual gauge theory i.e the dimensionally reduced LST is very minimal. Thus we think that the gravity dual \eqref{defkoan} may be the 
only way to extract non-trivial informations about the gauge theory. 

\begin{table}[tb]
  \centering
\renewcommand{\arraystretch}{1.5}
  \begin{tabular}{|c||c|c|c|c|c|c|c|c|}
    \hline
   {Original (I$'$)} & \multicolumn{2}{c||}{${\widetilde G}'_{\mu\nu}, \phi_{(c)}$} 
& ${\cal H}^{'\rm NS}_{\theta_1\theta_2\phi_1}$ & \multicolumn{1}{c||}{${\cal H}^{'\rm NS}_{\theta_1\theta_2\phi_2}$} 
& ${\widetilde G}^{(4)}_{\theta_k\phi_k r\psi}$ & \multicolumn{1}{c||}{${\widetilde G}^{(4)}_{\theta_k\phi_k\theta_i\psi}$}
&${\widetilde F}^{'(2)}_{\theta_1\phi_1}$ 
&\multicolumn{1}{c|}{${\widetilde F}^{'(2)}_{\theta_2\phi_2}$}\\ 
    \hline 
{Mirror (I)} & \multicolumn{4}{c||}{$G^{'\rm NK}_{\mu\nu}, -{\widetilde\phi}$} & ${\cal H}^{\rm RR}_{\theta_k\phi_j r}$ 
&\multicolumn{1}{c||}{${\cal H}^{\rm RR}_{\theta_k\theta_i\phi_l}$}
& ${\cal H}^{\rm RR}_{\theta_1\phi_2\psi}$ & \multicolumn{1}{c|}{${\cal H}^{\rm RR}_{\theta_2\phi_1\psi}$}\\
    \hline
{U-dual (Het)} & \multicolumn{4}{c||}{$G^{\rm NK}_{\mu\nu}, {\widetilde\phi}$} & ${\cal H}_{\theta_k\phi_j r}$ 
& ${\cal H}_{\theta_k\theta_i\phi_l}$ 
& ${\cal H}_{\theta_1\phi_2\psi}$ & \multicolumn{1}{c|}{${\cal H}_{\theta_2\phi_1\psi}$}\\
    \hline
  \end{tabular}
\renewcommand{\arraystretch}{1}
  \caption{The field content we obtain in the heterotic theory by following the final stages of the duality chain 
in {\bf figure \ref{hetGT}}, starting from the type I$'$ theory.}
  \label{fieldc2}
\end{table}

To complete the story, we should perform a few tests on the type IIB background to check that the background describes a dual to the 
four-dimensional confining theory.  One important consistency check is to find confining strings, which we examine in section \ref{sec:confining}, but before that we further discuss the geometry.

\subsubsection{Geometry and instanton transitions}

\begin{figure}[tb]
\begin{center}
\includegraphics[width=0.5\textwidth]{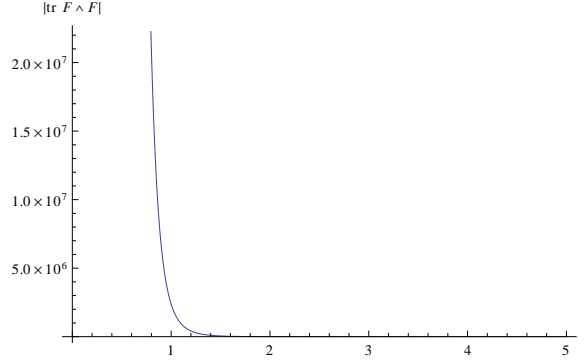}
\end{center}
\caption{{The behavior of $\vert {\rm tr} (F \wedge F)\vert$
plotted near the origin for case I. Note that the singularity at $r = 0$ signifies the presence of zero size instantons. After
geometric transition, we expect that the singularity in the vector bundles to be absent. This way the branes will convert to geometry and fluxes. As before we have 
chosen $e_0 = 100$ in units of $\alpha'$ with $\alpha' = 1$.}}
\label{FFcaseI}
\end{figure}  

Looking at the metric \eqref{defkoan}, we see that the two two-spheres are now given by ($\theta_1, \phi_1$) and 
($\theta_2, \phi_2$).  The fact that the two-spheres have asymmetrical coefficients suggests that the metric describes a warped resolved-deformed 
conifold, consistent with the expectation from the three cases studied earlier in {\bf section \ref{sec:hetslns}}. 
We can also compute the torsion ${\cal H}$, using the dilaton $\widetilde{\phi}$, 
using the standard formula:
\bg\label{torb} {\cal H} = - e^{2{\widetilde\phi}}\ast d\left(e^{-2{\widetilde\phi}}J\right) , \nd
which we expect this to satisfy the Bianchi identity \eqref{torclosed} without a source term, therefore ${\rm tr} (F \wedge F)$ should be regular. 

In the heterotic frame before the geometric transition, we saw in case I, for example, that ${\rm tr} (F \wedge F)$ had a singularity at $r\to 0$ (see {\bf figure \ref{FFcaseI}}), signifying the presence of the $k$ heterotic NS5-branes.\footnote{This singularity can be easily seen from \eqref{muksin} using the values for $\mathfrak{f}_i$ in 
\eqref{joishila} at $r \to 0$.}  
Now, after the 
transition, the branes convert to fluxes \eqref{torb} and the singularity in the vector bundle should vanish. 
A puzzling aspect however is what happens to the small instanton singularities? This can be answered from the 
M-theory picture that we have. Before the flop the small instanton singularities become multi centered TN space. Thus after the flop, the vector bundle singularities 
become singularities of the manifold (as there are no moduli for the small instanton singularities to blow up). 
In other words, before transition we have a vector bundle that has 
two parts: one singular piece from the small instantons and another non-singular piece that is supported by the underlying non-K\"ahlerity of the background. After 
transition, the singular piece becomes part of the manifold (so it is no longer part of the vector bundle) 
and the non-singular piece gets further deformed by the flop operation. This is almost like a small instanton transition, and 
this is why we assume that geometric transitions in heterotic theories to be closely related to small instanton transitions.

In this way, a heterotic background with branes is converted into another heterotic background 
without branes but with fluxes instead. Thus the little string background \eqref{bgmetDf} is converted to a closed string background \eqref{defkoan} where the 
singularities of the branes are transformed to the manifold (and also to the torsion to satisfy the Bianchi identity), and the vector bundles become smooth. 
The whole duality chain depicted
in {\bf figure \ref{hetGT}} transforms the brane background of \eqref{bgmetDf} to a closed string background \eqref{defkoan}. In the decoupling limit of the LST, 
the metric given in \eqref{defkoan} will be the gravity dual.

\subsubsection{In search of the confining strings}
\label{sec:confining}

Our next check is to look for the confining strings and potentials from the heterotic gravity dual, studying the behavior of a string connecting two points on the boundary. This is easier said than done for two reasons:
\begin{itemize}
\item Taking a string to the boundary means that we need to know the $r\to \infty$ behavior of the background --- i.e., we need to know the 
UV completion of the background whose IR physics is captured by \eqref{defkoan}.  
\item Looking at the metric \eqref{defkoan}, we see that the spacetime part of the metric is flat and comes with no warp factor. This suggests that even if we 
know the UV behavior, a Nambu--Goto action for the string will not produce a nontrivial result. 
\end{itemize}
The second concern is more subtle because the string coupling is also typically strong on the heterotic side.  The fact that it is strong on the heterotic side means we could instead ask this question on the S-dual, type I side where it will be weak.
So we will consider fundamental strings in the type I frame which appears to be an appropriate candidate for the confining string. This clearly shows that the confining 
string from the heterotic frame is a more exotic object.

To proceed, let us    
define $g_{rr} = e^{2{\widetilde\phi}} \widetilde{g}_{rr}$ in \eqref{defkoan}.\footnote{This is consistent with our convention described in {\bf footnote \ref{consconv}}.}  
With this the type I metric becomes:
\bg\label{timetu}
ds^2 = e^{-{\widetilde\phi}} (-dt^2 + dx_idx^i) + e^{\widetilde\phi} \widetilde{g}_{uu} du^2 + e^{-\phi} ds^2_{\rm internal}\nd
where $u = 1/r$  and $ ds^2_{\rm internal}$ is the internal metric that can be read off from \eqref{defkoan}.  
Note that there are no NS fluxes in the type I frame, which will be
useful below.

\begin{figure}[htb]
\begin{center}
\includegraphics[width=\textwidth]{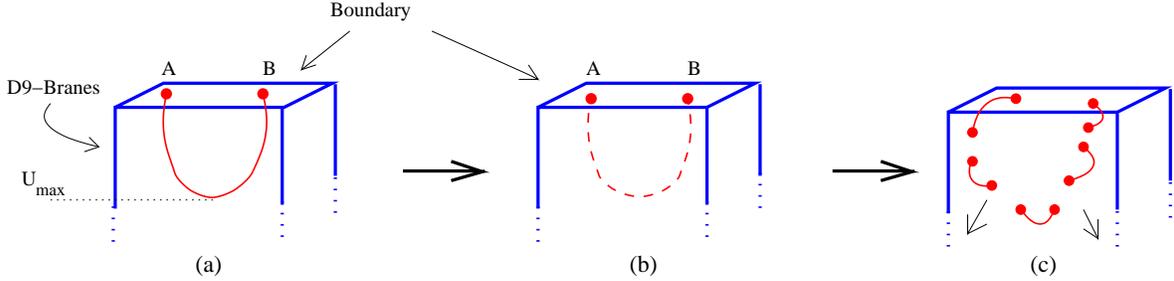}
\end{center}
\caption{{A long fundamental string in type I theory is unstable and is prone to breaking into small strings on the D9-brane. In the dual theory this  
captures the phenomena of a confining string breaking into small strings connecting quark-antiquark pairs.}}
\label{quarks}
\end{figure}

Now we can address the first concern regarding UV completion. We will assume that for large $r$, i.e., for $u \to 0$, the four-dimensional gauge theory becomes 
almost conformal. This means, without loss of generality, that we can take\footnote{Note that, although we are using a specific asymptotic expansion, the confining 
nature of the string will {\it only} depend on the IR physics! This will be clarified below. The UV completion is done only to have renormalized results.}:
\bg\label{hinsha}
e^{-{\widetilde\phi}} = \sum_{n=0}^\infty c_n u^{n-2}, ~~~~~~~~~~~~~~ \widetilde{g}_{uu} = \sum_{n=0}^\infty b_l u^{l-4} , \nd
where $c_n$ and $b_l$ are independent of $u$ and of the four spacetime directions (for simplicity, we will just take them to be constants with $c_1 = c_2 = 0$). 
With these choices, the type I 
metric becomes:
\bg\label{timbwhat}
ds^2 = c_n u^{n-2} (-dt^2 + dx_i dx^i) + {b_l u^l\over c_m u^{m+2}} du^2 + c_p u^{p-2}{ds^2_{\rm internal}} . \nd
Next, we want to study a long fundamental string in this background, but we must first discuss one thing: type I theory has no stable fundamental strings.  
Instead, the only stable string is the 
type I D-string.  Of course, 
to describe fluctuations of the D9-background, type I should allow some form of fundamental strings otherwise duality to type IIB will not work properly.
So this means that if we construct a long open 
fundamental string, it will break into small
open strings and other radiation.  This is exactly what we 
want from the gravity dual of a
confining theory: the fundamental string represents the gravity dual of the confining flux tube which, as we know, are susceptible to 
breaking up into quark-antiquark pairs.  These quark-antiquark pairs are precisely the little open strings that our big fundamental string breaks into as 
depicted in {\bf figure \ref{quarks}}.  
Note that this also tells us that the matter sector of our theory contains quarks, which was not easy 
to decipher directly from the dimensional reduction of the LSTs.  Note that this was also discussed in \cite{caris}. 

Now, consider a fundamental string connecting two boundary points so that it projects into the five-dimensional space ($x^{0, 1, 2, 3}, u$), and let 
$X^\mu: (\sigma, \tau) ~\to ~ (x^{0, 1, 2, 3}, u, \psi, \theta_i, \phi_i)$ be the embedding of the worldsheet into spacetime. 
If we choose a parameterization $\tau \equiv t$ and $\sigma \equiv x$ and the following configuration:
\bg\label{stequiva}
&&X^0 = t, ~~~~~X^1= x, ~~~~~ X^2 = X^3 = 0, ~~~~~ X^8 = u(x), ~~~~~ X^9 = \psi = 0,\nonumber\\
&& (X^4, X^5) = (\theta_1, \phi_1) = \left({\pi\over 2}, 0\right), ~~~~~
(X^6, X^7) = (\theta_2, \phi_2) = \left({\pi\over 2}, 0\right),\nd
then this will traverse a path in the five-dimensional space provided we equip this with an additional boundary condition:
\bg\label{ghoti}
u(\pm d/2) ~ = ~ 0\nd
where $x=\pm d/2$ is the value of $x$ at the two endpoints of the string. We then have to consider the equation of motion for $u(x)$.

The Nambu--Goto action for the string is 
given by:
\bg\label{namgo}
\textstyle S_{\rm NG} = {T_0\over 2\pi}\int d\sigma d\tau \left[\sqrt{-{\rm det}[(g_{\mu\nu} + \del_\mu \phi \del_\nu\phi)\del_aX^\mu \del_bX^\nu]} 
+ \epsilon^{ab} \del_aX^m \del_bX^n \bar{\Theta} \Gamma_m\Gamma^{pqs}\Gamma_n \Theta {\cal H}_{pqs}\right]\nonumber\\ \nd  
where ${\cal H}_{pqs}$ is the type I RR three-form field strength, $T_0$ is the fundamental string tension, and $a, b = 1, 2,~\del_1 \equiv {\del\over \del \tau}, 
~\del_2 \equiv {\del\over \del \sigma}$.  Note that once we switch off the fermions $\bar{\Theta} = \Theta = 0$, the 
RR three-form decouples from the Nambu--Goto action. Furthermore the absence of type I NS three-form fields simplify things further. Plugging in \eqref{hinsha} in 
\eqref{namgo}, we get:
\bg\label{namgoti}
S_{\rm NG} = {T_0\over 2\pi}\int_{-d/2}^{+d/2}~{dx\over u^2}\sqrt{(c_nu^n)^2 + g_lu^l\left({\del u/\del x}\right)^2}\nd
with $g_lu^l = b_ku^k +$ dilaton contributions that can be derived from \eqref{namgo} and \eqref{hinsha}. 

Now the analysis is very similar to that performed in \cite{andreev, mia2, mia3}, so the reader may want to look there for more details.  
{}From \cite{mia2, mia3} 
we know that the string will take a $U$-shape because of symmetry.  Letting $u_{\rm max}$ be the maximum value of $u$, one can then show that for $d$ to 
be real, we should satisfy \cite{mia2, mia3}:
\bg\label{miaco}
{1\over 2} (m+1)c_{m+3} u^{m+3}_{\rm max} ~ \le ~ 1 . \nd
There will generally be multiple solutions for $u(x)$ that satisfy this bound (and, likely, solutions that saturate it as well). Define ${\bf u}_{\rm max}$ to be the largest value of $u_{\rm max}$ attained by solutions $u(x)$ satisfying the bound \eqref{miaco} (which is possibly just the value of $u_{\rm max}$ that saturates the bound), and define ${\bf M}$ to
be:
\bg\label{Adel}
{\bf M} ~ \equiv ~ {n^2 c_n {\bf u}^n_{\rm max}\over c_m {\bf u}^m_{\rm max}} -4 \nd
It is then easy to see from \eqref{namgoti} that the Nambu--Goto action has the following dominant behavior \cite{mia2, mia3}:
\bg\label{dombe}
d ~ = ~ \lim_{\epsilon \to 0} {2\sqrt{g_n {\bf u}^n_{\rm max}}{\bf u}_{\rm max}\over c_m {\bf u}^m_{\rm max}} ~{{\rm log}~\epsilon\over \sqrt{\bf M}}, ~~~~~
S_{\rm NG} =  \lim_{\epsilon \to 0} {T_0\over \pi} {\sqrt{g_n {\bf u}^n_{\rm max}}\over {\bf u}_{\rm max}} ~{{\rm log}~\epsilon\over \sqrt{\bf M}}\nd
which means that both $d$ and $S_{\rm NG}$ have identical logarithmic behavior. Thus, the finite quantity is the ratio between the two, i.e., the action per unit length, which is 
\bg\label{shoksa}
{S_{\rm NG}\over T_0} ~ \equiv ~ V_{Q\bar{Q}} ~ \equiv \left({c_m {\bf u}^m_{\rm max}\over \pi {\bf u}^2_{\rm max}}\right) d.\nd
Then $V_{Q\bar{Q}}$ is the expected confining potential between the quark and the antiquark. 

Note that while the above computation tells us where to look for the confining string, it doesn't tell us how the string breaks into smaller strings.  For the latter, we would presumably require more machinery, such as string field theory. Furthermore, the result \eqref{shoksa} is independent of the UV completion as the details of the UV 
completion appears on both $d$ and $S_{\rm NG}$ via the coefficient $c_m$ but {\it not} via $b_l$. The details of the full series $\{c_m\}$ appearing in the 
dilaton \eqref{hinsha} are not required to study the IR confining behavior. Even if the UV completion is not made, the result \eqref{shoksa} will continue to hold.

\subsubsection{New exotic states in LST from the gravity dual}

Another interesting direction is to look for exotic states in the dimensionally reduced LST using exotic stable states in the type I and heterotic theories. One candidate 
is the stable non-BPS D0-brane state in type I $SO(32)$ theory that is constructed by switching on a tachyonic kink solution on a D1-anti-D1 
pair in type IIB \cite{sentac}. The
heterotic dual of this construction is the lightest spinor state of the $SO(32)$ theory \cite{sentac}.  Unfortunately, this state was constructed in a flat background, but quantization of open strings in a curved background 
in not well understood, so it could be difficult to properly analyze the spectrum of a type IIB open string with one endpoint on the D1 and the other on the anti-D1. 

What we can try, however, is to go to a regime where the metric is approximately flat.\footnote{For example if we define $R \equiv \int_\infty^u {du\over \sqrt{G(u)}}$ 
where $G(u)$ is as given in \eqref{5ddual}, then in the local coordinate system ($x^{0, 1, 2, 3}, R$) the five-dimensional metric is flat. 
On the other hand, the internal space parametrized by ($\theta_i, \phi_i, \psi$) 
in \eqref{defkoan} however is now much more complicated.}
Thus the configuration of a 
D1-anti-D1 pair with one endpoint at the boundary and the other endpoint at $u \to \infty$ in \eqref{5ddual} (or in \eqref{defkoan}) 
should appear as a stable non-BPS state in the dual gauge theory.  From the 
heterotic $SO(32)$ point of view, we could refine this even further by taking the non-trivial $T^3$-fibration 
of the metric \eqref{defkoan} and looking for the stability 
region of this state on a $T^3$. This has recently been discussed in \cite{seotac}, where the stability region was shown to 
satisfy the inequalities:
\bg\label{nonbps}
16R_a^2 + {1\over R_b^2} + {1\over R_c^2} ~ > ~ 8, ~~~~~~~~~ R_a^2 + R_b^2 + R_c^2 ~ > ~ {1\over 2} , \nd
where $R_a$, $R_b$, and $R_c$, are the radii of the three cycles of $T^3$ measured in units of $\sqrt{\alpha'}$. Thus in our five-dimensional space, 
all regimes where the internal tori parametrized by ($\phi_1, \phi_2, \psi$) satisfy the inequalities \eqref{nonbps}, should have corresponding stable 
non-BPS states that should be visible in the dual gauge theory at all energy scales (including the far IR). The question of what this state corresponds to in the $E_8$ theory 
will be discussed elsewhere.

\section{Discussions and Conclusions}

In this paper, we have identified gravity duals for little string theories that have been dimensionally reduced to four dimensions --- these theories have permanent confinement in the deep infrared. 
Our constructions focused on heterotic $SO(32)$ and $E_8\times E_8$ theories, where we wrapped heterotic NS5-branes on a ${\bf P}^1$ of a non-K\"ahler resolved conifold.  The form of the metric we considered was:
\bg\label{disc1}
&& ds^2 = ds^2_{0123} + {2\Delta \over r \sqrt{a_3}} dr^2 + {2 \Delta \sqrt{a_3}\over r}(d\psi + \cos \theta_1 d\phi_1 + \cos \theta_2 d\phi_2)^2 \nonumber\\
&& ~~~~~~~~~~~~ + (\Delta + a^2)\left(d\theta_1^2 + \sin^2\theta_1 d\phi_1^2\right) + \Delta \left(d\theta_2^2 + \sin^2\theta_2 d\phi_2^2\right)\nd
where $\Delta$ and $a_3$ were determined in {\bf section \ref{sec:hetslns}}, along with the values for the torsion and the dilaton. We also discussed the global 
and local symmetries in {\bf section \ref{sec:hetbundle}}. 

We determined the respective gravity duals through a series of dualities depicted in {\bf figures \ref{hetGT}} and
{\bf \ref{varyTau}}.  The gravity duals take the following form:
\bg\label{disc2} 
ds^2 & = & \textstyle ds^2_{0123} + {2\Delta\over r\sqrt{a_3}} dr^2 + {1\over {\widetilde H}_1 + {\cal A}}\left[d\psi 
+ {\widetilde \Delta}_1 {\rm cos}~\theta_1 \left(d\phi_1 + \alpha_1 d\theta_1\right) 
+ {\widetilde \Delta}_2 {\rm cos}~\theta_2\left(d\phi_2 + \alpha_2 d\theta_2\right)\right]^2 \nonumber\\
&& + \left[g_{\theta_1\theta_1} d\theta_1^2 + g_{\phi_1\phi_1}\left(d\phi_1 + \alpha_1 d\theta_1\right)^2\right]  
+ \left[g_{\theta_2\theta_2} d\theta_2^2 + g_{\phi_2\phi_2}\left(d\phi_2 + \alpha_2 d\theta_2\right)^2\right]\nonumber\\   
&& ~~~~~~~~ + g_{\theta_1\theta_2} d\theta_1 d\theta_2 + g_{\phi_1\phi_2} \left(d\phi_1 + \alpha_1 d\theta_1\right)\left(d\phi_2 + \alpha_2 d\theta_2\right)\nd
where the components of the metric are given in {\bf subsection \ref{sec:hetgravity}}, along with the other background fields. The gravity duals have non-compact two- and 
three-cycles and resemble a non-K\"ahler deformed conifold but with an interesting fibration structure for $\phi_1$ and $\phi_2$.  Also, note that the 
spacetime directions are flat, so the Einstein-frame background does not contain an AdS space (because of nontrivial dilaton).  It 
will be interesting to see if aspects of the linear dilaton background \cite{ahasaab} could be recovered in some regime of parameter space. 

It would also be informative to study a proper decoupling limit of our
backgrounds and check whether the final result will differ significantly
from  
\eqref{disc2}.\footnote{For example one of the simplest decoupling limit generically applied is the so-called 
double scaling limit where $\alpha'\to 0, r \to 0$ such that 
$u \equiv \alpha'/r$ is kept fixed. Using this we can easily check that the form of \eqref{disc2} do not change. However we haven't verified whether the double 
scaling limit is effective in decoupling the bulk degrees of freedom from the brane for our case.}   
One possible ansat\"ze for the five-dimensional part of the dual metric, in the decoupling limit, is:
\bg\label{decmeto}
ds^2 ~=~ ds^2_{0123} + {du^2\over G(u)}\nd
where $u = \alpha'/r$ and $G(u)$ can be extracted from ${2\Delta\over r\sqrt{a_3}}$ in \eqref{disc2}. The various choices of $G(u)$ will determine the gravity duals for 
various configurations of LSTs studied here.  
Note also that the duality sequences given in {\bf figures \ref{hetGT}} and {\bf \ref{varyTau}} are possibly the only way to 
determine the gravity duals for the configurations studied in {\bf subsection \ref{sec:hetslns}}, in part because 
our knowledge of the dual gauge theory side is extremely limited. 
Of course, this means that the gravity dual is particularly useful as a way to discover interesting information about the gauge theory.

One such interesting piece of information is to look for a candidate confining string in the gauge theory, which we discussed in {\bf section \ref{sec:confining}}.  The most natural candidate seems to be the type I fundamental string, which is prone to breaking into smaller strings, similar to the breaking of a confining string into quark-anti-quark pairs.  Our simple analysis showed a 
linear potential between quarks and anti-quarks, but a full understanding of the string breaking remains a challenge. Additionally, it is not clear whether the 
theory that we study using our gravity dual has a simple mass gap, implying that the matter content of the gauge theory is nontrivial, as can also be 
seen from the possible
existence of stable, uncharged, non-BPS states in some region of the moduli space of the gauge theory.  A detailed study of this is left for future work.

Another future direction that could improve our understanding of gauge/gravity duality in the context of heterotic string theory stems from the fact that all the background fluxes are from the NS sector.  Thus, one might hope to gain a better understanding of the backgrounds, and perhaps even the geometric transition itself, through RNS worldsheet techniques, similar to those in \cite{caris,Adams:2006kb,Adams:2009zg}, where a non-perturbative (in $\alpha'$) understanding of a class of non-K\"ahler backgrounds was given.  We hope to return to this in the future.

\vskip.2in

\centerline{\bf Acknowledgments}

\vskip.1in

\noindent  We warmly thank Ori Ganor, Sheldon Katz, and Ashoke Sen for many helpful discussions. 
The work of F.C. is supported in part
by the National Science Foundation under Grant No. NSF PHY11-25915.
The work of  
K.D., J.M.L., and J.S., is supported in part by the National Science and Engineering Research Council of Canada.  The work of R.T. is supported partially by STFC.

\newpage

\appendix

\section{Explicit forms for ${\cal H}$ and $dJ$ \label{app2}}

In this section, we will work out the explicit forms for ${\cal H}$ that we used in earlier sections.  One can also see from here that $dJ \neq 0$. The expression for ${\cal H}$ is given by:
\begin{eqnarray}
{\cal H} &=& - e^{2\phi}\ast d(e^{-2\phi}J)= 2\ast d\phi\wedge J - \ast\, dJ
\end{eqnarray}
The term $\ast\, dJ$ has the following form:
\begin{eqnarray}
\ast\, dJ&=&\frac{(H_{3}c_1)_r+\sqrt{H_1H_2}c_{4\theta_1}}{H_3\sqrt{H_2}c_1}e_3\wedge e_4\wedge e_5
+\frac{(H_{4}c_5)_r+\sqrt{H_1H_2}c_{7\theta_2}}{H_4\sqrt{H_2}c_5}e_5\wedge e_1\wedge e_2\nonumber\\
&& -\frac{(H_{3}c_1)_{\theta_2}}{\sqrt{H_4}H_3c_1}(\cos\psi_2e_4\wedge e_5\wedge e_6+\sin\psi_2e_5\wedge e_6\wedge e_3)\nonumber\\
&& -\frac{(H_1H_2)_{\theta_2}}{2H_1H_2\sqrt{H_4}}(\cos\psi_2e_1\wedge e_2\wedge e_4+\sin\psi_2e_1\wedge e_2\wedge e_3)\nonumber\\
&&+\frac{\sqrt{H_1}c_{4\theta_2}}{\sqrt{H_3H_4}c_1}(\cos\psi_2\sin\psi_1e_2\wedge e_4\wedge e_5-\cos\psi_1\cos\psi_2e_1\wedge e_4\wedge e_5\nonumber\\
&&~~~~~~~~~~~~+\sin\psi_1\sin\psi_2e_2\wedge e_3\wedge e_5-\sin\psi_2\cos\psi_1e_1\wedge e_3\wedge e_5)
\end{eqnarray}
while the first term, involving the dilaton $\phi$, has the following form:
\begin{eqnarray}
\ast d\phi\wedge J&=&\Big(\frac{\phi_{\phi_1}\sin\psi_1}{\sqrt{H_3}c_1}+\frac{\phi_{\theta_1}\cos\psi_1}{\sqrt{H_3}}-\frac{\phi_{\psi}\sin\psi_1c_4}{\sqrt{H_1H_3}c_1}\Big)e_2\wedge (e_5\wedge e_6+e_3\wedge e_4)\nonumber\\
&&+\Big(\frac{\phi_{\phi_1}\cos\psi_1}{\sqrt{H_3}c_1}-\frac{\phi_{\theta_1}\sin\psi_1}{\sqrt{H_3}}-\frac{\phi_{\psi}\cos\psi_1c_4}{\sqrt{H_1H_3}c_1}\Big)e_1\wedge
(e_5\wedge e_6+e_3\wedge e_4)\nonumber\\
&&+\Big(\frac{\phi_{\phi_2}\sin\psi_2}{\sqrt{H_4}c_5}+\frac{\phi_{\theta_2}\cos\psi_2}{\sqrt{H_4}}-\frac{\phi_{\psi}\sin\psi_2c_7}{\sqrt{H_1H_4}c_5}\Big)e_4\wedge
(e_5\wedge e_6+e_1\wedge e_2)\nonumber\\
&&+\Big(\frac{\phi_{\phi_2}\cos\psi_2}{\sqrt{H_4}c_5}-\frac{\phi_{\theta_2}\sin\psi_2}{\sqrt{H_4}}-\frac{\phi_{\psi}\cos\psi_2c_7}{\sqrt{H_1H_4}c_5}\Big)e_3\wedge
(e_1\wedge e_2-e_5\wedge e_6)\nonumber\\
&&+\frac{\phi_{\psi}}{\sqrt{H_1}}e_6\wedge (e_1\wedge e_2-e_3\wedge e_4)+\frac{\phi_r}{\sqrt{H_2}}e_5\wedge (e_1\wedge e_2+e_3\wedge e_4)
\end{eqnarray}
These are the components of ${\cal H}_{NS}$ presented in \eqref{torsa}.

\newpage

\section{Background solutions using geometric transitions}
\label{bocksol}

Our starting solution in the heterotic theory before geometric transition, i.e with wrapped heterotic five-branes, takes the following form:
\begin{eqnarray}\label{mca} 
ds^2=ds_{0123}^2+H_2dr^2+H_1(d\psi+c_4d\phi_1+c_7d\phi_2)^2+H_3(d\theta_1^2+c_1^2d\phi_1^2)+H_4(d\theta_2^2+c_5^2d\phi_2^2)\nonumber\\
\end{eqnarray}
where we haven't taken the decoupling limit. For the duality chain that we will follow, taking the decoupling limit at this stage is not necessary. The metric \eqref{mca} contains the 
backreactions of the wrapped five-branes. The background also has a non-trivial dilaton $\phi$, a three-form ${\cal H}$ given by:
\begin{eqnarray}\label{murga}
{\cal H} &=&(2B_6-A_1)H_4\sqrt{H_1}c_5d\theta_2\wedge d\phi_2\wedge d\psi\nonumber\\ &&-\Big((A_1-2B_6)H_4\sqrt{H_1}c_5c_4\nonumber\\ &&-2(B_1\cos\psi_1+B_2\sin\psi_1)H_4\sqrt{H_3}c_1c_5\Big)d\theta_2\wedge d\phi_2\wedge d\phi_1\nonumber\\ &&-(A_2-2B_6)H_3\sqrt{H_1}c_1d\theta_1\wedge d\phi_1\wedge d\psi\nonumber\\ &&-\Big((A_2-2B_6)H_3\sqrt{H_1}c_1c_7-(A_4 +2B_3\cos\psi_2\nonumber\\ &&-2B_4\sin\psi_2)H_3\sqrt{H_4}c_1c_5\Big)d\theta_1\wedge d\phi_1\wedge d\phi_2\nonumber\\ &&-2\sqrt{H_1H_2H_4}(B_4\cos\psi_2+B_3\sin\psi_2)d\theta_2\wedge (d\psi+c_4d\phi_1+c_7d\phi_2)\wedge dr\nonumber\\ &&-2\sqrt{H_1H_2H_4}(B_4\sin\psi_2+B_3\cos\psi_2)c_5d\phi_2\wedge (d\psi+c_4d\phi_1)\wedge dr\nonumber\\ &&-2\sqrt{H_1H_2H_3}(B_1\sin\psi_1-B_2\cos\psi_1)d\theta_1\wedge (d\psi+c_4d\phi_1+c_7d\phi_2)\wedge dr\nonumber\\ &&+2\sqrt{H_1H_2H_3}(B_1\cos\psi_1-B_2\sin\psi_1)c_1d\phi_1\wedge (d\psi+c_7d\phi_2)\wedge dr\nonumber\\ &&-2H_3\sqrt{H_4}(B_3\cos\psi_2-B_3\sin\psi_2)c_1d\theta_1\wedge d\theta_2\wedge d\phi_1\nonumber\\ &&-2H_4\sqrt{H_3}(B_1\sin\psi_1-B_2\cos\psi_1)c_1d\theta_1\wedge d\theta_2\wedge d\phi_2\nonumber\\ &&+B_5\sqrt{H_2}dr\wedge (c_1 d\theta_1\wedge d\phi_1-c_5 d\theta_2\wedge d\phi_2)\nonumber\\&&-\frac{H_1c_5c_{4\theta_2}}{c_1}d\theta_1\wedge d\phi_2\wedge (d\psi+c_4d\phi_1+c_7d\phi_2)\end{eqnarray}
where the coefficients $A_i$'s and $B_i$'s appearing in \eqref{murga} are given in \eqref{torsacoef} and \eqref{torsacoef2}, and a non-abelian vector bundle $F$, whose simple $U(2)$ form is given as:
\begin{eqnarray}\label{takla}
F=(\mathfrak{f}_1e_1\wedge e_2+\mathfrak{f}_2e_3\wedge e_4+\mathfrak{f}_3e_5\wedge e_6)I+(\mathfrak{f}_4e_1\wedge e_2+\mathfrak{f}_5e_3\wedge e_4)\sigma^1
\end{eqnarray}
where $\sigma^1$ is the first Pauli matrix and $\mathfrak{f}_i$'s are given in \eqref{joishila}. One may note that a simplified form of the above background can be 
presented as \eqref{fulbag} where the coefficients of the metric \eqref{mca}, the three-form flux \eqref{murga} and the dilaton $\phi$ take a specific form. The values of
these coefficients can be inserted in \eqref{joishila} to determine the vector bundle for the background \eqref{fulbag}.

For the simple case of \eqref{fulbag}, one may determine the metric for the gravity dual in the heterotic theory. The full set of transformations and the duality chain
required to describe the final result is depicted in details in the text. The result is:
\bg\label{jacsin}
ds^2 & = & \textstyle ds^2_{0123} + {2\Delta\over r\sqrt{a_3}}dr^2 + {1\over {\widetilde H}_1 + {\cal A}}\left[d\psi 
+ {\widetilde \Delta}_1 {\rm cos}~\theta_1 \left(d\phi_1 + \alpha_1 d\theta_1\right) 
+ {\widetilde \Delta}_2 {\rm cos}~\theta_2\left(d\phi_2 + \alpha_2 d\theta_2\right)\right]^2 \nonumber\\
&& + \left[g_{\theta_1\theta_1} d\theta_1^2 + g_{\phi_1\phi_1}\left(d\phi_1 + \alpha_1 d\theta_1\right)^2\right]  
+ \left[g_{\theta_2\theta_2} d\theta_2^2 + g_{\phi_2\phi_2}\left(d\phi_2 + \alpha_2 d\theta_2\right)^2\right]\nonumber\\   
&& ~~~~~~~~ + g_{\theta_1\theta_2} d\theta_1 d\theta_2 + g_{\phi_1\phi_2} \left(d\phi_1 + \alpha_1 d\theta_1\right)\left(d\phi_2 + \alpha_2 d\theta_2\right)\nd
Note that there is no coefficient in front of the space-time part. The coefficient, which appears in \eqref{resanswer}, is in fact the dilaton 
in the type I/heterotic frame and is related to the dilaton $\phi_{(c)}$ in type I$'$ frame by mirror/U-duality transformation. Therefore in the heterotic frame 
we see no factor of dilaton in front of the space-time part.

Our claim then is that the gravity dual of the wrapped heterotic five-branes in the heterotic $E_8 \times E_8$ theory can be extracted from 
\eqref{jacsin} by performing a geometric transition 
on the metric \eqref{mca} or, in the simplified form, \eqref{mopla}. 
In the decoupling limit of \eqref{mopla}, \eqref{jacsin} will provide the precise gravity dual. Defining $u = {\alpha'\over r}$, 
we see that the five-dimensional part of \eqref{jacsin} takes the following form:
\bg\label{bhayanak}
ds^2 ~ = ~ ds^2_{0123} + {du^2\over G(u)}\nd
where $G(u)$ can be extracted from ${2\Delta\over r\sqrt{a_3}}$ in \eqref{jacsin} in the decoupling limit. The spacetime part is flat, 
so we conclude that, in the string frame, a flat 
Minkowski space is dual to LST compactified to 4 dimensions on a two-cycle of 
the non-K\"ahler resolved conifold. Additionally, irrespective of the details of the internal space, 
the five-dimensional dual metric will always be of the form \eqref{bhayanak}.


\newpage
\section{The torsional connection $\Omega^m_{np}$}
 \label{app3}

The connection that we use throughout this paper is the so-called $+$-connection, constructed from the Christoffel symbol and ${\cal H}$ as:
\bg\label{tgop}
\Omega_{+\mu\nu}^\rho \equiv \Omega_{\mu\nu}^\rho = \Gamma^\rho_{\mu\nu} + {1\over 2} {\cal H}^\rho_{\mu\nu} . \nd
The various components of $\Omega_{\mu\nu}^\rho$ are given as follows:
\bg\label{t1}
&&\Omega^r_{\psi\psi}=-\frac{H_1'}{2 \sqrt{H_2} \sqrt{H_1}}, \qquad  \Omega^{\psi}_{\psi r}=\frac{H_1'}{2 \sqrt{H_2} \sqrt{H_1}}, \qquad  
\Omega^{\phi_1}_{\psi\theta_1}=\frac{1}{2} B \sqrt{H_1}-\frac{H_1}{2 H_3}\nonumber\\
&&\Omega^{\phi_2}_{\psi\theta_2}=\frac{1}{2} A \sqrt{H_1}-\frac{H_1}{2 H_4}, \qquad  \Omega^{\theta_1}_{\psi\phi_1}=\frac{H_1-B \sqrt{H_1} H_3}{2 H_3}, \qquad  
\Omega^{\theta_2}_{\psi\phi_2}=\frac{H_1-A \sqrt{H_1} H_4}{2 H_4}\nonumber\\
&& \Omega^r_{\phi_1\psi}=-\frac{\cos\theta_1 H_1'}{2 \sqrt{H_2} \sqrt{H_1}}, \qquad  \Omega^r_{\phi_1\phi_1}=-\frac{\sin\theta_1 H_3'}{2 \sqrt{H_2} \sqrt{H_3}}, \qquad  
\Omega^{\psi}_{\phi_1 r}=\frac{\cos\theta_1 H_1'}{2 \sqrt{H_2} \sqrt{H_1}}\nonumber\\
&& \Omega^\psi_{\phi_1\theta_1}=-\frac{\left(\sqrt{H_1}+B H_3\right) \sin\theta_1}{2
\sqrt{H_3}}, \qquad  \Omega^\psi_{\phi_1 r}=\frac{\sin\theta_1 H_3'}{2 \sqrt{H_2} \sqrt{H_3}}\nonumber\\
&& \Omega^{\phi_1}_{\phi_1\theta_1}=\frac{1}{2} \cos\theta_1 \left(2+B \sqrt{H_1}-\frac{H_1}{H_3}\right), \qquad  
\Omega^{\phi_2}_{\phi_1\theta_2}=\frac{1}{2} A \cos\theta_1 \sqrt{H_1}-\frac{\cos\theta_1
H_1}{2 H_4}\nonumber\\
&&\Omega^{\theta_1}_{\phi_1\psi}=\frac{\left(\sqrt{H_1}+B H_3\right) \sin\theta_1}{2
\sqrt{H_3}}, \qquad  \Omega^{\theta_1}_{\phi_1\phi_1}=\frac{\cos\theta_1 \left(H_1-2 H_3-B \sqrt{H_1}
H_3\right)}{2 H_3}\nonumber\\
&& \Omega^{\theta_2}_{\phi_1\phi_2}=-\frac{1}{2} A \cos\theta_1 \sqrt{H_1}+\frac{\cos\theta_1
H_1}{2 H_4}, \qquad  \Omega^r_{\phi_2\psi}=-\frac{\cos\theta_2 H_1'}{2 \sqrt{H_2} \sqrt{H_1}}\nonumber\\
&& \Omega^r_{\phi_2\phi_2}=-\frac{\sin\theta_2 H_4'}{2 \sqrt{H_2} \sqrt{H_4}}, \qquad  \Omega^\psi_{\phi_2r}=\frac{\cos\theta_2 H_1'}{2 \sqrt{H_2} \sqrt{H_1}}, \qquad  
\Omega^{\psi}_{\phi_2\theta_2}=-\frac{\left(\sqrt{H_1}+A H_4\right) \sin\theta_2}{2\sqrt{H_4}}\nonumber\\
&& \Omega^{\phi_1}_{\phi_2\theta_1}=\frac{1}{2} B \cos\theta_2 \sqrt{H_1}-\frac{\cos\theta_2 H_1}{2 H_3}, \qquad  
\Omega^{\phi_2}_{\phi_2 r}=\frac{\sin\theta_2 H_4'}{2 \sqrt{H_2} \sqrt{H_4}}, \nonumber \\
&&\Omega^{\phi_2}_{\phi_2\theta_2}=\frac{1}{2} \cos\theta_2 \left(2+A \sqrt{H_1}-\frac{H_1}{H_4}\right) , \qquad \Omega^{\psi}_{\theta_1\phi_1}=\frac{\sqrt{H_1}+B H_3}{2 \sqrt{H_3}}\nonumber\\
&& \Omega^{\theta_1}_{\phi_2\phi_1}=-\frac{1}{2} B \cos\theta_2 \sqrt{H_1}+\frac{\cos\theta_2 H_1}{2 H_3}, \qquad  \Omega^{\theta_2}_{\phi_2\psi}
=\frac{\left(\sqrt{H_1}+A H_4\right) \sin\theta_2}{2\sqrt{H_4}}\nonumber\\
&& \Omega^{\theta_2}_{\phi_2\phi_2}=\frac{\cos\theta_2 \left(H_1-2 H_4-A \sqrt{H_1}
H_4\right)}{2 H_4}, \qquad  \Omega^r_{\theta_1\theta_1}=-\frac{H_3'}{2 \sqrt{H_2} \sqrt{H_3}}  \nonumber\\
&& \Omega^{\phi_1}_{\theta_1\psi}=\frac{-\sqrt{H_1}-B H_3}{2 \sqrt{H_3}}, \qquad  \Omega^{\theta_1}_{\theta_1 r}=\frac{H_3'}{2 \sqrt{H_2} \sqrt{H_3}}, \qquad  
\Omega^r_{\theta_2\theta_2}=-\frac{H_4'}{2 \sqrt{H_2} \sqrt{H_4}}\nonumber\\
&& \Omega^{\psi}_{\theta_2\phi_2}=\frac{\sqrt{H_1}+A H_4}{2 \sqrt{H_4}}, \qquad  \Omega^{\phi_2}_{\theta_2\psi}=\frac{-\sqrt{H_1}-A H_4}{2 \sqrt{H_4}}, \qquad  
\Omega^{\theta_2}_{\theta_2 r}=\frac{H_4'}{2 \sqrt{H_2} \sqrt{H_4}}\nd

\newpage

\section{Proofs for eq. \eqref{dilh1} and eq. \eqref{a3ans} \label{proof}}

Let us now clarify the functional forms for the dilaton and $h_1$ that we used in \eqref{dilh1}.  In the process we will also justify \eqref{a3ans}. 
 
A simple dimensional analysis will tell us that all three warp factors 
in \eqref{manipul}, namely $h_1a_3$, $a_2$, and $h_1 a_5$, are proportional 
to the square of the radius $r^2$.\footnote{Recall that we are working in the limit where $h_2 = 1$, unless mentioned otherwise. Of course, the
dimensional analysis remains unaffected even for non-constant $h_2$.} 
Additionally, the generic form for the 
heterotic metric \eqref{manipul} should be:\footnote{We are following the convention of \cite{DRS} for assigning the dilaton factor in the metric.  
Other conventions are simply related to ours by a redefinition of the dilaton.  See also {\bf footnote \ref{convention}}. Alternatively we can write 
the internal part of the metric in \eqref{gfhet} in the standard way as $e^{2\phi} d{\widetilde s}^2$ where $d{\widetilde s}^2 = 
e^{-\phi} {ds}^2_{\rm internal}$ implying that the internal part of the metric is intrinsically non-K\"ahler. This way even in the regime where the effect of the 
wrapped five-branes is negligible, one would require non-zero torsion to balance the non-K\"ahlerity of the metric. This is indeed the case as will be seen from the 
torsion plots depicted in {\bf figures \ref{Hcase1tot}, \ref{Hcase2}} and {\bf \ref{Hcase3}}.\label{consconv}}
\bg\label{gfhet} ds^2 = ds^2_{0123} +   e^{\phi} ds^2_{\rm internal}\nd
where the five-branes wrap the spacetime $x^{0, 1, 2, 3}$ and the two-cycle ($\theta_1, \phi_1$); and  
$ds^2_{\rm internal}$ is the internal non-K\"ahler resolved conifold metric that we will typically demand takes the following form:
\bg\label{rescdema}
 ds^2_{\rm internal} \equiv && {dr^2\over {\cal F}_3(r)} +  {r^2 {\cal F}_3(r)\over 4} (d\psi + {\rm cos}~\theta_1 d\phi_1 + {\rm cos}~\theta_2 d\phi_2)^2\nonumber\\
&& ~~~~~~~~~ + {r^2 + a^2e^{-\phi} \over 4}\left(d\theta_1^2 + {\rm sin}^2\theta_1 d\phi_1^2\right) + {r^2\over 4} \left(d\theta_2^2 + {\rm sin}^2\theta_2 d\phi_2^2\right)\nd
where ${\cal F}_3(r)$ will turn out to be ${2\sqrt{a_3}\over r}$, a fact that will be justified below. Thus if we isolate the metric of the two-sphere on which we will have
the wrapped five-branes from \eqref{rescdema}, 
we will get the metric of the four-dimensional internal space which is locally an ALE space.
This means, looking at \eqref{manipul} and \eqref{akhta}, that we find $h_1 a_5 = \Delta$, so using 
\eqref{gfhet} and dimensional analysis, the dilaton becomes:
\bg\label{dilbeco} e^\phi ~=~ {4\Delta\over r^2}  + {\cal O}(a^2)\nd
where we have inserted a factor of 4 just for convenience to comply with the non-K\"ahler resolved conifold metric \eqref{rescdema} 
that we consider here.  Interestingly, one may also verify \eqref{dilbeco} using the torsion classes that we derived earlier. Using the normalization of the 
dilaton factor that we consider here, the relation is $d\phi = 2W_4$, as discussed in {\bf footnote \ref{convention}}. 
Thus from the $W_4$ computed in \eqref{torkada}, we get the following equation for the 
dilaton:
\bg\label{dilkhush}
\phi(r) = {1\over 2}\int dr \left(1+{\Delta\over \Delta + a^2}\right) {d\over dr} {\rm log}\left({4\Delta\over r^2}\right), ~~~~~~ r > 0 \nd
In the limit that $a^2$ is much smaller than other scales, we immediately reproduce \eqref{dilbeco}. For finite $a^2$, \eqref{dilkhush} can give us the correction to the 
dilaton due to the resolution parameter of the underlying non-K\"ahler resolved conifold metric.

From the dimensional analysis,
since the dilaton \eqref{dilbeco} is the simplest possible ansatz, one might ask whether a more generic ansatz, with a dimensionless function $f_0(r)$, could 
be envisioned for the dilaton, i.e.:
\bg\label{dildeke}
e^\phi = {4\Delta \over r^2 f_0(r)}, ~~~~~~~~~ f_0(r) \equiv 1 + \sum_{n \ge 1} c_n \left({r\over \sqrt{\alpha'}}\right)^n + 
\sum_{m \ge 1} d_m \left({\sqrt{\alpha'}\over r}\right)^m . \nd
Without doing any computations, we might say that $c_n = 0$ because for large $r$, where the effect of the wrapped heterotic five-branes is negligible, we expect 
the standard metric of a resolved conifold. The rest of $f_0$ could then be absorbed into the definition of $\Delta$, implying that the form of the dilaton \eqref{dilbeco}
is generic enough. The above conclusions are {\it almost} true, but we will consider a case where positive powers of $r$ will sum up asymptotically to yield a 
result that will behave as {\it inverse} powers of $r$ as $r$ tends to a large value. Furthermore, the underlying manifold may not exactly become a Calabi--Yau resolved 
conifold at large $r$. Despite these subtleties, 
we can generically absorb $f_0(r)$ into the definition of the dilaton as long 
as we have good asymptotic behavior. This is also consistent with the torsion class analysis \eqref{dilkhush}.    

The dimensional analysis now tells us that the coefficient $a_2$ in \eqref{manipul} should be $a_2 \equiv {e^\phi r^2\over 4} + a^2$, which is of course consistent with 
$a_2 = \Delta + a^2$ in \eqref{bgmetDf}. The other two coefficients in \eqref{manipul} now take the form:
\bg\label{otcoef2} h_1 a_3 ~ = ~  r^2 e^\phi f_1(r), ~~~~~~~~~~~ h_1 ~ = ~ e^\phi f_2(r) \nd
where $f_1(r)$ and $f_2(r)$ are dimensionless functions of $r$. 
Only $a_3$ could carry a dimensionful factor of $r^2$. The question now is: how are $f_1$ and $f_2$ related?  

To find the relation between $f_1$ and $f_2$, let us go back to the metric ansatz \eqref{alu} where $a_3$ first appears. Dimensionally, $a_3$ should be $r^2$, so 
we can only attach a dimensionless factor to it. Motivated by the form of the other term in \eqref{alu}, let us then make the following choice for $a_3$:
\bg\label{a3ans2}
a_3 ~ \equiv ~ {4r^2f_1^2} \nd
where again we put a factor of $4$ for later convenience. Plugging this into \eqref{otcoef2}, we immediately get the required value for $h_1$ in \eqref{dilh1} and 
\bg\label{f2def} f_1 = {1\over 4 f_2} = {\sqrt{a_3}\over 2 r} . \nd
Thus everything can be expressed either in terms of $\Delta$ or $a_3$ or both, provided \eqref{a3ans2} holds. Of course, this is not the only ansatz since 
we can always express $a_3$ as a function proportional to $r^2 f_1^n$ or to $r^2 g_1^m$ with ($n, m$) being any number (integer or fractional). However in this paper
we will stick with the simplest choice \eqref{a3ans2}. 

\newpage

 \section{Supersymmetry Condition Revisited}
\label{conventioncheck}

 The supersymmetry condition for the conventions in this paper is \eqref{suco}, i.e.,
 $2W_4 = W_5$, as our choice of the dilaton is minus the choice of the dilaton in \cite{carluest2}. 

However it is interesting to see, just for the sake of curiosity, 
how the backgrounds change if we taken the supersymmetry condition to be $2W_4 = - W_5$, without worrying about the sign of the dilaton. To be 
 concrete, let us just consider the scenario depicted in case I. The equation for $a_3$ changes from \eqref{difsim} to the following:
\bg\label{difsimnow}
{da_3\over dr} - {\sqrt{a_3}\over 4} + {5a_3\over 12 r}\left({17 r^2 + 8e_0\over r^2 + e_0}\right) + {\cal O}(a^2) = 0 \nd
This can again be exactly solved, and the solution for $a_3$ will become:
\bg\label{solfa3}
a_3(r) = \left({3\over 6649}\right)^2 {\left[45 e_0^2 + 106 e_0 r^2 + 61 r^4 - 45 e_0^2 \left({e_0 + r^2 \over e_0}\right)^{1/16} {\cal J}_{1,2}\right]^2 \over 
r^2(e_0 + r^2)^2}\nd
which can be compared to \eqref{a3r1}. The hypergeometric function now, denoted ${\cal J}_{1,2}$, is:
\bg\label{hyperbeta}
{\cal J}_{1,2} \equiv {}_2F_1\left({1\over 16}, {1\over 3}, {4\over 3}, -{r^2\over e_0}\right) . \nd
The behavior of $a_3$ at the origin and at infinity 
 is:
 \bg\label{dhongsho}
 &&a_3\vert_{r\to 0}~ = ~ {9r^2\over 4096} - {405 r^4\over 229376 e_0} + {\cal O}(r^6), ~~~~
 a_3\vert_{r\to \infty} ~ = ~ {9r^2\over 11881} + {810 e_0\over 724741} + {\cal O}\left({1\over r^2}\right)\nd

 \begin{figure}[htb]
 \begin{center}
 \includegraphics[width=0.5\textwidth]{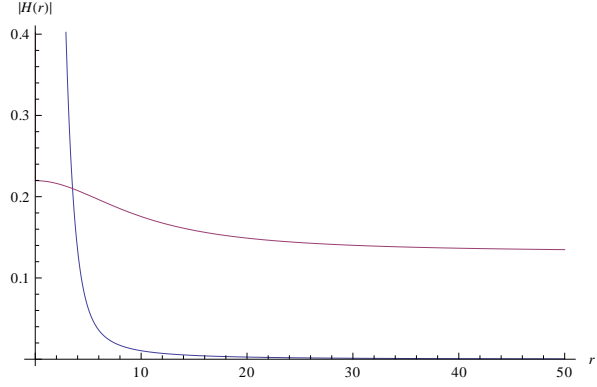}
 \end{center}
 \caption{{The plot of the torsion coefficient $\vert H(r)\vert$ for $r\to 0$ and the
 intermediate region. Near the origin the torsion blows up, signifying the presence of the 
 sources, as depicted by the blue curve using \eqref{difsimchangee}. For large $r$ the story is captured by the red curve plotted using \eqref{difsimnow}.
 Notice that for $r \to \infty$ the torsion becomes finite to maintain zero energy in the system \eqref{potcar} as for the three cases studied earlier.
 We have chosen $e_0 = 100$ in units of $\alpha'$.}}
 \label{Hcase4}
 \end{figure}

Having obtained $a_3$, we can now work out $\Delta$ from the warp factor. This is easy to write down since $a_3$ is expressible again as a square (similar to the 
 scenario we encountered in \eqref{delu}). 
 The functional form for 
 $\Delta$ then is:
 \bg\label{dhono}
 \Delta(r) ~ = ~ {3\over 13298 r^2} \left[45 e_0^2 + 106 e_0 r^2 + 61 r^4 - 45 e_0^2 \left({e_0 + r^2 \over e_0}\right)^{1/16} {\cal J}_{1,2}\right]\nd
 which immediately tell us us that the boundary values for this will be:
 \bg\label{dhonne}
 \Delta\vert_{r\to \infty} ~ = ~ {3r^2\over 218} + {159 e_0\over 6649} + {\cal O}\left({1\over r^2}\right), ~~~~~~~~ 
 \Delta\vert_{r\to 0} ~ = ~ {3e_0\over 128} + {201 r^2 \over 14336} + {\cal O}(r^4) \nd
 consistent with the parallel story in \eqref{delu}, although the numerical coefficients are different from \eqref{bval}. 

 The picture is not complete until we figure out the $r\to 0$ behavior as the small $r$ behavior in \eqref{dhonne} doesn't take into account the subtlety 
 that the wrapped five-branes behave like three-branes. Therefore
 as before, near the origin, we will assume that the warp factor is given by $1+\alpha'e_0/r^4$ 
 with $\alpha' =1$. The $a_3$ equation now changes from \eqref{difsimnow} to
 \bg\label{difsimchangee}
 {da_3\over dr} - {\sqrt{a_3}\over 4} + {5a_3\over 12 r}\left({17 r^4 - e_0\over r^4 + e_0}\right) + {\cal O}(a^2) = 0 \nd
 mirroring the change from \eqref{difsim} to \eqref{case1change}. The solution for $a_3$ now is similar to \eqref{solfa3}:
 \bg\label{a3bonda}
 a_3(r) ~ = ~ \left({3r\over 2071}\right)^2 {\left[19(e_0 + r^4) + 90 e_0 \left(1+{r^4\over e_0}\right)^{1/16} \widetilde{\cal J}_{1,2}\right]^2\over 
 (e_0 + r^4)^2}\nd 
 with $ \widetilde{\cal J}_{1,2}$ being another hypergeometric function. On the other hand $\Delta$ becomes:
 \bg\label{churaman}
 \Delta(r) ~ = ~ \left({3\over 4142 r}\right)\left[19(e_0 + r^4) + 90 e_0 \left(1+{r^4\over e_0}\right)^{1/16} \widetilde{\cal J}_{1,2}\right]\nd
 which is similar to \eqref{dhono}, as one might have expected.
 We now only care about the $r\to 0$ behaviors of $a_3$ and $\Delta$ since the large $r$ behavior should be captured by \eqref{solfa3} and 
 \eqref{dhono}, respectively. They are given by:
 \bg\label{rto0be}
 a_3 ~ \to ~ {9r^2\over 361} - {324 r^6\over 8303 e_0} + 
 {2142936 r^{10}\over 40294459} + {\cal O}(r^{14}), ~~~~~~ 
 \Delta ~ \to ~ {3 e_0\over 38 r^2} + {15 r^2\over 874} 
 -{162 r^6\over 92207 e_0} + {\cal O}(r^{10})\nonumber\\ \nd
 One final computation needed to complete the story is to determine the torsions ${\cal H}$ for both the cases \eqref{difsimnow} and \eqref{difsimchangee}. Near $r = 0$, the torsion
 is given by:
 \bg\label{hnear0}
{\cal H} ~  & = & \textstyle {9\over 34312328 e_0 r^2}\left({e_0\over e_0 + r^4}\right)^{31/16}\Bigg[ 361 e_0^2 \left(1 + {r^4\over e_0}\right)^{31/16} + 
 285\left(121 e_0^2 + 230 e_0 r^4 + 109 r^8\right) \widetilde{\cal J}_{1,2} \nonumber\\
 && ~~~~~~~~ + 1350 e_0 \left(1 + {r^4\over e_0}\right)^{1/16} (115 e_0 + 109 r^4) \widetilde{\cal J}^2_{1,2}\Bigg] \left(\Omega_1 + \Omega_2\right) \wedge e_\psi. \nd
 This is plotted as the blue curve in 
 {\bf figure \ref{Hcase4}}. The behavior is no different from the three cases studied earlier. This confirms the statement that we made earlier: we can go from
 one convention to another by redefinitions of the variables in the problem.

\bibliographystyle{JHEP}
\bibliography{SWcurve}

\end{document}